\documentclass[sigconf]{acmart}

\usepackage{
  color,
  float,
  epsfig,
  wrapfig,
  graphics,
  graphicx
}
\usepackage{bm}

\usepackage{amssymb}
\usepackage{amsmath}

\let\subfigure\relax

\usepackage[ruled]{algorithm2e}

\usepackage{appendix}
\usepackage{xparse} 
\usepackage{xspace}

\usepackage{makecell}
\usepackage{multirow}
\usepackage{hyperref}
\usepackage{balance}
\usepackage{subfigure}
\usepackage{graphicx}
\usepackage{dblfloatfix}

\renewcommand\appendix{\par
    \setcounter{section}{0}
    \setcounter{subsection}{0}
    \gdef\thesection{Appendix. \\ \Alph{section}}}

\newcommand{\name}{WiKI-Eve\xspace}
\newcommand{\model}{\textsf{o}-IKI\xspace}

\ifodd 0
\newcommand{\newrev}[1]{{\color{black}#1}} 
\newcommand{\needrev}[1]{{\color{black}#1}} 
\newcommand{\hjy}[1]{{\color{blue}#1}} 
\else

\newcommand{\newrev}[1]{#1} 
\newcommand{\needrev}[1]{#1} 
\newcommand{\hjy}[1]{#1} 
\fi

\def\BibTeX{{\rm B\kern-.05em{\sc i\kern-.025em b}\kern-.08em
    T\kern-.1667em\lower.7ex\hbox{E}\kern-.125emX}}

\copyrightyear{2023}
\acmYear{2023}
\setcopyright{rightsretained}
\acmConference[CCS'23]{Proceedings of the 2023 ACM SIGSAC Conference on Computer and Communications Security}{November 26--30, 2023}{Copenhagen, Denmark}
\acmBooktitle{Proceedings of the 2023 ACM SIGSAC Conference on Computer and Communications Security (CCS'23), November 26--30, 2023, Copenhagen, Denmark}\acmDOI{10.1145/3576915.3623088}
\acmISBN{979-8-4007-0050-7/23/11}

\begin{document}
\title{Password-Stealing without Hacking: Wi-Fi Enabled Practical Keystroke Eavesdropping}

\ifodd 1
\author{Jingyang Hu}
\authornote{Both authors contributed equally to this research; this work is done when Jingyang Hu works as a CSC visiting scholar at Nanyang Technological University (NTU).}
\affiliation{%
  \institution{Hunan University}
  \country{China}
}
  \email{fbhhjy@hnu.edu.cn}
\author{Hongbo Wang}
\authornotemark[1]
\affiliation{%
  \institution{Nanyang Technological University}
  \country{Singapore}
}
  \email{hongbo001@ntu.edu.sg}
\author{Tianyue Zheng}
\affiliation{%
  \institution{Nanyang Technological University}
  \country{Singapore}
}
  \email{tianyue002@ntu.edu.sg}
\author{Jingzhi Hu}
\affiliation{%
  \institution{Nanyang Technological University}
  \country{Singapore}
}
  \email{jingzhi.hu@ntu.edu.sg}
\author{Zhe Chen}
\affiliation{%
  \institution{Fudan University}
  \country{China}
}
  \email{zhechen@fudan.edu.cn}
\author{Hongbo Jiang}
\affiliation{%
  \institution{Hunan University}
  \country{China}
}
  \email{hongbojiang@hnu.edu.cn}
\author{Jun Luo}
\affiliation{%
  \institution{Nanyang Technological University}
  \country{Singapore}
}
\email{junluo@ntu.edu.sg}
\renewcommand{\shortauthors}{Jingyang Hu, Hongbo Wang, Tianyue Zheng, Jingzhi Hu, ... \& Jun Luo}
\else
\author{{\Large
    Jingyang Hu$^{1*}$\quad Hongbo Wang$^{2*}$\quad  Tianyue Zheng$^{2}$\quad Jingzhi Hu$^{2}$\quad Zhe Chen$^{3}$\quad Hongbo Jiang$^{1}$\quad Jun Luo$^{2}$
    }}
    \thanks{*~Both authors contributed equally to this research; this work is done when Jingyang Hu works as a CSC visiting scholar at NTU\vspace{-1ex}}
\affiliation{
   \institution{{\normalsize
        $^1$College of Computer Science and Electronic Engineering, Hunan University\country{China} \\
	$^2$School of Computer Science and Engineering, Nanyang Technological University (NTU) \country{Singapore} \\
	$^3$Intelligent Networking and Computing Research Center and School of Computer Science, Fudan University \country{China} \\
    Email: \{fbhhjy, hongbojiang\}@hnu.edu.cn,~~~ \{hongbo001, tianyue002, jingzhi.hu, junluo\}@ntu.edu.sg,~~~ zhechen@fudan.edu.cn
        }}
    }
    \renewcommand{\authors}{J. Hu, H. Wang, T. Zheng, J. Hu, Z. Chen, H, Jiang, and J. Luo}
    \renewcommand{\shortauthors}{J. Hu, H. Wang, T. Zheng, J. Hu, Z. Chen, H, Jiang, and J. Luo}
\fi 

\begin{abstract}
%

The contact-free sensing nature of Wi-Fi has been leveraged to achieve privacy breaches, yet existing attacks relying on Wi-Fi CSI (\textit{channel state information}) demand hacking Wi-Fi hardware
to obtain desired CSIs. Since such hacking
has proven prohibitively hard due to compact hardware, its feasibility in
keeping up with fast-developing Wi-Fi technology
becomes very questionable. 
To this end, we propose \name to eavesdrop keystrokes on smartphones without the need for hacking. \name exploits a new feature, BFI (\textit{beamforming feedback information}), offered by latest Wi-Fi hardware: since BFI is transmitted from a smartphone to an AP in clear-text, it can be overheard (hence eavesdropped) by any other Wi-Fi devices switching to \textit{monitor} mode. 
As existing keystroke inference methods offer very limited generalizability,
\name further innovates in an adversarial learning scheme to enable its inference generalizable towards unseen scenarios.
We implement \name and conduct extensive evaluation on it;
the results demonstrate that \name achieves 88.9\% inference accuracy for individual keystrokes and up to 65.8\% top-10 accuracy
for stealing passwords of mobile applications (e.g., WeChat).
%
\end{abstract}



\maketitle

\section{Introduction}
\label{sec:intro}
%

Mobile devices (e.g., smartphones and tablets), along with their software applications, have been increasingly adopted to identify human users in modern societies~\cite{cover2015digital, id4d}. Consequently, stealing passwords from these devices becomes almost like \textit{identify theft}, hence attracting diversified eavesdropping attacks, either \textit{direct}~\cite{maggi2011fast, yue2014my} or \textit{indirect}~\cite{ali2015keystroke, liu2015snooping, zhou2018patternlistener, shukla2014beware, yue2014blind, liu2015good, marquardt2011sp}. Bearing no need to have a visual on the target screen, the indirect attacks often incur a much higher risk as they leverage side-channels to infer passwords in a stealthy manner.
Typical side-channels considered by prior works include acoustic~\cite{liu2015snooping, zhou2018patternlistener}, electromagnetic emission~\cite{jin2021periscope}, indirect vision~\cite{shukla2014beware, yue2014blind, sun2016visible, chen2018eyetell}, and motion sensors~\cite{liu2015good, marquardt2011sp, cai2011touchlogger}. Though they have demonstrated successes for particular scenarios, these successes often rely on strong assumptions~\cite{li2016csi}, including i) eavesdropping devices are in proximity to \needrev{the victim device} (e.g., in centimeter scale or within line-of-sight region)~\cite{liu2015snooping, liu2015good, marquardt2011sp, jin2021periscope, shukla2014beware, chen2018eyetell}, ii) rogue software have been implanted to \needrev{the victim device}~\cite{liu2015good, cai2011touchlogger, zhou2018patternlistener}, and iii) the content to eavesdrop has linguistic structure~\cite{ali2015keystroke, marquardt2011sp, yue2014blind, liu2015good}.
\begin{figure}[t]
\centerline{\includegraphics[width=0.8\columnwidth]{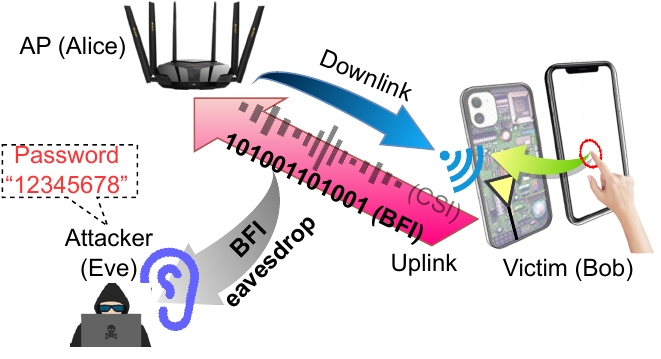}}
\caption{Vision of \name: eavesdropping clear-text BFI (representing downlink channel states) transmitted to the AP, Eve can readily infer the Bob's password typing that physically ``hits'' the Wi-Fi channel.}
\label{fig:teaser}
\end{figure}

Among all side-channels, Wi-Fi CSI (\textit{channel state information}) stands out as it appears to be void of all aforementioned weaknesses~\cite{li2016csi, yang2022wink}. Essentially, since keystrokes affect wireless channels as shown in Figure~\ref{fig:teaser}, the ``twisted'' CSIs can be used to infer individual keys involved in typing a password. The practical significance of this type of attack is also backed by the wide adoption of Wi-Fi infrastructure and extensive reach of Wi-Fi signals (thus CSIs).
Nonetheless, this seemingly plausible attack actually bears one fatal issue: though CSI was hacked\footnote{\hjy{Instead of high-level software hacking, here we refer to a \textit{low-level hacking}, including firmware patching~\cite{nexmon:project} and driver modification~\cite{jiang2021eliminating}, on Wi-Fi hardware.}} from Wi-Fi hardware more than a decade ago~\cite{csi}, only a handful of such hardware have been hacked by far while Wi-Fi standards/technologies are constantly getting upgraded every two or three years.\footnote{As a matter of fact, most research proposals driven by Wi-Fi CSIs are still leveraging the 15-year-old Wi-Fi~4 hardware~\cite{5300}.} Therefore, it is highly questionable if CSI-based side-channel attacks are able to keep up with the technology developments, hence our passwords appear to remain secure.

Unfortunately, 
technology developments of Wi-Fi also introduce new vulnerability, as new Wi-Fi hardware (starting from Wi-Fi~5~\cite{survivalguide}) piggybacks BFI (\textit{beamforming feedback information}), a compressed \textit{digital} version of \textit{analog} CSI, in clear-text onto control frames. Basically, BFIs are used to feed downlink channel states back to an access point (AP), for the sake of guiding AP beamforming~\cite{5677290}. Though they only account for part of the downlink CSIs concerning the AP side, 
the fact that on-screen typing directly impacts the Wi-Fi antennas (hence channels) right behind the screen (see Figure~\ref{fig:teaser}) allows BFIs to contain sufficient information about keystrokes.
Consequently, any device capable of overhearing Wi-Fi traffic (under the \textit{monitor} mode~\cite{bullock2017wireshark}) may obtain
BFIs for free. 
As shown in Figure~\ref{fig:teaser}, 
our proposal 
aims to take advantage of this new vulnerability, in order to achieve keystroke eavesdropping without the need for hacking the constantly evolving Wi-Fi hardware.



However, we still face two challenges for realizing this idea. On one hand, passwords lack linguistic structure in natural languages (e.g.,  word structure and occurrence frequency of letters) to serve as prior information and features; this has forced existing password inference methods to either rely on independent keystroke features~\cite{li2016csi} or leverage transition features between two keystrokes~\cite{yang2022wink}. Nonetheless, as these features have strong environment dependency, the resulting inference methods can hardly be generalized to unseen scenarios.
Although supervised learning techniques may address this issue with a dataset containing sufficient training data, gathering such a labeled dataset can be prohibitively difficult due to diversified smartphone models and human typing habits. 
%
On the other hand, BFIs, carried by control traffic, can be sparse and sporadic. This relatively minor issue, if not properly addressed, may exacerbate the data deficit challenge for training a password inference model.
%




To tackle these challenges, we propose \name\ to steal \textit{numerical} passwords by eavesdropping on keystroke-induced BFI variations.  
\hjy{Thanks to BFI's clear-text nature, no low-level hacking is needed on Wi-Fi hardware.}
Given the lack of linguistic structure in passwords, we follow the canonical way of identifying individual keystrokes, but we leverage a deep learning model with a natural segmentation as input to get rid of the artifacts introduced by \newrev{rule-based} segmentation and environment interference. 
We exploit \textit{adversarial learning}~\cite{gan} to extract features relevant only to individual keystrokes; such a cross-domain training is capable of
%
%
generalizing keystroke inference to unseen scenarios with limited amount of training data, making \name achieve practical inference without having to gather a prohibitively large dataset. 
Furthermore, we design a sparse recovery algorithm to address the data deficiency issue for training the keystroke inference model. Finally, we implement a prototype of \name\ using a laptop or a rooted smartphone, and conduct extensive experiments on it to evaluate the performance of \name. In summary, our main contributions are:


\begin{itemize}
    \item We propose \name as the first WiFi-based \textit{hack-free} keystroke eavesdropping system; leveraging the clear-text BFI, it allows a wide range of Wi-Fi devices to eavesdrop on confidential passwords at ease.
    %
    \item We innovate in leveraging adversarial learning to remove environment dependencies, rendering \name's inference model generalizable to unseen scenarios.
    \item We design a sparse recovery algorithm to address the sparsity issue of BFI, handling the data deficiency issue for training the keystroke inference model.
    \item We conduct extensive evaluations; 
    the results indicate that \name\ achieves 88.9\% accuracy for identifying single numerical keys, and a top-100 accuracy of 85.0\% for inferring a 6-digit numerical password.
\end{itemize}


The paper is structured as follows. Section~\ref{sec:bg_moti} introduces the background and motivation of our work. Section~\ref{sec:design} presents the attack design of \name in detail. Sections~\ref{sec:implementation} and~\ref{sec:evaluation} respectively explain \name's implementation and report the extensive evaluations on \name, followed by a discussion on extension from numerical to general keystroke inference. \hjy{In Section~\ref{sec:Discussion}, we study the impact of different background traffic on BFI/CSI data flow and discuss defense strategies against \name.} Related works are briefly captured in Section~\ref{sec:rela_work}. Finally, we conclude our paper in Section~\ref{sec:conclusion}.	

\setcounter{figure}{2}
\begin{figure*}[b]
    \setlength\abovecaptionskip{8pt}
		\centering
    	\subfigure[Time series of the same keys.]{
            \centering
    		\label{sfig:time_bfi_same}
    		\includegraphics[width = 0.23\textwidth]{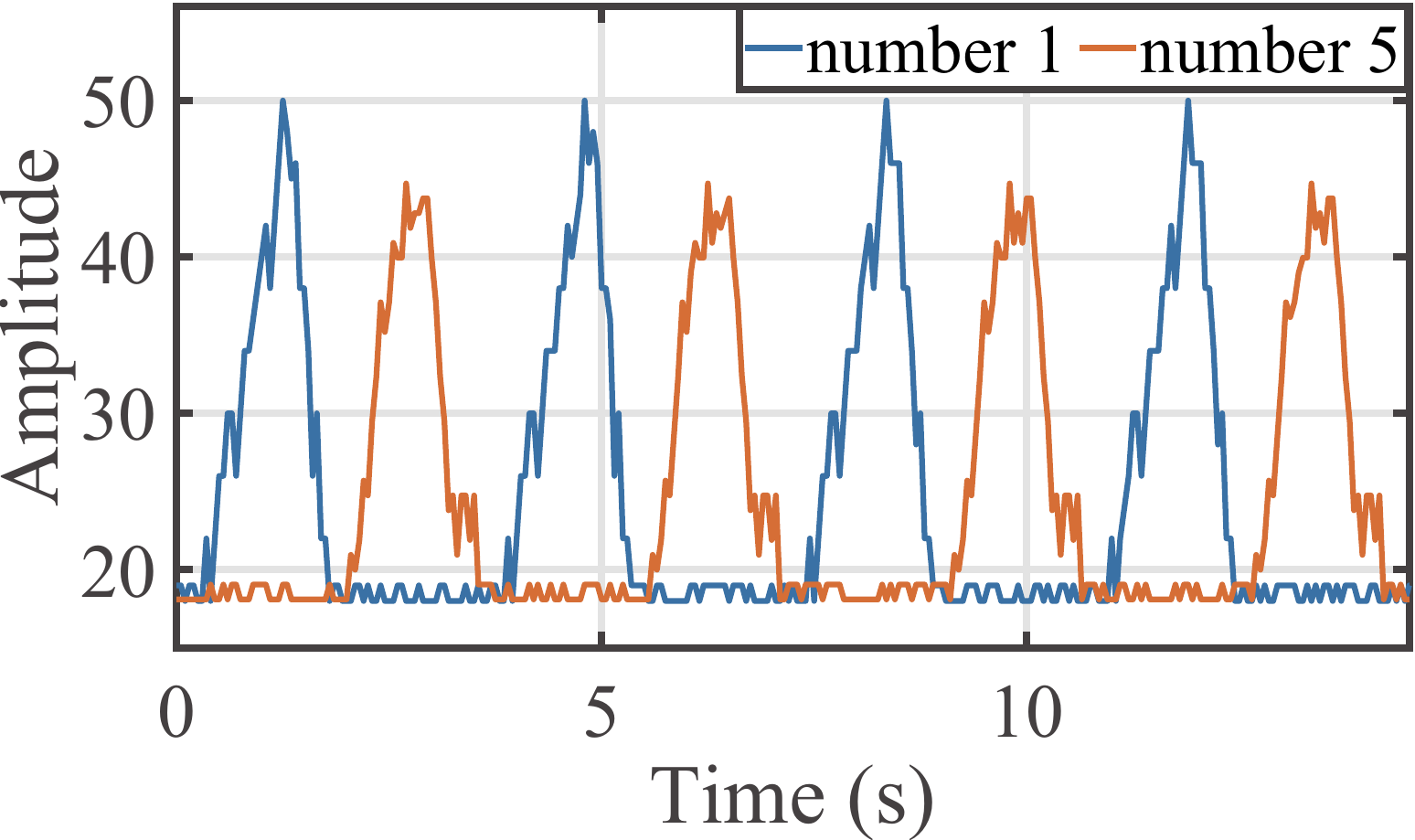} 
    		\vspace{-8ex}
    	}
            \hfill
    	\subfigure[Spectrogram of the same key.]{
            \centering
    		\label{sfig:freq_bfi_same}
    		\includegraphics[width = 0.23\textwidth]{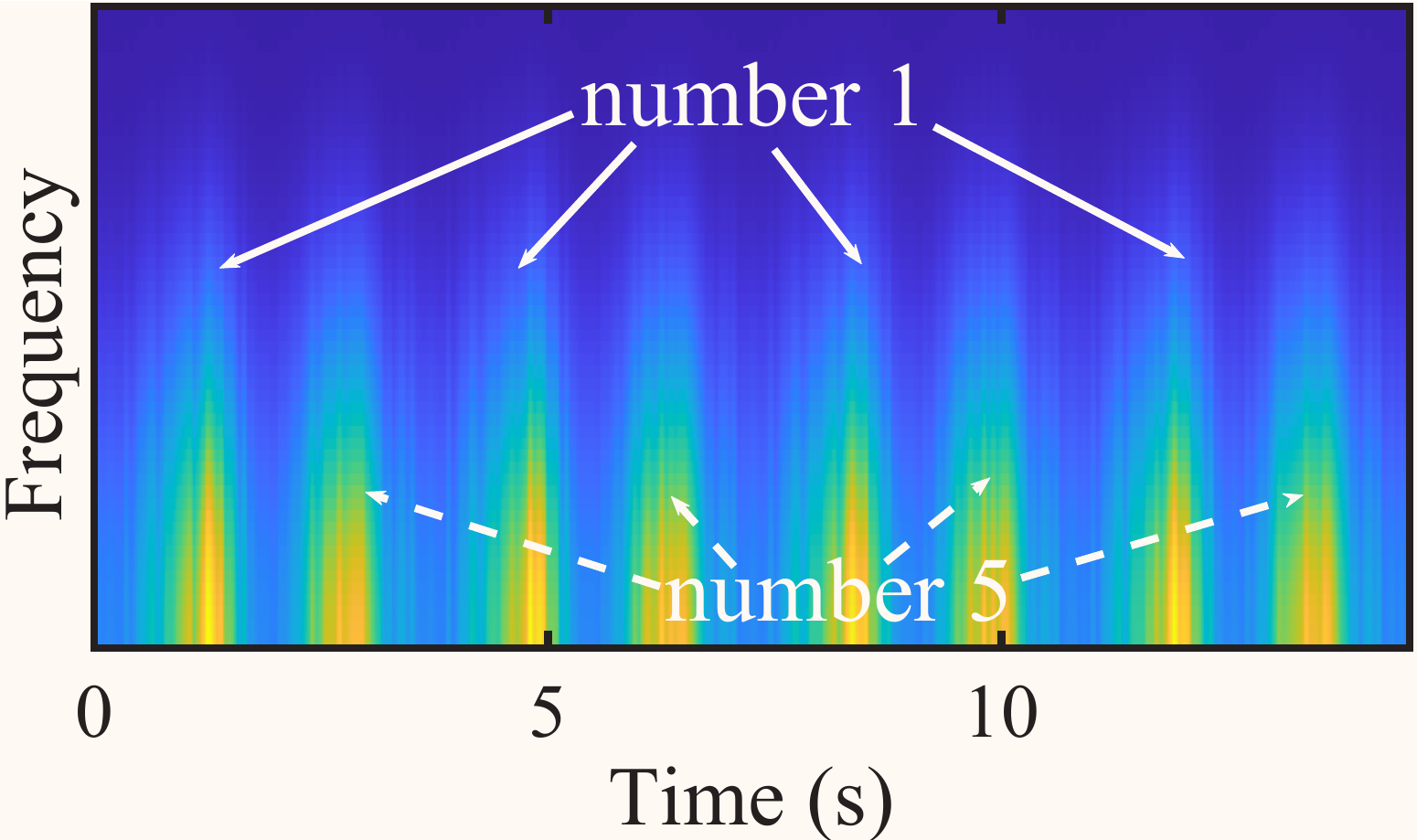}
    		\vspace{-8ex}
    	}
            \hfill
    	\subfigure[Time series of different keys.]{
            \centering
    		\label{sfig:time_bfi_diff}
    		\includegraphics[width = 0.23\textwidth]{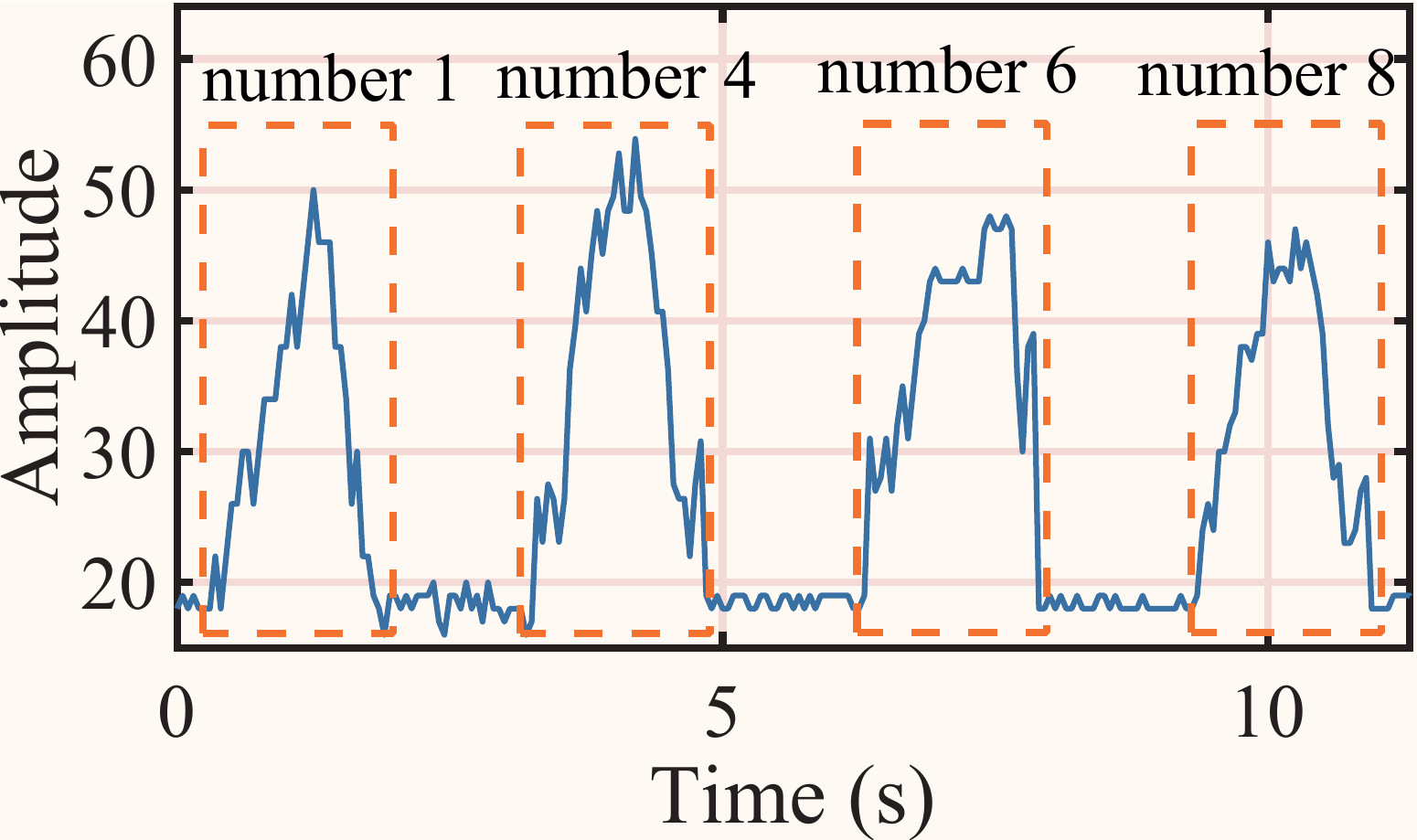}
    		\vspace{-8ex}
    	}
            \hfill
    	\subfigure[Spectrogram of different keys.]{
            \centering
    		\label{sfig:freq_bfi_diff}
    		\includegraphics[width = 0.23\textwidth]{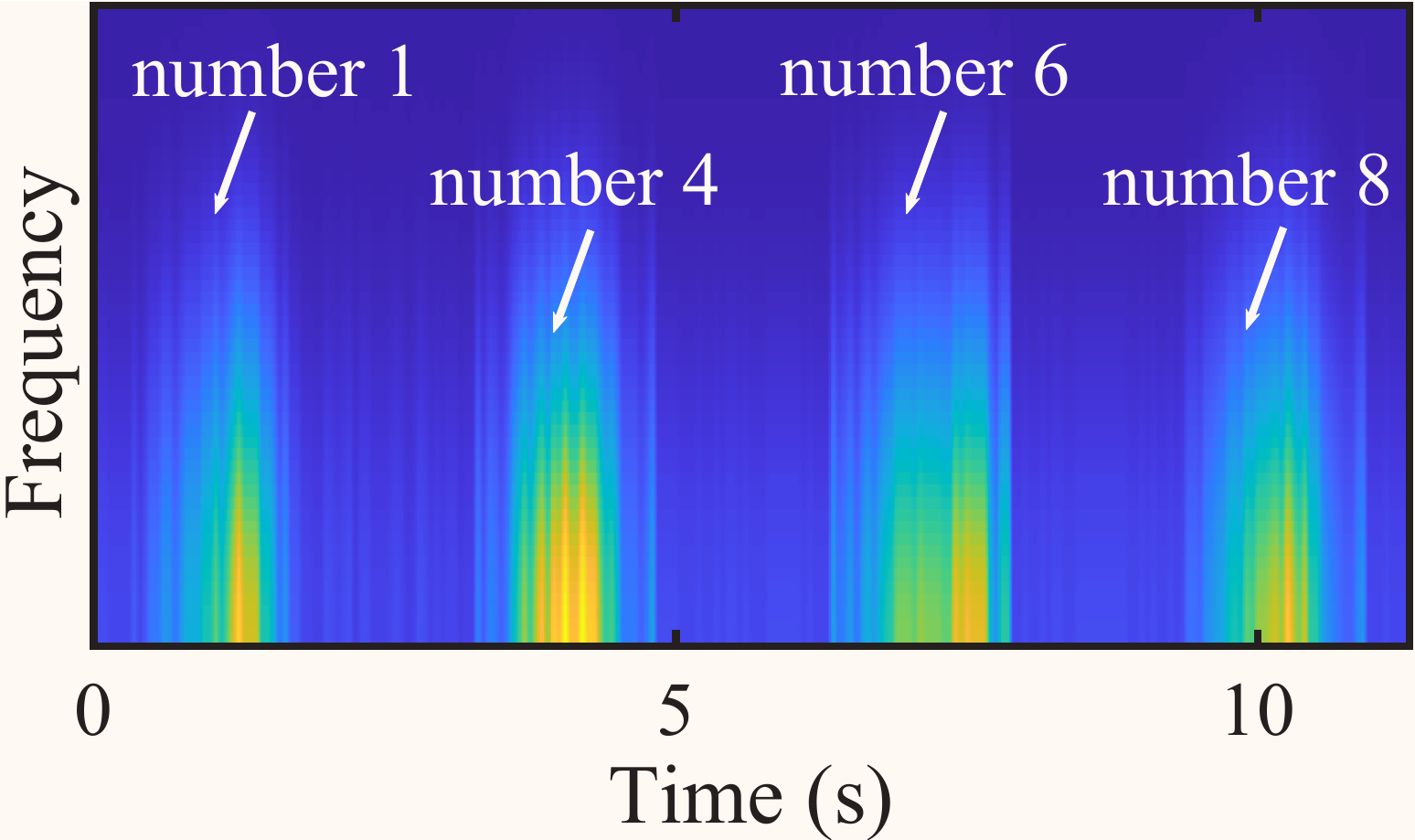}
    		\vspace{-8ex}
    	}
    	\\ 
    	\subfigure[Time series of the same keys.]{
            \centering
    		\label{sfig:time_csi_same}
    		\includegraphics[width = 0.23\textwidth]{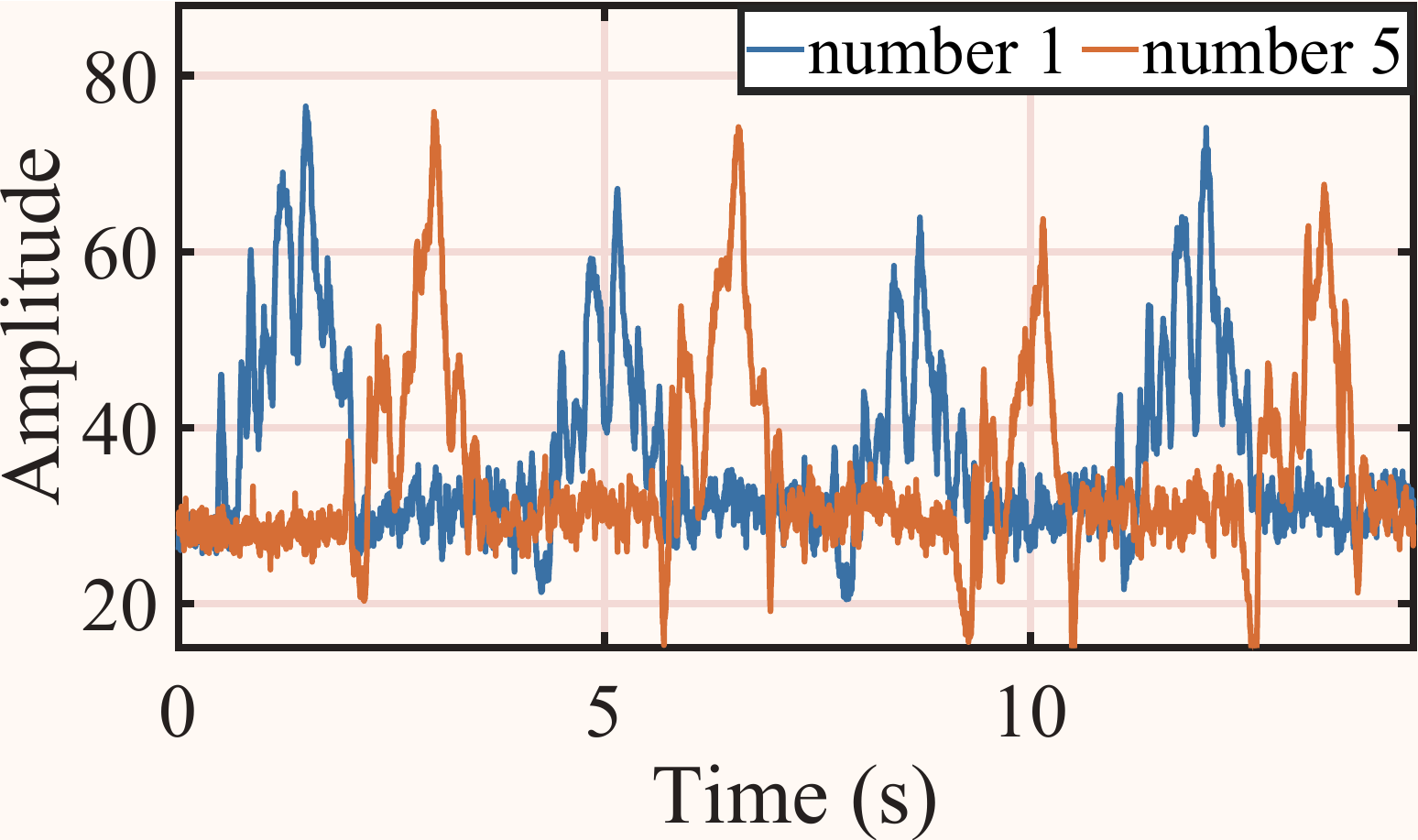}
    		\vspace{-8ex}
    	}
            \hfill
    	\subfigure[Spectrogram of the same key.]{
            \centering
    		\label{sfig:freq_csi_same}
    		\includegraphics[width = 0.23\textwidth]{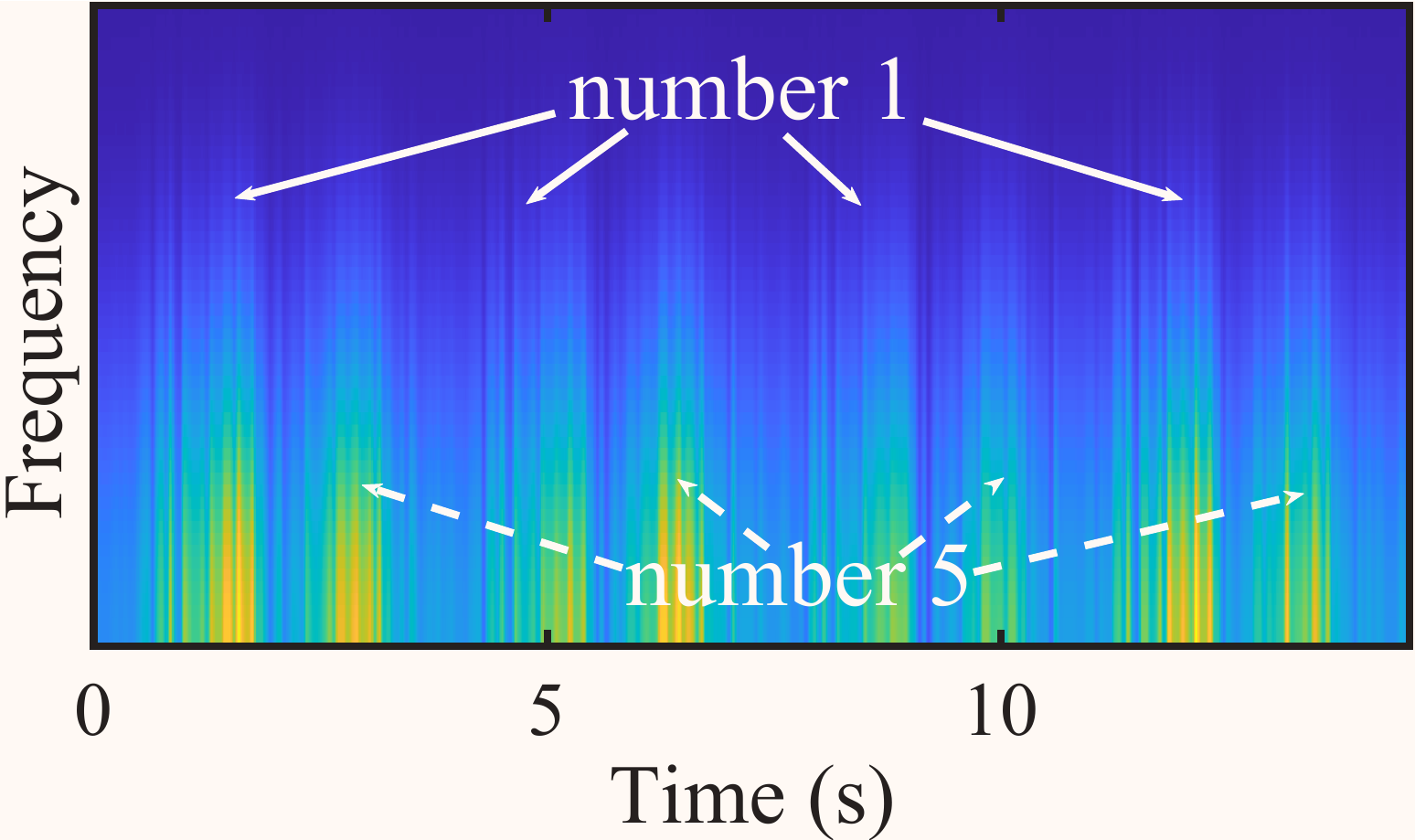}
    		\vspace{-8ex}
    	}
            \hfill
    	\subfigure[Time series of different keys.]{
            \centering
    		\label{sfig:time_csi_diff}
    		\includegraphics[width = 0.23\textwidth]{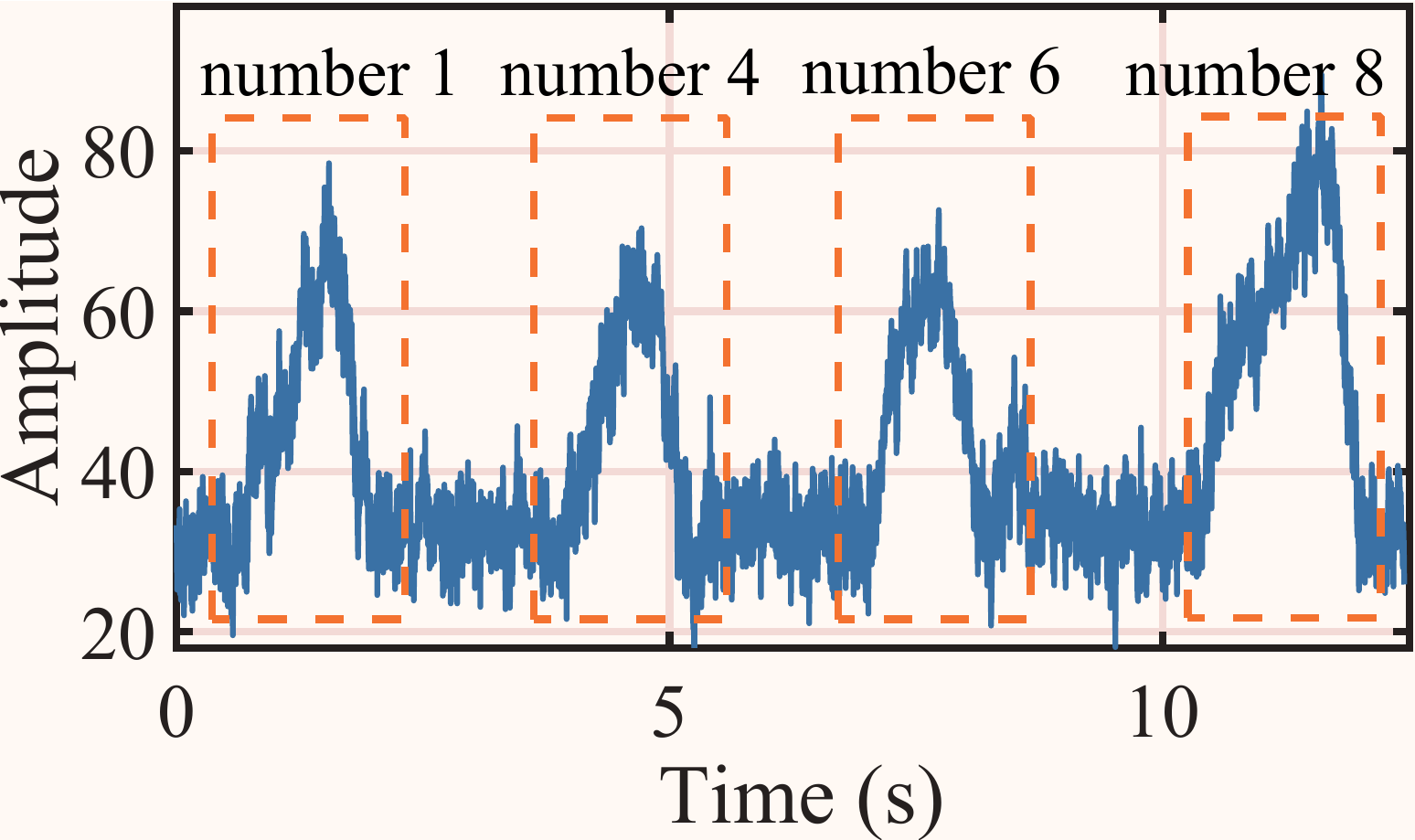}
    		\vspace{-8ex}
    	}
            \hfill
    	\subfigure[Spectrogram of different keys.]{
            \centering
    		\label{sfig:freq_csi_diff}
    		\includegraphics[width = 0.23\textwidth]{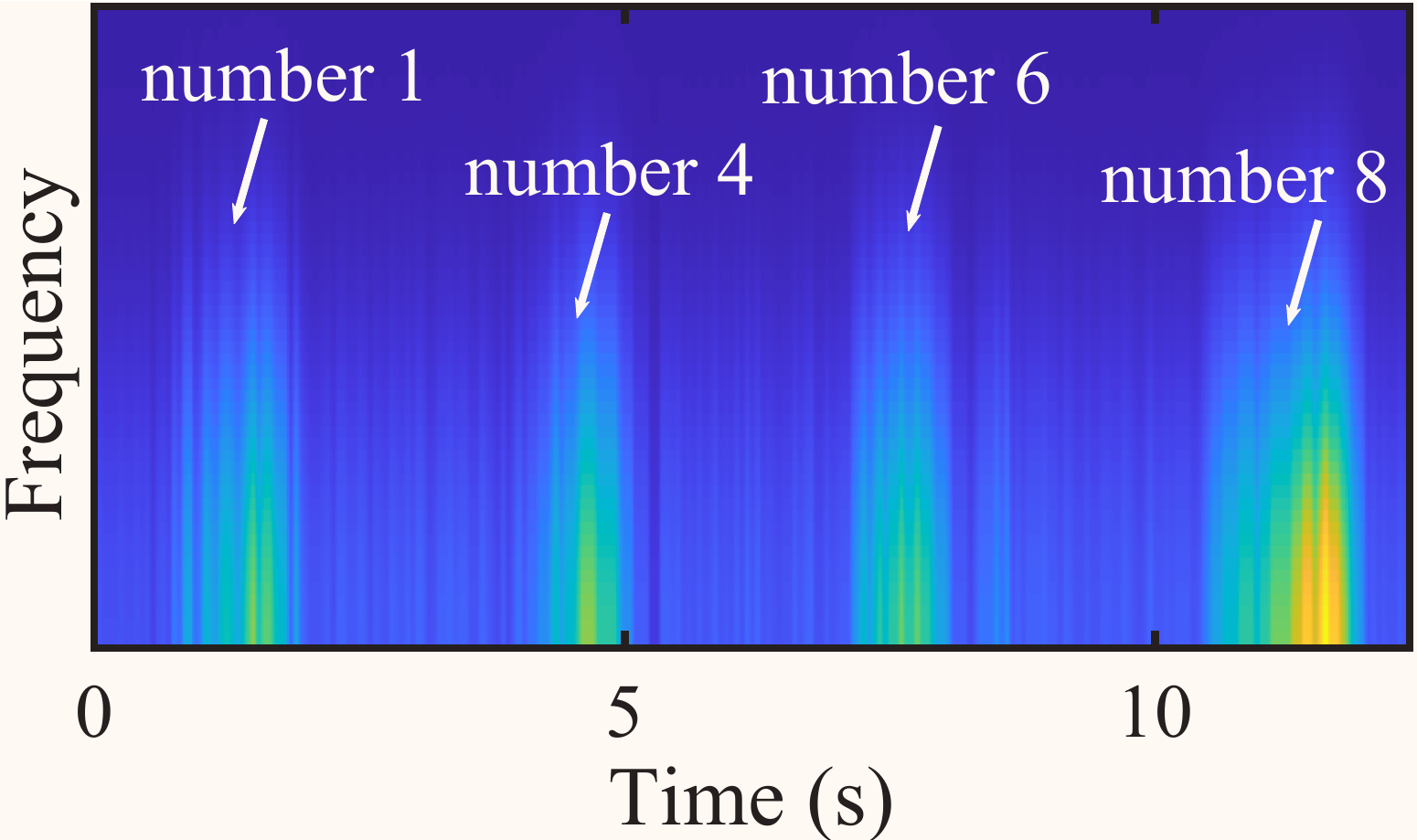}
    		\vspace{-8ex}
    	}
		\caption{BFI-KI (a)-(d) vs. CSI-KI (e)-(h): whereas BFIs exhibit both consistency for the same key and distinction for different keys, CSI's irregular patterns may cause ambiguities for keystroke inference. }
		\label{fig:bfi_csi_compare}
\end{figure*}

\section{Background and Motivation}
\label{sec:bg_moti}
In this section, we first introduce our \textit{keystroke inference} (KI) attack scenario, contrasting it to those considered by existing Wi-Fi CSI-based proposals. Then we demonstrate the advantages of BFI over CSI for realizing the keystroke eavesdropping.
\setcounter{figure}{1}
\begin{figure}[t]
    \setlength\abovecaptionskip{8pt}
    \centering
    \subfigure[In-band KI (IKI).]{
	    \includegraphics[width=.5\linewidth]{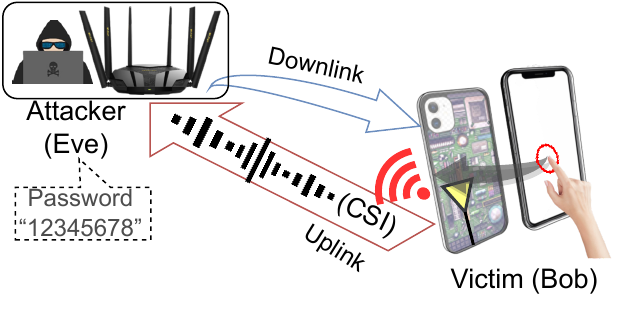}
	    \label{sfig:iki}
    }
    \hfill
	\subfigure[Out-of-band KI (OKI).]{
		\includegraphics[width=.448\linewidth]{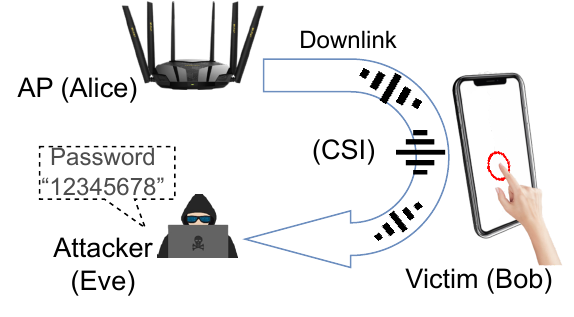}
		\label{sfig:oki}
	}
\caption{CSI-based keystroke inference (KI) methods.}
\label{fig:cis_ki}
\vspace{-1em}
\end{figure}

\subsection{Attack Scenarios and Methods}
\label{ssec:atk_scen_mth}

\hjy{We consider a scenario where a victim, Bob, uses his mobile device (smartphone or tablet) to connect to a Wi-Fi access point (AP) with a shared password or even no password protection; this is a reasonable assumption in public places such as shopping malls, office buildings, airports, and restaurants, because such a Wi-Fi access is often provided for the convenience to customers or visitors.} After connecting to the AP for accessing the Internet, 
Bob happens to have the need to access a sensitive account (e.g., online payment) protected by a \textit{password},
which makes him a target of attack launched by Eve (see Figure~\ref{fig:teaser}). \newrev{We follow the convention~\cite{li2016csi, yang2022wink} to mainly focus on numerical passwords, but we also consider an extension to general KI in Section~\ref{sssec:qwerty}.}
From here on, our method diverges from existing ones that either demand a rogue AP to trick Bob into using its service~\cite{li2016csi,fang2018no}, or require setting up extra Wi-Fi communication links to ``sense'' Bob's typing~\cite{ali2015keystroke, yang2022wink}.


Essentially, \name's attack method allows Eve to launch an attack on Bob regardless of which AP Bob is connected to. It leverages only a laptop equipped with a network interface card (NIC);
in fact, \name may even use a mobile device, as far as its Wi-Fi NIC can be switched to the monitor mode~\cite{bullock2017wireshark}.
We term our method \model (\textit{overhearing in-band keystroke inference}), named after the IKI method proposed by Li~\textit{et al.}~\cite{li2016csi} where the Wi-Fi link (actually its CSI) between Bob and the AP is exploited for password-stealing, as shown in Figure~\ref{sfig:iki}. However, \name innovates in getting rid of the need for hacking a Wi-Fi NIC and tricking Bob to use it as an AP. This improvement of \model over IKI is significant because, while the feasibility of hacking the continuous evolving Wi-Fi NICs is questionable (see Section~\ref{sec:intro}), effectively deploying rogue AP has been made extremely challenging due to the increasing alerts raised by individuals and companies on such attacks~\cite{beyah2011rogue,wang2016locating,cisco}.
%


Another method known as \textit{out-of-band keystroke inference} (OKI)~\cite{ali2015keystroke,yang2022wink}, shown in Figure~\ref{sfig:oki}, requires Eve to create a separate channel irrelevant to Bob, using Eve's Wi-Fi NIC and another device (e.g., an AP). Eve then infers Bob's keystrokes by observing the CSIs of this channel. Compared with OKI relying on analog CSIs, the digital nature of \model eavesdropping BFI leads to a significantly larger sensing range,
while the in-band sensing for KI ensures a sufficiently high signal-to-noise ratio (SNR). 
Unlike IKI having Eve directly observing data traffic via its rogue AP~\cite{li2016csi}, both \model and OKI require Eve to be able to identify Bob's device: whereas this has proven very difficult to achieve under realistic scenarios for OKI's analog CSI sensing (given the low spatial resolution of Wi-Fi sensing~\cite{zhang2022quantifying, MUSE-Fi-MobiCom23}), we shall demonstrate in Section~\ref{ssec:identify} that there exists a natural solution for \model's digital BFI eavesdropping.
%

\vspace{-1ex}
\subsection{Why BFI instead of CSI?}
\label{ssec: motiv bfi}
%
%
%
%
BFI actually offers other advantages over CSI in terms of KI attack, apart from its easy acquirement explained earlier. To be specific, BFI behaves \textit{less sensitive} to channel variation than CSI, rendering the sensing outcome more \textit{stable} especially upon IKI's close impact (from on-screen keystrokes) on Wi-Fi channels. This stability stems from the way BFI is generated. Given the downlink CSI represented as $H = Y/X$, where $X$ and $Y$ respectively denote the transmitted (Tx) and received (Rx) signals~\cite{li2016csi}, BFI is generated by partitioning $H$ (hence the channels it represents) into separated Tx and Rx components; only the Tx component is fed back to the AP for guiding AP beamforming~\cite{5677290}.
%
%
Thanks to this ``channel splitting'', BFI becomes less susceptible to channel variations caused by IKI's on-screen keystrokes, which otherwise leads to significant ambiguities in CSI-enabled KI.

To showcase the superiority of BFI over CSI in KI, we conduct a series of experiments, leveraging iPerf~\cite{tirumala1999iperf} to generate saturated traffic and collecting only raw BFI and CSI samples; this temporarily neglects the sample sparsity issue to be elaborated in Section~\ref{ssec:recovery}. 
%
In particular, Figure~\ref{sfig:time_bfi_same} and Figure~\ref{sfig:freq_bfi_same} respectively depict the BFI time series and spectrograms for clicking numerical keys `1' and `5' four times. One may readily observe that the BFI patterns remain consistent for clicking the same keys at different times, while the distinctions between two keys are also pronounced.
Additionally, Figure~\ref{sfig:time_bfi_diff} and Figure~\ref{sfig:freq_bfi_diff}, presenting BFI time series and spectrograms for clicking four different keys, again confirm the remarkable distinctions across these keys. 
In short, BFI is well-suited for KI with minimal preprocessing. 
As a comparison, Figures~\ref{sfig:time_csi_same} and~\ref{sfig:freq_csi_same}, with the same contents as those for Figures~\ref{sfig:time_bfi_same} and~Figures~\ref{sfig:freq_bfi_same} but for CSIs collected simultaneously with the aforementioned BFIs, fail to indicate either remarkable consistence for the same key or pronounced distinctions between two different keys. Meanwhile, the four-key tests shown in Figures~\ref{sfig:time_csi_diff} and~\ref{sfig:freq_csi_diff} also suggest the need for some heavy denoising before using CSIs for KI, as the distinctions between certain keys (e.g., `4' and `6') appear to be overwhelmed by noises. We suspect that such noises cannot be easily eliminated using conventional signal processing techniques, since their wide spectrum may confuse themselves with CSI features, \newrev{as confirmed by the following KI test with denoised CSI and raw BFI.}

\setcounter{figure}{3}
\begin{figure}[t]
\setlength{\abovecaptionskip}{8pt}
\centering
\subfigure[BFI-KI.]{
    \begin{minipage}[t]{0.473\linewidth}
    \label{BFI—classification}
    \centering
    \includegraphics[width=1\textwidth]{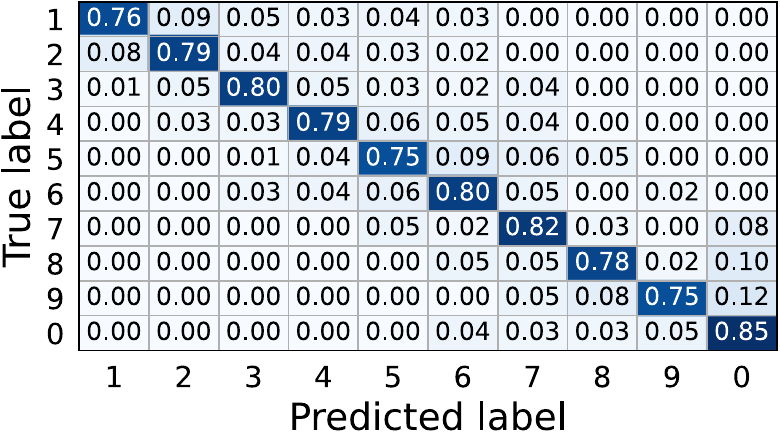}
    \end{minipage}
} \hfill
\subfigure[CSI-KI.]{
    \begin{minipage}[t]{0.473\linewidth}
    \label{CSI—classification}
    \centering
    \includegraphics[width=1\textwidth]{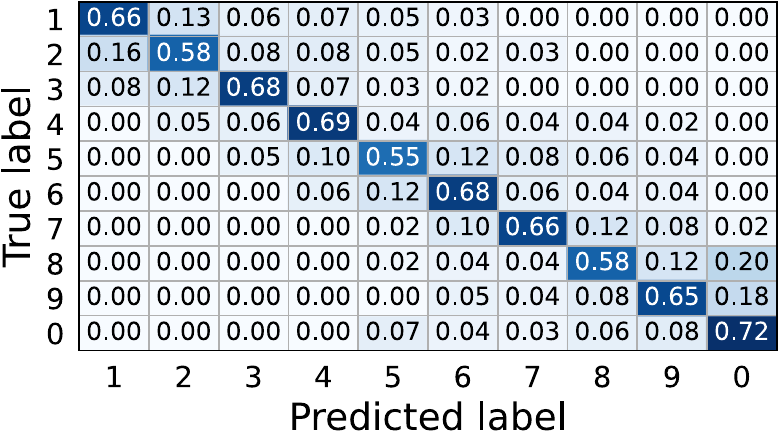}
    \end{minipage}
}
\centering
\caption{Confusion matrices for BFI- vs. CSI-based keystroke inference, demonstrating the superiority of BFI over CSI in completing this task.}
\label{fig:Confusionmatrices}
\vspace{-1em}
\end{figure}

%



%
We collect both BFI and CSI samples from 20 subjects typing numerical keys `0' to `9'. The same denoising technique in~\cite{li2016csi} is applied to the CSI samples, while the BFI samples are kept raw. We then use a one-dimensional convolutional neural network (1-D CNN)~\cite{kiranyaz20191} to perform classification for the sake of KI and evaluate the KI accuracy by cross-validation. Figures~\ref{BFI—classification} and~\ref{CSI—classification} present the confusion matrices for BFI- and CSI-based KIs, respectively; these results evidently demonstrate that BFI achieves higher accuracy for individual keys, as indicated by the diagonal of the confusion matrix. Overall, the average accuracy achieved by BFI is 78.9\%, notably higher than 64.5\% achieved by CSI, 
confirming the benefit of BFI's stability over even denoised CSI in terms of realizing KI.

\section{The Design of \name}
\label{sec:design}
\begin{figure}[t]
\setlength{\abovecaptionskip}{8pt}
\centerline{\includegraphics[width=0.46\textwidth]{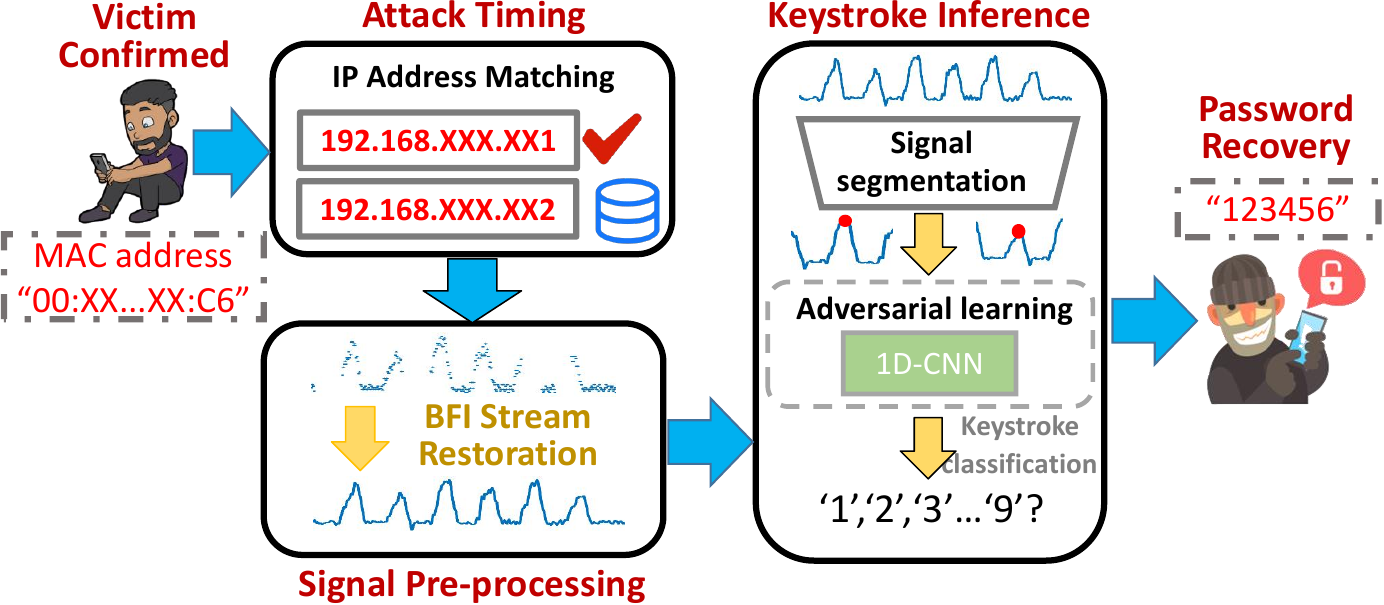}}
\caption{The workflow of \name's attack strategy.}
\label{fig:system_design}
\vspace{-1em}
\end{figure}

%
In this section, we introduce the attack strategy of \name. As shown in Figure~\ref{fig:system_design}, the whole workflow consists of five steps: i) identifying the victim, ii) determining the attack time when the victim accesses the targeted application service, iii) capturing the victim-associated BFI time series, iv) 
parsing and restoring 
the (possibly) sparse BFI series,
and finally v) segmenting the BFI series and performing KI to recover the intended password.
Key contributions in iv) and v) are respectively presented in Sections~\ref{ssec:recovery} and~\ref{ssec:inference}.


\subsection{Victim Identification and Attack Timing} \label{ssec:identify}
%
Following an implicit assumption of~\cite{li2016csi}, we also allow Eve to have prior knowledge of Bob's device identity (e.g., MAC address). In reality, Eve can acquire this information beforehand by conducting visual and traffic monitoring concurrently:
correlating network traffic originating from various MAC addresses with users' behaviors should allow Eve to link Bob's physical device to his digital traffic, thereby identifying Bob's MAC address. It is worth noting that victim identification is only possible through IKI, since the analog nature of OKI~\cite{ali2015keystroke,yang2022wink} forbids the use of header information to differentiate multiple subjects.


Once locked onto Bob's MAC address, Eve waits for the right time (when Bob is about to enter his password) to launch attack. \hjy{This timing issue can be readily addressed if visual hints are presented (e.g., Bob scan the WeChat Pay QR code or Bob's screen shows the payment page)}; otherwise, Eve can inspect the requests made to a payment service. Consider the case of WeChat~\cite{wechat}, though most of its traffic is secured via application-layer encryption~\cite{wu2017forensic}, IP addresses are not encrypted \hjy{for the public Wi-Fi networks targeted by \name.}
To exploit this vulnerability, Eve creates a database of IP addresses associated with the payment service: though such IP addresses can be dynamic, our experiments reveal that users from the same region are directed to the same IP address within a certain period. 
Subsequently, upon detecting an IP address recorded in the database, the attack can be launched; the recording of BFI series will be stopped once no more requests to the IP can be observed. 




\begin{figure}[b]
\setlength{\abovecaptionskip}{8pt}
\centering
\centerline{\includegraphics[width=0.38\textwidth]{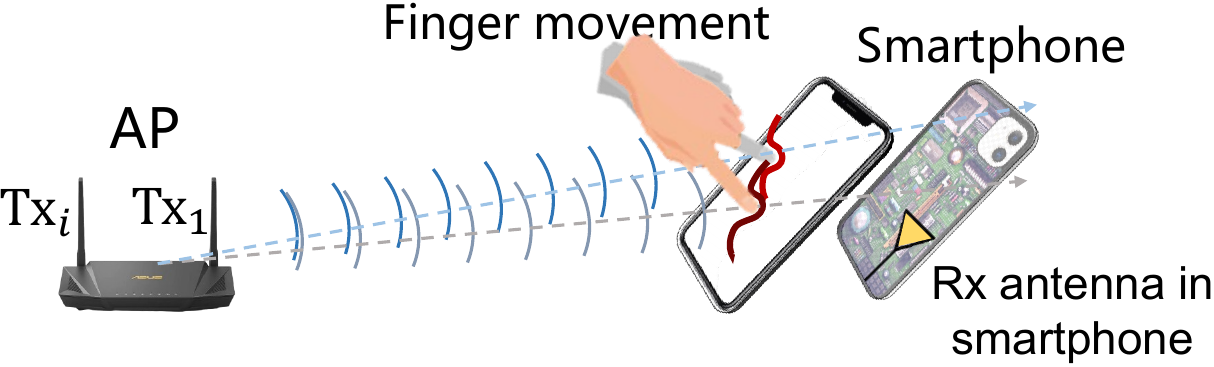}}
\caption{Finger movements cause diffraction on the downlink path, which 
is manifested in BFI variations.}
\label{fig:BFI_SENSING}
\vspace{-.5ex}
\end{figure}

\subsection{BFI Analysis and Parsing}
\label{ssec:keystroke_sensing}
%
%
%
%
We hereby provide more details on how password typing can manifest in BFI to facilitate later developments, by first explaining how BFI is generated.
As explained in Section~\ref{ssec: motiv bfi}, BFI is the Tx component of CSI $H$ and is fed back to guide AP beamforming. 
This is accomplished by SVD (singular value decomposition)~\cite{stewart1993early} that decomposes the channel as $H=USV$. Among these components, only the right matrix $V$ is chosen as BFI, while the other two matrices $U$ and $S$ (representing Rx beamforming and channel gains, respectively) are not.  As illustrated in Figure~\ref{fig:BFI_SENSING}, Bob's password typing affects the diffraction pattern of the Wi-Fi signals around the phone body. This altered pattern is then reflected in downlink CSI that is in turn decomposed with SVD to obtain BFI $V$.


As BFI is transmitted in clear text, Eve can easily intercept it using a Wi-Fi device in monitor mode, along with Wireshark~\cite{orebaugh2006wireshark}. The frame structure of 802.11ac can be followed to locate BFI in the ``VHT beamforming report'' field within the Wi-Fi Action frames~\cite{5677290}. To completely extract BFI, the length of the field can be calculated based on the number of Tx and Rx antennas, as outlined in~\cite{survivalguide}.
By continuously recording the BFIs in the Wi-Fi frames from Bob during the time window of Bob's password typing, Eve can obtain a time series of BFI samples 
correlated with the password. 
If the BFI time series is too sparse due to a low control frame rate during the time window, \name\ tries to restore it. To remain focused on the key component, we first explain the KI function in Section~\ref{ssec:inference}, and postpone the discussion on BFI restoration to Section~\ref{ssec:recovery}.
\subsection{Keystroke Inference}
\label{ssec:inference}
In this section, we elaborate on how \name\ conducts BFI-KI. We first discuss the drawbacks of previous proposals and explain possible improvements upon them. After that, we specify the signal segmentation on BFI series to kick off KI, which is then followed by the design of the KI neural model and its adversarial learning framework to generalize KI towards unseen scenarios. 

\vspace{-1ex}
\subsubsection{What's Wrong with Prior Art?} \label{ssec:why_fail}
\begin{figure}[b]
    \setlength\abovecaptionskip{3pt}
    \vspace{-1em}
    \centering
    \subfigure[Segmentation ambiguity.]{
	    \includegraphics[width=.47\linewidth]{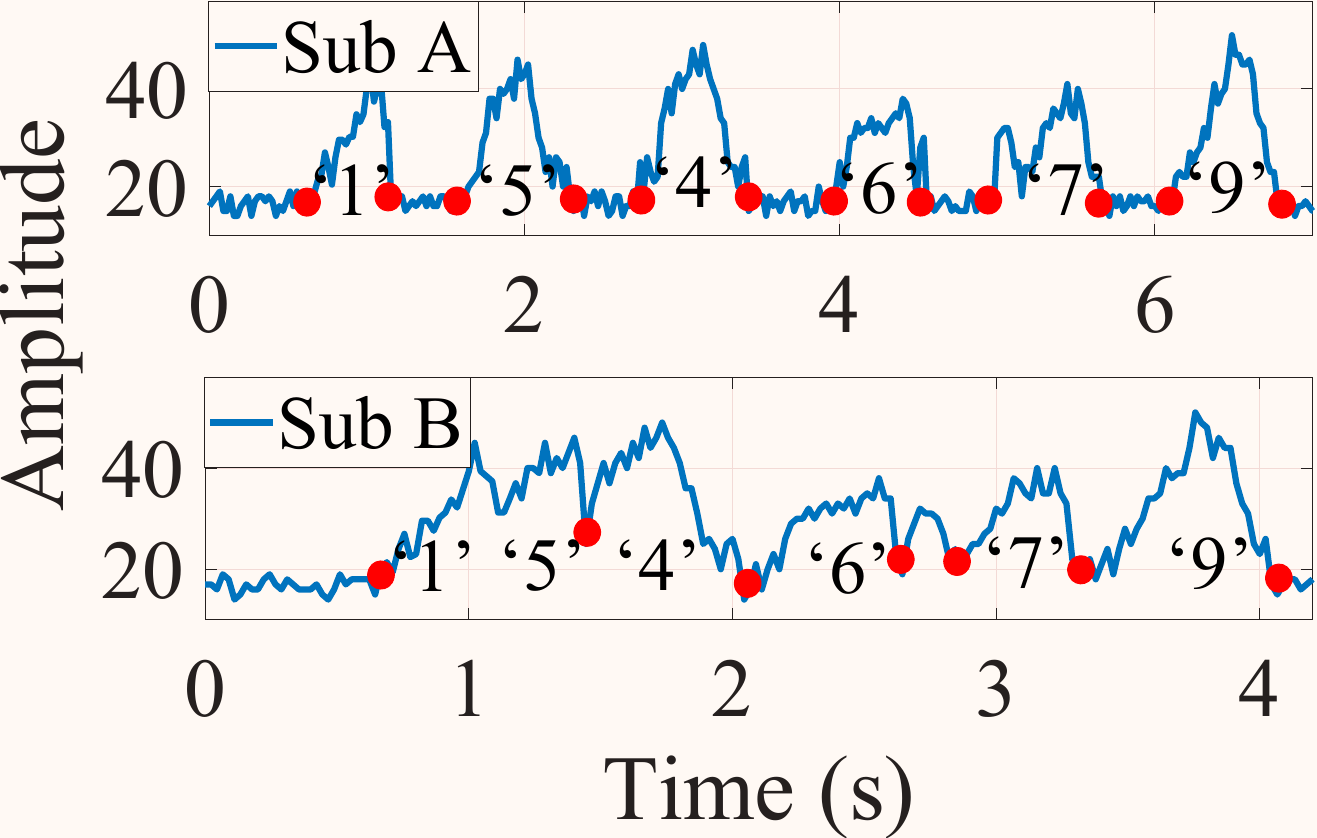}
	    \label{sfig:user_diversity}
    }
    \hfill
    \subfigure[Variance in transition features.]{
		\includegraphics[width=.47\linewidth]{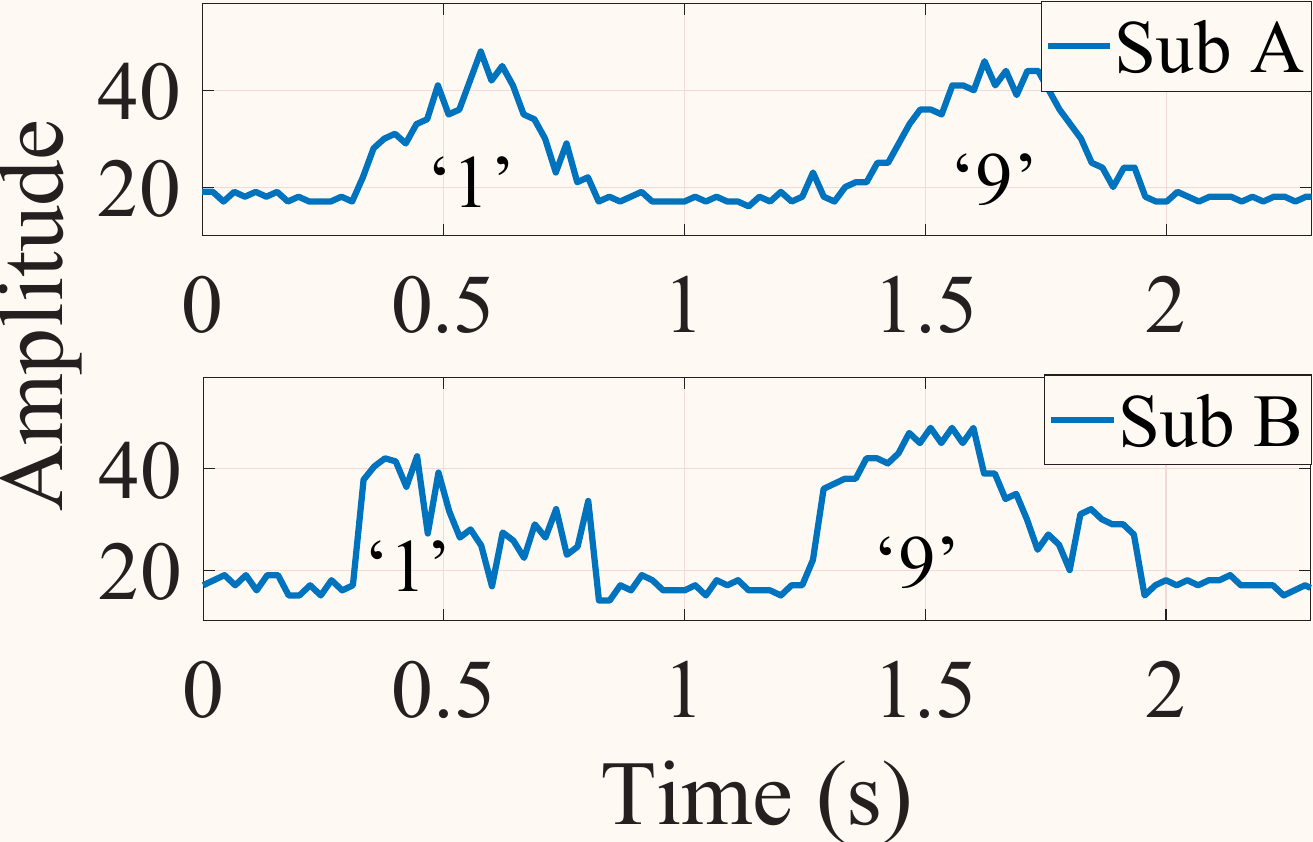}
		\label{sfig:transition}
    }
    \vspace{-.5ex}
\caption{Two cases where previous methods fail.}
\label{fig:previous_fail}
\vspace{-.5ex}
\end{figure}

Only two major contributions exist for leveraging Wi-Fi side-channels to steal passwords. The seminal proposal of WindTalker~\cite{li2016csi} performs classification upon individual keystrokes \newrev{with rule-based CSI series segmentation. Intuitively, such} segmentation should not perform well because it can result in information loss or introduce artifacts. To confirm this suspicion, we ask two subjects to type passwords on their respective smartphones, and Figure~\ref{sfig:user_diversity} shows their corresponding CSI series. Apparently, the duration of the keystrokes and the amount of overlap between them vary significantly due to the subjects' distinct typing habits. While \newrev{rule-based} segmentation may be effective for Subject A who types more steadily, it most likely fails for Subject B whose inter-keystroke patterns appear rather messy. In attempting to forcibly assign different sections of the BFI series to individual keys, the segmentation process introduces artifacts (e.g., clipping) to each keystroke, potentially harming the KI performance.

A recent proposal WINK~\cite{yang2022wink} claims to improve the KI performance via series learning. However, it inherits the \newrev{rule-based} segmentation adopted by~\cite{li2016csi} and hence the same weakness too. Additionally, as linguistic structure cannot be exploited for series learning, WINK argues that transition features between keystrokes may serve as replacements for improving KI accuracy. Unfortunately, factors such as typing habits and smartphone types can affect CSI during the transition period, resulting in different features for the same password. To illustrate this, we ask two subjects to \newrev{type} two keys `1' and then `9' on their phones, and Figure~\ref{sfig:transition} shows significant morphological and temporal differences in these two transitions. Therefore, it is very questionable if transition features can ever replace linguistic structure.

To overcome the disadvantages in previous proposals, the \newrev{rule-based} segmentation needs to be replaced with a more sensible method, preferably a data-driven one. Also, as using transition features to replace linguistic structure cannot be reliable, \name\ falls back to the canonical approach of inferring individual keystrokes as executed by WindTalker. To prevent information loss in segmentation, \name\ deems the environment-dependent transition periods as different ``domains'' of the same numerical keystroke. Consequently, an adversarial learning is exploited to train the KI model, aiming to remove domain interference (i.e., environment dependency) and hence generalize KI to unseen scenarios. Note that the data-driven nature of \name\ also prevents it from taking a series learning perspective, as it would otherwise
demand a prohibitively large training dataset whose size grows exponentially with the password length.
\subsubsection{Signal Segmentation} \label{ssec:segmentation}
\begin{figure}[t]
\setlength{\abovecaptionskip}{3pt}
\centering
\centerline{\includegraphics[width=0.4\textwidth]{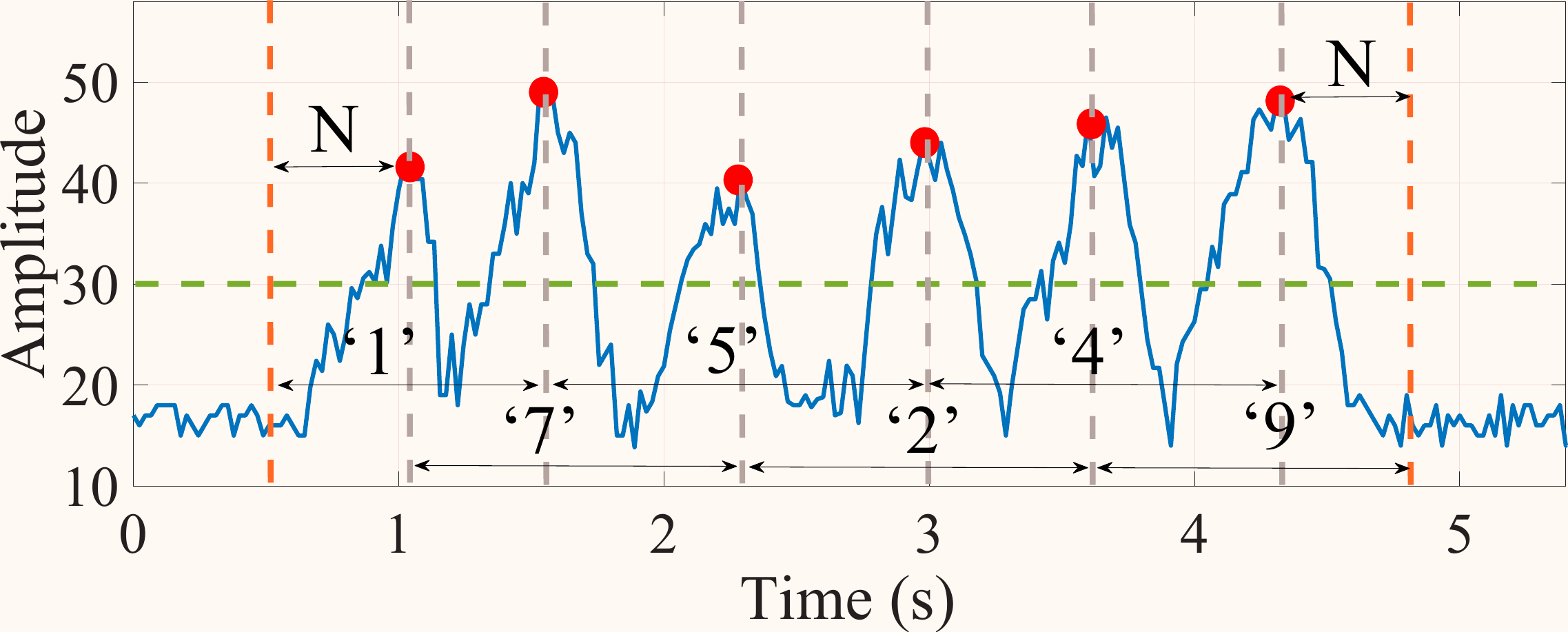}}
\caption{Signal segmentation with overlaps.}
\label{fig:Segmentation}
\vspace{-1em}
\end{figure}

%
In reality, BFI series may not show distinct boundaries between consecutive keystrokes, significantly complicating signal segmentation. Figure~\ref{fig:Segmentation} provides an example for such a case, where the BFI series displays prominent peaks corresponding to Bob's finger hitting the screen, as well as fluctuations between two peaks representing the transition movement of his fingers. Since the transitions carry 
information about both the preceding and succeeding keystrokes, segments of neighboring keystrokes should contain the transition. Therefore, we propose to employ an overlapping segmentation method that incorporates all data samples located between two consecutive peaks, from the preceding to the succeeding peaks, instead of the non-overlapping segmentation achieved by windows of 
\newrev{rule-defined} sizes~\cite{li2016csi, yang2022wink}.




Our segmentation method starts with utilizing the Constant False Alarm Rate (CFAR) algorithm~\cite{nitzberg1972constant} to identify peaks in a BFI series. Suppose Bob typing a $K$-digit numerical password to produce a BFI series after sparse recovery (will be discussed in Section~\ref{ssec:recovery}) of length $L$, the CFAR algorithm conducts statistical analysis on the series to determine an adaptive threshold and selects the peaks exceeding this threshold as candidates. Among these candidate peaks, we further eliminate minor ones within a distance of $W$ sampling points from a major peak. We then select the top-$K$ peaks corresponding to the $K$ numbers in the password, assisted by an inter-peak distance of $W$ sampling points, where 
$W = \alpha \times \frac{L}{K}$. For each peak, we include all the data samples between itself and its two neighboring peaks into the segment corresponding to a \newrev{single keystroke}; since the first and last numbers in the password have no preceding and succeeding numbers, we choose to extend $N$ points before and after as the segment boundaries, where $N = \beta \times \frac{L}{K}$. We shall empirically determine the values of $\alpha$ and $\beta$ in Section~\ref{sec:implementation}. 
As demonstrated in Figure~\ref{fig:Segmentation}, this approach effectively partitions a BFI series (for password ``175249'') into segments corresponding to individual keystrokes, while preserving the feature-rich transitions between keystrokes caused by finger movements.

\vspace{-.5ex}
\subsubsection{Adversarial Learning Framework}
\label{sssec:keystroke_inference}

This section explains how adversarial learning is employed to generalize KI to unseen domains. Prior to that, we briefly describe the basic design of KI network. The classification of time series is a well-established task that can be effectively addressed using a 1-D CNN. However, as discussed in Section~\ref{ssec:segmentation}, the BFI segments may differ in length, posing a challenge \newrev{to conventional 1-D CNNs.} To overcome this issue, we employ an adaptive average pooling layer~\cite{he2015spatial} to enhance the flexibility of 1-D CNNs. To be specific, this layer automatically calculates the appropriate kernel size required to yield a fixed-size output feature map, thus enabling 1-D CNNs to accommodate inputs of varying lengths. 

\begin{figure}[b]
    \setlength\abovecaptionskip{3pt}
    \vspace{-1.5ex}
    \centering
    \subfigure[BFI segments.]{
	    \includegraphics[width=.468\linewidth]{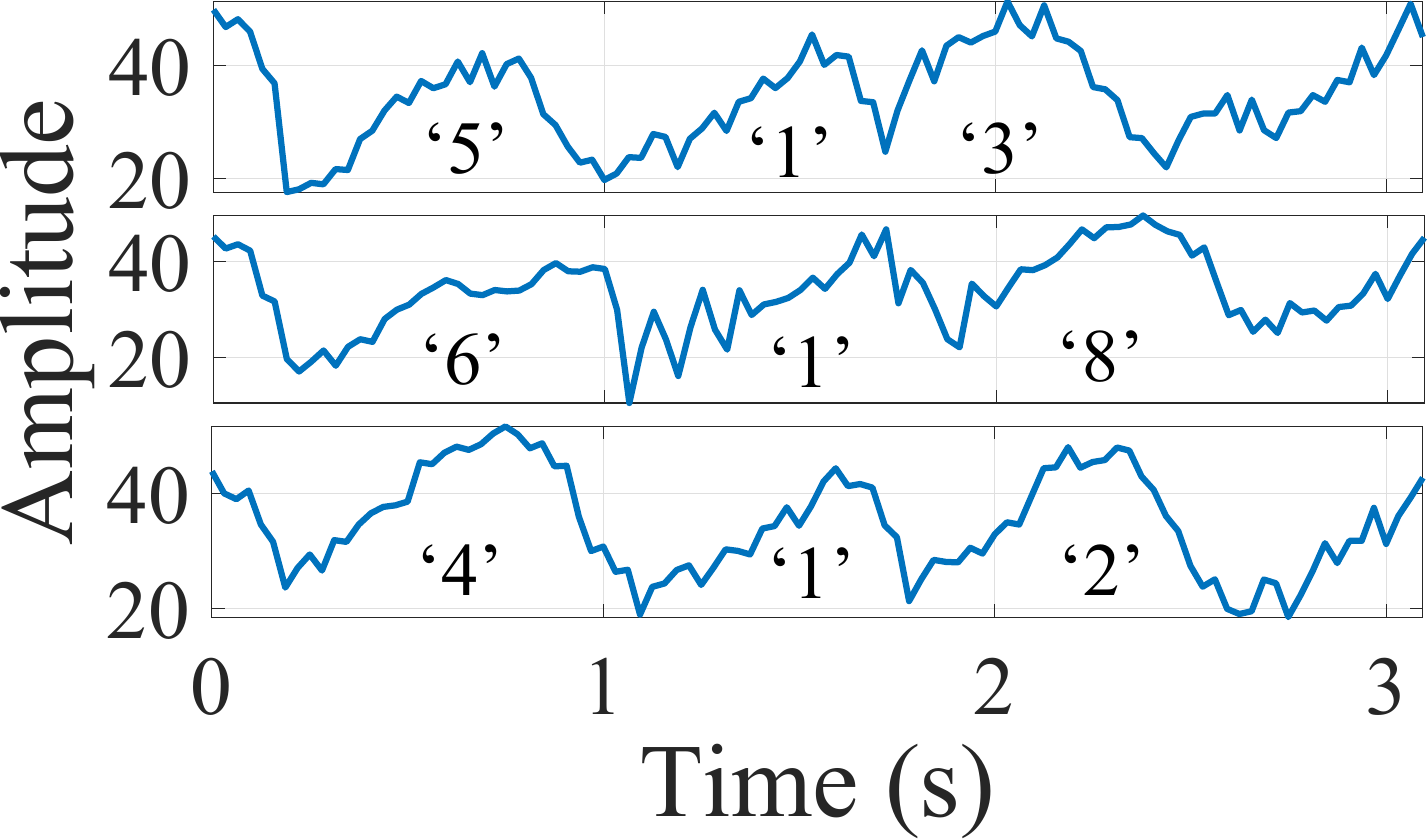}
	    \label{sfig:adv_series}
    }
    \hfill
    \subfigure[Feature maps.]{
		\includegraphics[width=.458\linewidth]{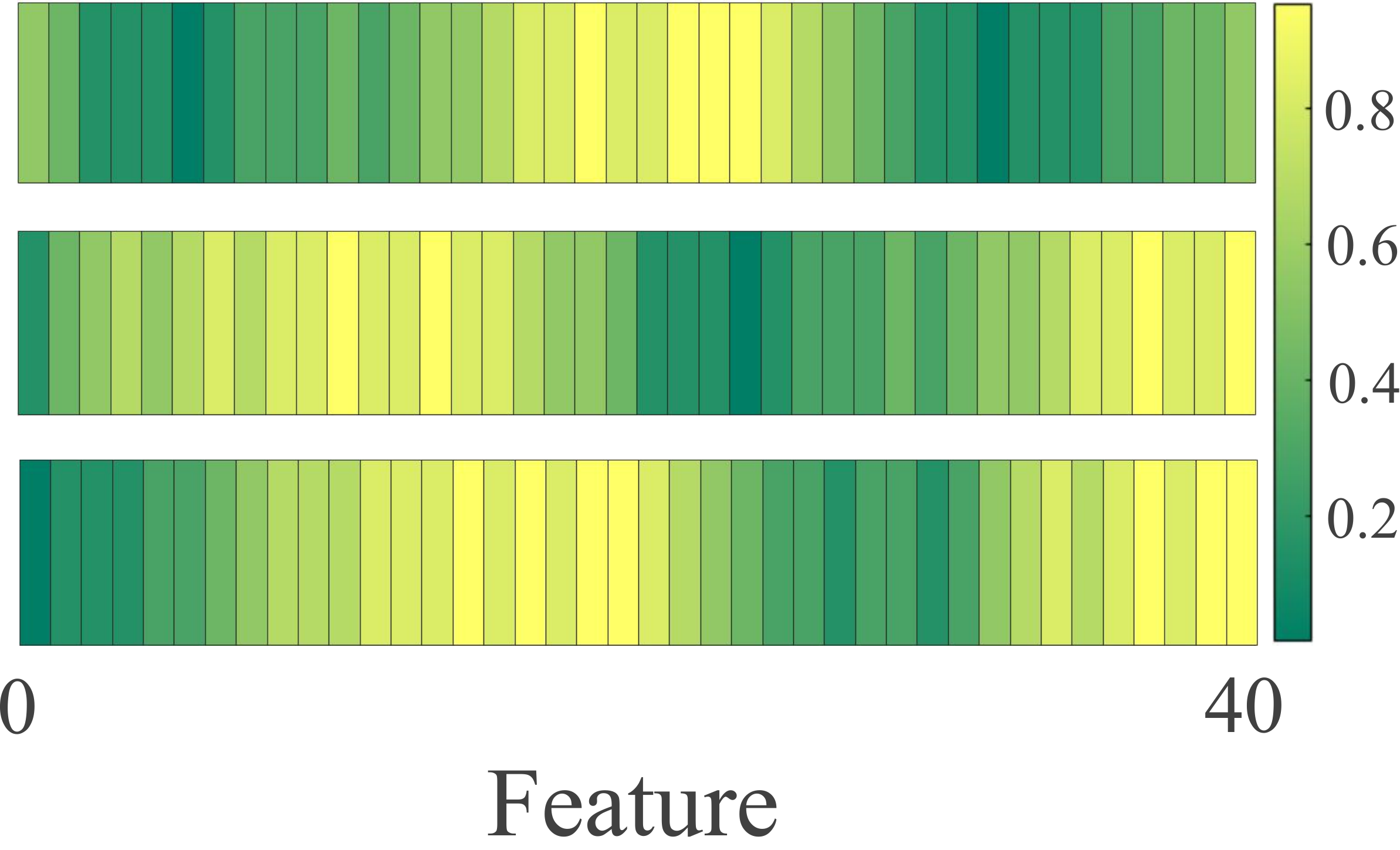}
		\label{sfig:adv_feature}
    }
\caption{Difference in BFI segments and features maps of key `1' indicates the domain dependency of KI.}
\vspace{-1ex}
\label{fig:adv_motivation}
\vspace{-.5ex}
\end{figure}

In fact, the direct deep learning approach mentioned above overlooks the impact of the domain on each keystroke. Here \textit{domain} refers to the context arising from the diversified transitions from the preceding and to the succeeding keystrokes; it includes the distinctions caused by, for example, typing speed, inter-typing irregularities, and the adjacent keystrokes.
To illustrate this, we consider the numerical key `1' in three different domains: `5-1-3', `6-1-8', and `4-1-2', and present their segments and corresponding feature maps in Figure~\ref{fig:adv_motivation}. Although the segments of key `1' under different domains, in Figure~\ref{sfig:adv_series}, exhibit a high degree of similarity near the peak, the `1' in `6-1-8' displays drastic fluctuations during transitions between neighboring keystrokes, while those in `5-1-3' and `4-1-2' have rather smooth transitions. Such differences can be attributed to larger channel variations induced by finger movements over greater distances between the keys in `6-1-8 '. 
Additionally, we show the feature heatmaps for different `1's after the adaptive average pooling layer in Figure~\ref{sfig:adv_feature}:
the same key `1' in different domains exhibit distinct feature maps, thus posing significant challenges to the subsequent keystroke classifier. 

\begin{figure}[t]
    \setlength{\abovecaptionskip}{8pt}
    \centering
    \centerline{\includegraphics[width=0.92\columnwidth]{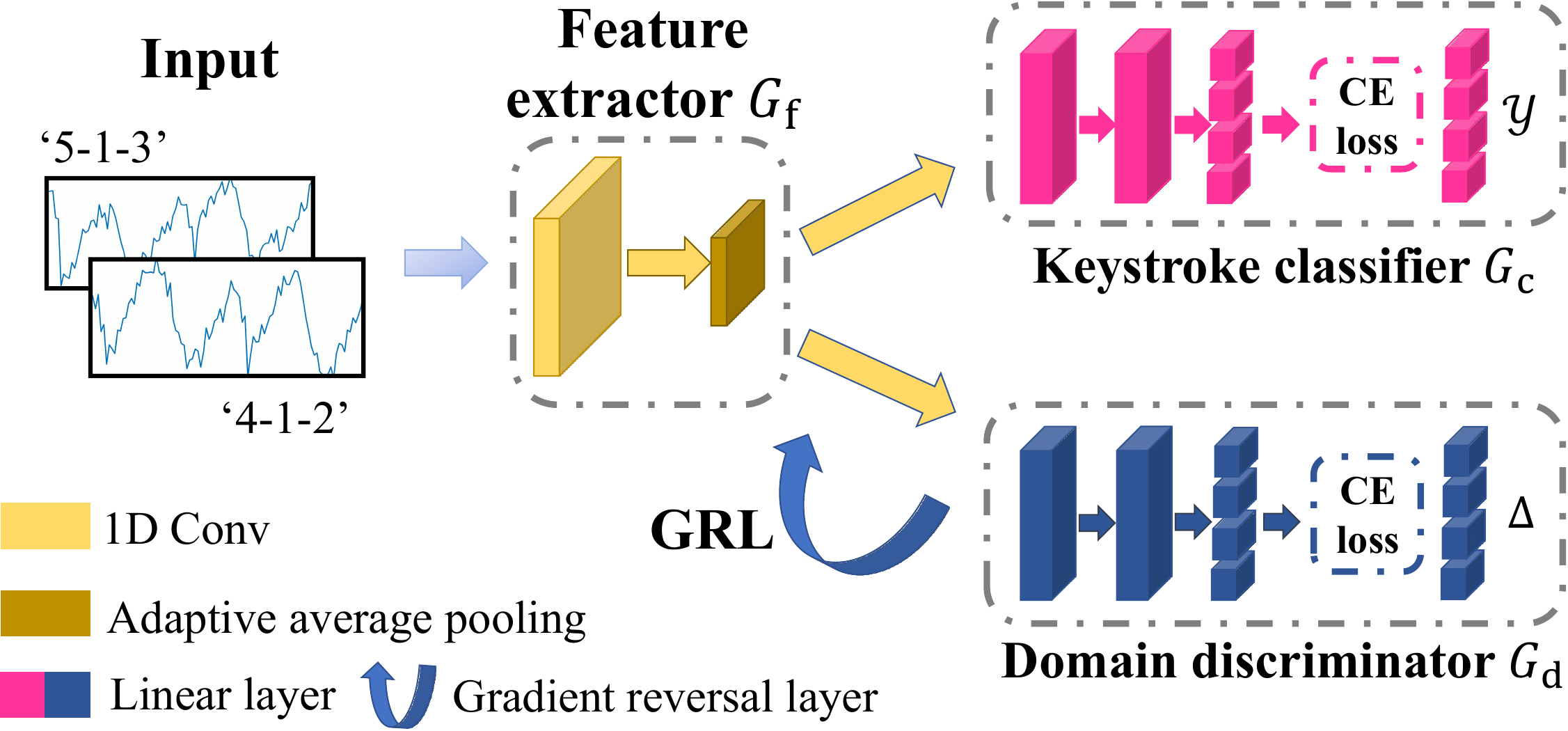}}
    \caption{The training strategy enabled by adversarial learning removes domain-specific information.}
    \label{fig:adv_strategy}
    \vspace{-1em}
\end{figure}

The aforementioned domain interference entails the need for a method ensuring KI's invariance to such interference, so we employ the idea of domain adaptation~\cite{ben2010theory} to learn 
keystroke representations invariant across different domains. Given the complexity of BFI segment features due to the 
\newrev{diversity} of inter-keystroke transition patterns, employing an explicit feature space transformation as in~\cite{pan2010domain} could be challenging. Instead, \name\ aims to achieve a consistent feature space representation in different domains, by harnessing the power of \emph{adversarial learning}~\cite{gan} to integrate domain adaptation with KI in a unified training process. To incorporate adversarial learning, we revamp the training strategy of 1-D CNN as illustrated in Figure~\ref{fig:adv_strategy}, whose training and inference processes are introduced as follows.

During the \textit{training} phase, we first prepare a dataset consisting of \newrev{randomly \textit{paired} BFI segments corresponding to the same key} (e.g., `1') but under different domains, e.g., \newrev{two `6-1-8' from different passwords or a pair of `4-1-2' and `5-1-3'.} We concatenate the pair as input $\bm{x}$ and process them through the feature extractor $G_\mathrm{f}$. The resulting features are then fed into both the keystroke classifier $G_\mathrm{c}$ and domain discriminator $G_\mathrm{d}$: \newrev{$G_\mathrm{c}$ infers the key $y$ shared by both segments within the pair, and $G_\mathrm{d}$ predicts the \emph{domain discrepancy} $\Delta\in\{0,1\}$, with 0 and 1 denoting the keys from the two segments \emph{are} and \emph{are not} from the same domain, respectively.
While $G_\mathrm{d}$ aims to improve the accuracy of predicting $\Delta$, the adversarial learning strategy ``cheats'' $G_\mathrm{d}$ by inverting its loss
via reversing the gradient during backpropagation using the Gradient Reversal Layer (GRL)~\cite{ganin2015unsupervised}; this procedure tends to suppress domain-specific features from the output of $G_\mathrm{f}$ and thus allows the 1-D CNN to learn keystroke representations invariant across domains. Denoting the parameters of $G_\mathrm{f}$, $G_\mathrm{c}$, and $G_\mathrm{d}$ as $\theta_\mathrm{f}$, $\theta_\mathrm{c}$, and $\theta_\mathrm{d}$, respectively,} the above training procedure can be formulated as:
\begin{align}
    (\hat{\theta}_\mathrm{f}, \hat{\theta}_\mathrm{c}) = \textstyle{\arg\min_{\theta_\mathrm{f}, \theta_\mathrm{c}}}  \mathcal{L}(y,\Delta,\bm{x}), \quad
\hat{\theta}_\mathrm{d} = \textstyle{\arg\max_{\theta_\mathrm{d}}} \mathcal{L}(y,\Delta,\bm{x}), \nonumber
\label{eq:adv}
\end{align}
where  $\mathcal{L}(y,\Delta,\bm{x}) = \mathcal{L}_\mathrm{c}\left(y, G_\mathrm{c}(G_\mathrm{f}(\bm{x}))\right)-\lambda \mathcal{L}_\mathrm{d} \left(\Delta, G_\mathrm{d}(G_\mathrm{f}(\bm{x})\right)$, $\mathcal{L}_\mathrm{c}$ and $\mathcal{L}_\mathrm{d}$ are respectively the cross-entropy losses for $G_\mathrm{c}$ and $G_\mathrm{d}$, and 
\newrev{$\lambda$, a balance factor controlling the trade-off between  $\mathcal{L}_\mathrm{c}$ and $\mathcal{L}_\mathrm{d}$, should have its value empirically determined in Section~\ref{sec:implementation}.} 
$G_\mathrm{d}$ is discarded during the \textit{inference} phase, and the input of segment pair $\bm{x}$ is emulated by replicating the original BFI segment.

\subsection{Recovering Sparse BFI Time Series} \label{ssec:recovery}
\begin{figure}[t]
\setlength{\abovecaptionskip}{8pt}
\centering
\subfigure[Keystroke missed.]{
    \begin{minipage}[t]{0.45\linewidth}
    \label{BFI:traffic_missed}
    \centering
    \includegraphics[width=1\textwidth]{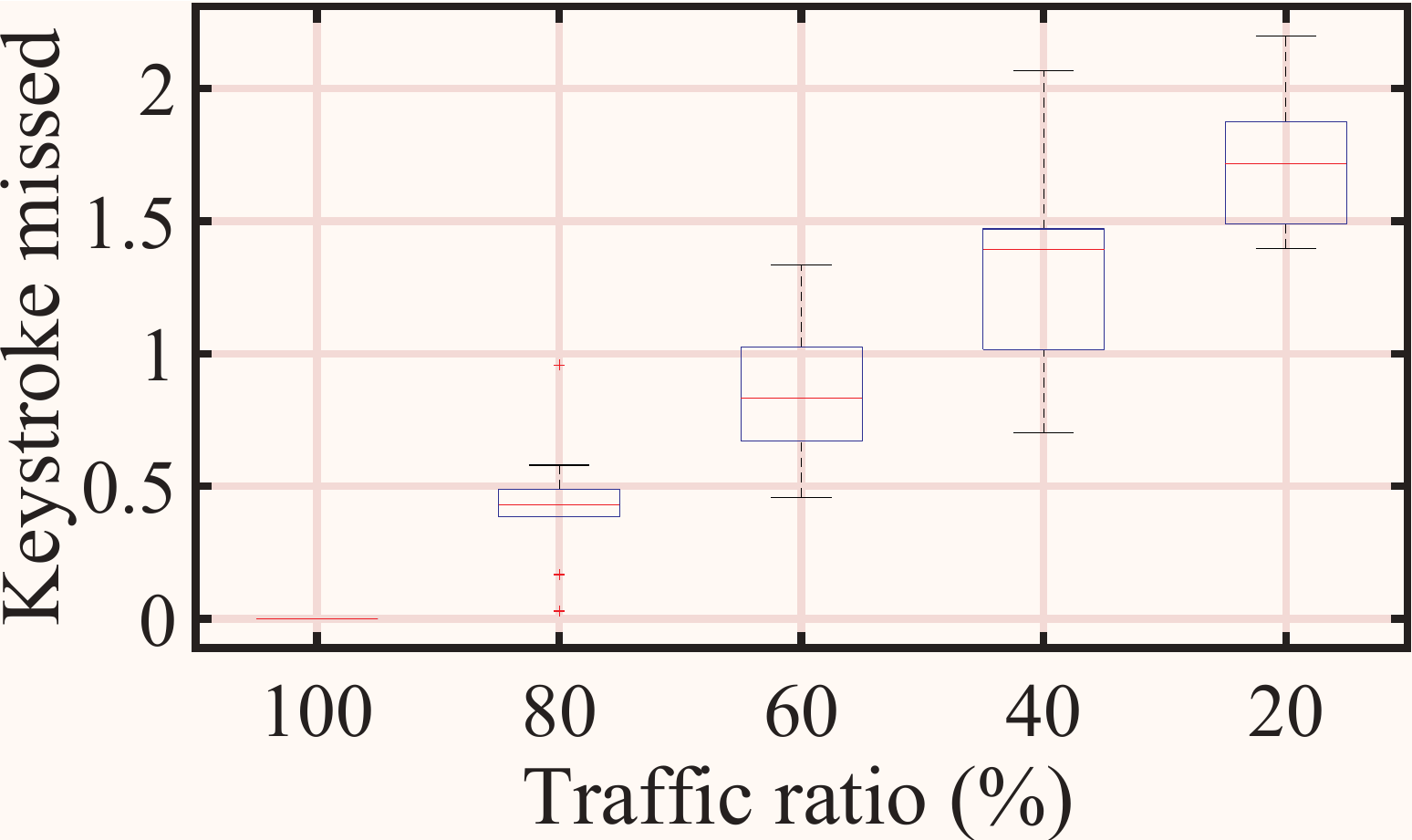}
    \end{minipage}
}
\subfigure[Classification accuracy.]{
    \begin{minipage}[t]{0.45\linewidth}
    \label{BFI:traffic_accuracy}
    \centering
    \includegraphics[width=1\textwidth]{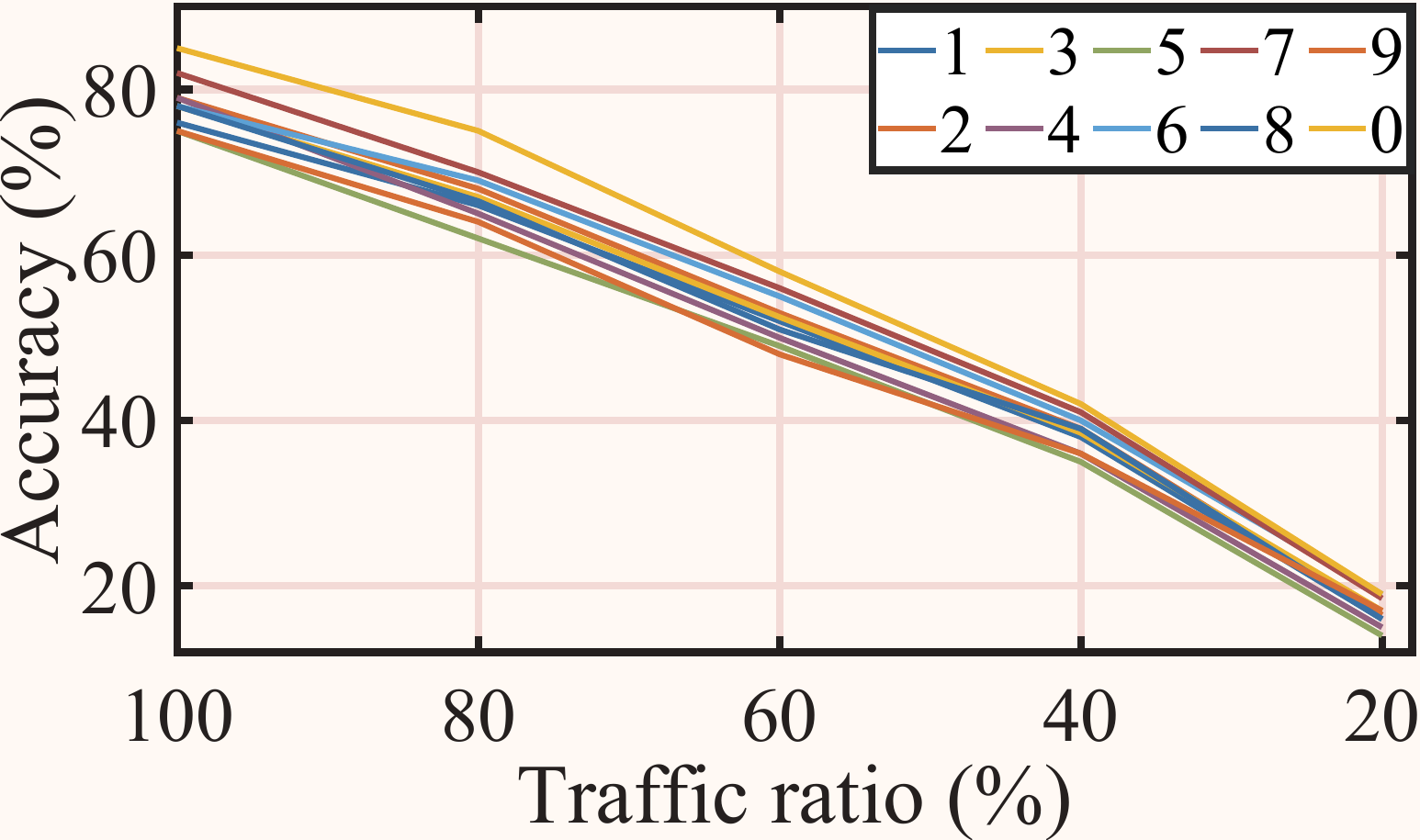}
    \end{minipage}
}
\centering
\caption{\hjy{Keystroke missing and affected classification accuracy under different traffic rates.}}
\label{Diss:traffic ratio}
\end{figure}


Another potential challenge to \name\ is traffic sparsity under extreme background traffic conditions:
the BFI series may hence become temporally sparse, containing discontinuous samples that negatively impact the KI. 
To study how sparse traffic affects keystroke missed and classification accuracy, we use iPerf~\cite{tirumala1999iperf} to generate 
data traffics constantly exchanged between the device and AP. 
We define 
five different traffic ratios as 100\% (saturated), 80\%, 60\%, 40\%, and 20\%. 
Given a certain ratio, the traffic generation follows a Poisson process~\cite{kingman1992poisson}.
Take the 6-digit password as an example, 
one may observe that the number of keystrokes missed almost increases linearly with sparsity as shown in Figure~\ref{BFI:traffic_missed}. When the traffic ratio is 20\%, up to 2 keystrokes can be missed. Even for those keystrokes not missed, as shown in Figure~\ref{BFI:traffic_accuracy}, the accuracy of classifying a single digit decreases from 80\% to less than 20\% when the traffic ratio decreases to 20\%. 
In Section~\ref{sec:Traffic_Analysis}, we study the impact of five different real-life background traffic on BFI sparsity. Unlike the emulated results in Figure ~\ref{BFI:traffic_missed}, real-life background traffic appears to be more benign so that we barely observe severe sparsity causing missed keystrokes.

\begin{figure}[b]
\setlength{\abovecaptionskip}{8pt}
\centerline{\includegraphics[width=0.48\textwidth]{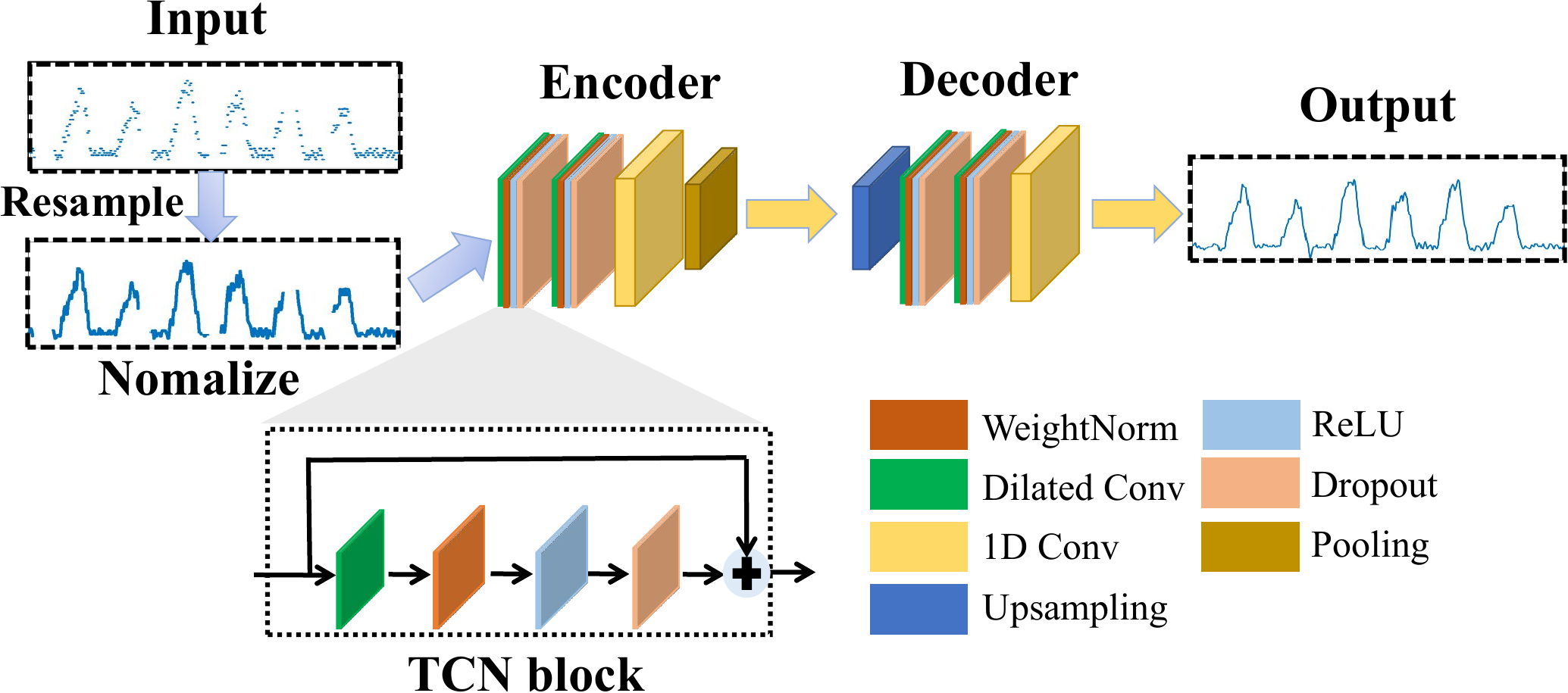}}
\caption{\newrev{The neural models of \name's SRA.}}
\label{fig:SpareOverview}
\vspace{-.5ex}
\end{figure}

To this end, we propose SRA (sparse recovery algorithm) for \name;
\hjy{it is invoked only if no keystroke is missing. Specifically, we use a sliding window of length $\Delta t = 1$~\!s to check whether sufficient samples are included; if there is a continuous 50\% period without BFI within the sliding window, the attack fails; otherwise, the SRA is initiated.}
As illustrated in Figure~\ref{fig:SpareOverview}, SRA starts with resampling the collected series to \newrev{make it evenly spaced} with a sampling frequency of $f_s$. Subsequently, \newrev{the series is normalized to the range of $[0, 1]$, and sparse segments void of data samples are further tagged} with -1 to denote their missing status. \hjy{After resampling, We represent the input data to SRA as the one-dimensional time series ${x_t}$ extracted from the BFI, where $t$ is the sampling time. SRA outputs ${y_t}$ as a uniform and densely sampled time series.}

\newrev{To generate the missing samples,} SRA employs a TCN \textit{(temporal convolutional network)} based AE \textit{(autoencoder)} network, consisting of an encoder and a decoder as depicted in Figure~\ref{fig:SpareOverview}. TCN is a more suitable choice over other types of neural networks, such as LSTMs~\cite{yu2019review}, as it uses convolutional layers with dilated kernels to capture long-range dependencies in samples while keeping the number of parameters manageable~\cite{bai2018empirical}. 
The encoder network maps the input BFI series to a latent representation containing the intrinsic features of the input.
Subsequently, the decoder network relies on this representation to
reconstruct a non-sparse series. 
The TCN-AE is trained in a self-supervised manner: we first generate non-sparse BFI series as ground truth using saturated traffic, then we randomly remove data samples to create sparse series that emulate realistic sparsity by following the \newrev{temporal} distribution of real-life BFI series generated \newrev{under sparse traffic.}


\section{Implementation and Setup}
\label{sec:implementation}
In this section, we elaborate on \name's implementation, as well as introduce the experiment setup and metrics.

\vspace{-1.5ex}
\paragraph{System Implementation}
Though \newrev{a rooted smartphone under the monitor mode} can act as Eve, \newrev{Android systems} offer minimal support in capturing Wi-Fi traffic at application layer. Therefore, we focus on a laptop implementation in our experiments. We use an Acer TravelMate laptop~\cite{acer} with an Intel AX210 Wi-Fi NIC~\cite{intel} supporting 802.11b/g/n/ac 
\newrev{as the basis; setting the
NIC to the monitor mode, we then use} WireShark to capture the BFI series contained in Action No-ACK frames. The captured BFIs are analyzed using Matlab and Python, with the neural \newrev{models} built upon PyTorch 1.7.1~\cite{pytorch}. For the segmentation, the two parameters $\alpha$ and $\beta$ are set to 0.6 and 0.5, respectively. In the adversarial learning framework, the balance factor $\lambda$ is set to 0.5. For sparse recovery, the sampling frequency $f_s$ is set to 40~\!Hz. \hjy{Our collected data and code for preprocessing the data are publicly available online~\cite{WiKI}.}

%

\vspace{-1.5ex}
\paragraph{Experiment Setup} \label{ssec:setup}
We recruit 20 subjects, of 12 males and 8 females, between the ages of 20 and 53. All subjects are right-handed and use their own smartphones of various models, including iPhone 13~\cite{iphone}, OnePlus 10T~\cite{oneplus}, Xiaomi 13 Pro~\cite{xiaomi}, Huawei P40 Pro~\cite{huawei}, Samsung Galaxy S20~\cite{samsung}, and Google Pixel 6a~\cite{google}. The subjects \needrev{type a total of 1,500 predefined passwords} of 4, 6, and 8 \newrev{digits}, with each length having 5,000 passwords. During typing, background apps remain active to \newrev{emulate daily smartphone usage.}
The subjects adopt different postures while typing on the smartphones, such as holding it with one or both hands or placing it on a stand or table. The typing speed of the subjects ranges from 0.5 to 2~\!cps (\textit{characters per second}). These experiments have strictly followed our IRB.

\begin{figure}[t]
    \setlength\abovecaptionskip{3pt}
    \centering
    \subfigure[Experiment scene.]{
	    \includegraphics[width=.57\linewidth]{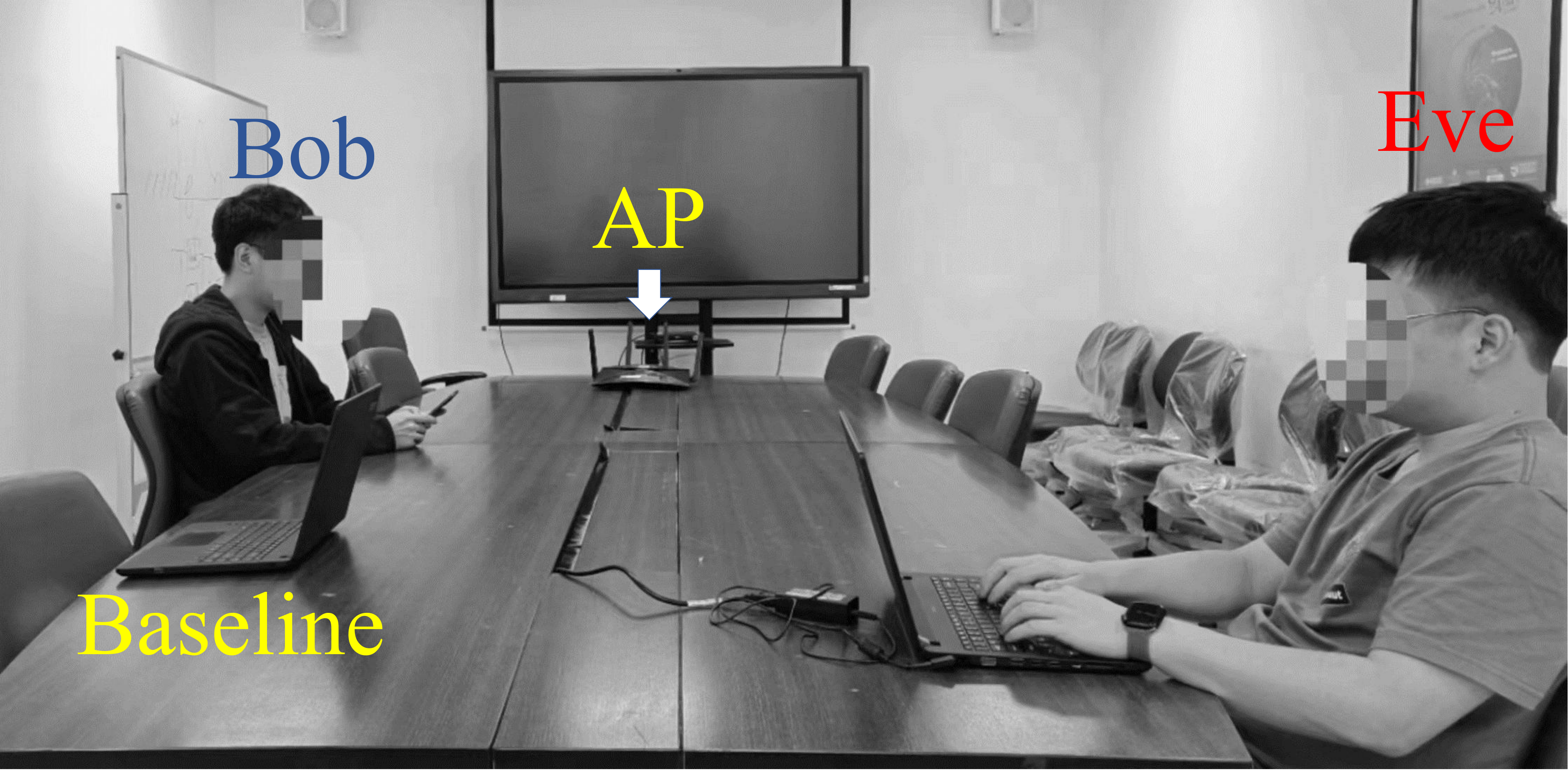}
	    \label{sfig:Layout}
    }
    \hfill
    \subfigure[Adopted hardware.]{
		\includegraphics[width=.37\linewidth]{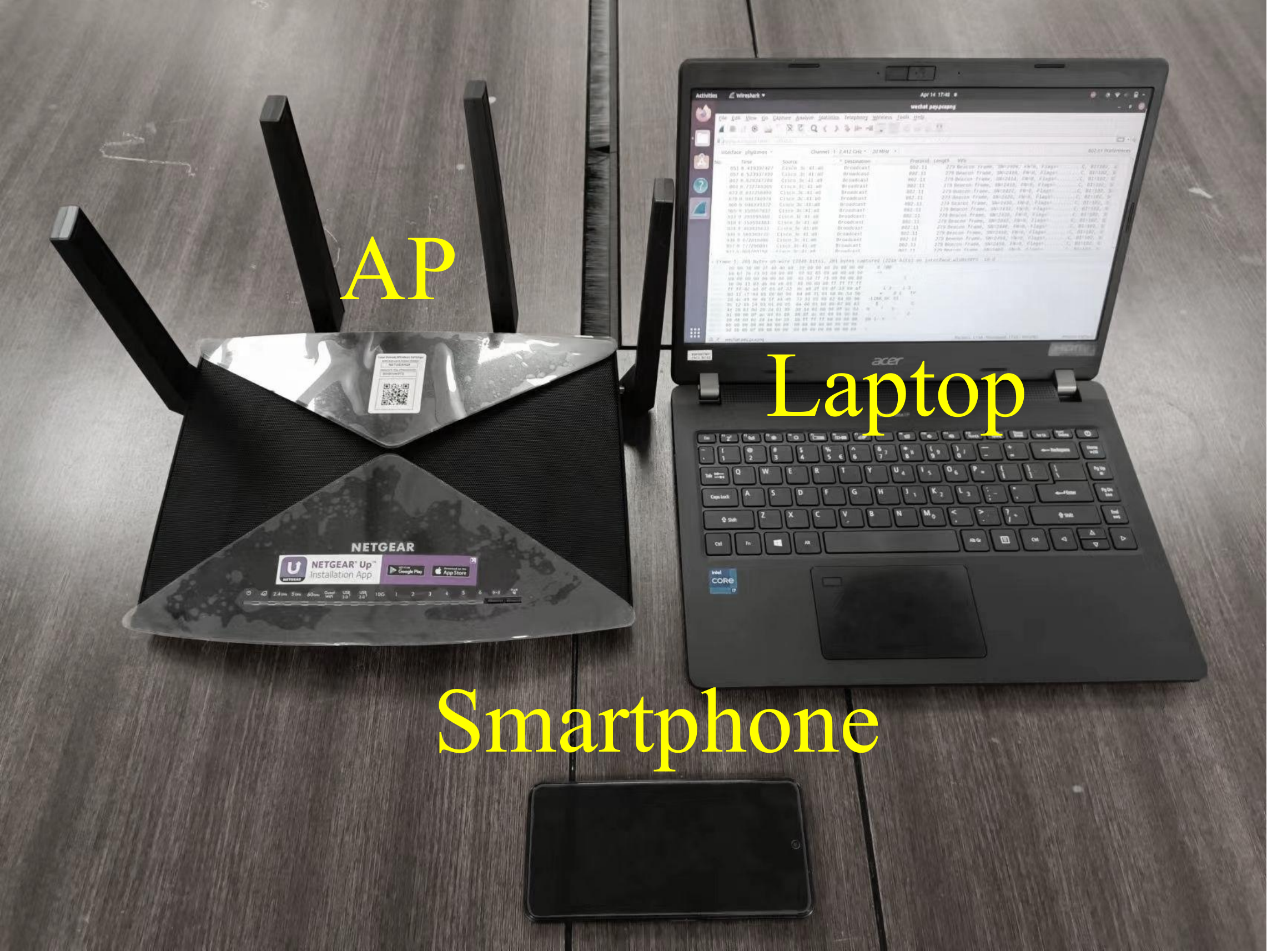}
		\label{sfig:Implementation}
    }
\caption{\newrev{Evaluative \name: (a) experiment scene in a conference room} and (b) hardware configurations.}
\label{fig:experiment_scene}
\end{figure}

We conduct experiments and collect BFI series in six environments, including a library, bookstore, auditorium, cafeteria, corridor, and conference room. In each environment, a Wi-Fi router working as an AP for the subjects to connect. Besides BFI collection, we simultaneously obtain CSIs from the AP and another laptop to respectively serve as comparison baselines of WindTalker~\cite{li2016csi} and WINK~\cite{yang2022wink}. \newrev{The distance between a subject and the AP ranges from 1 to 10~\!m, and the distance between the attacker and the subject ranges from 3 to 10~\!m.} Figure~\ref{fig:experiment_scene} shows an example experiment scene and the hardware we use. \name\ segments the BFI series using the overlapping scheme described in Section~\ref{ssec:segmentation}, while the baselines conduct segmentation according to their respective proposals~\cite{li2016csi, yang2022wink}. We use 70\% of the collected data for training and the remaining 30\% for testing.

\vspace{-1.5ex}
\paragraph{Metrics} \label{ssec:metric}
We adopt two metrics \newrev{for our evaluations,} namely keystroke \textit{classification accuracy} and top-$N$ password \textit{inference accuracy}. For single keystroke classification, \newrev{the classification accuracy measures} the percentage of correctly classified keystrokes. For password inference, since an attacker may try multiple passwords to increase the success rate, we adopt the top-$N$ accuracy as the evaluation metric: \newrev{the probability of a candidate password is computed as the product of the probability of each key present in the password, then the top-$N$ accuracy is measured by checking if any of the candidates within top-$N$ probability matches the true one.}
%


\section{Evaluation}
\label{sec:evaluation}
\newrev{We start with two micro-benchmark studies to demonstrate the effectiveness of \name's building blocks. These are followed by evaluations on overall performance and the impact of practical factors. Finally, we conduct real-world experiments to showcase how \name\ steals passwords of WeChat Pay, while also extending it to general KI on QWERTY keyboards.}

\subsection{Micro-benchmark Studies}

\vspace{.5ex}
\subsubsection{Domain Adaptation}
\begin{figure}[b]
    \setlength\abovecaptionskip{3pt}
    \centering
    \hspace{-0.5em}
    \subfigure[w/o adversarial learning.]{
	    \includegraphics[width=.48\linewidth]{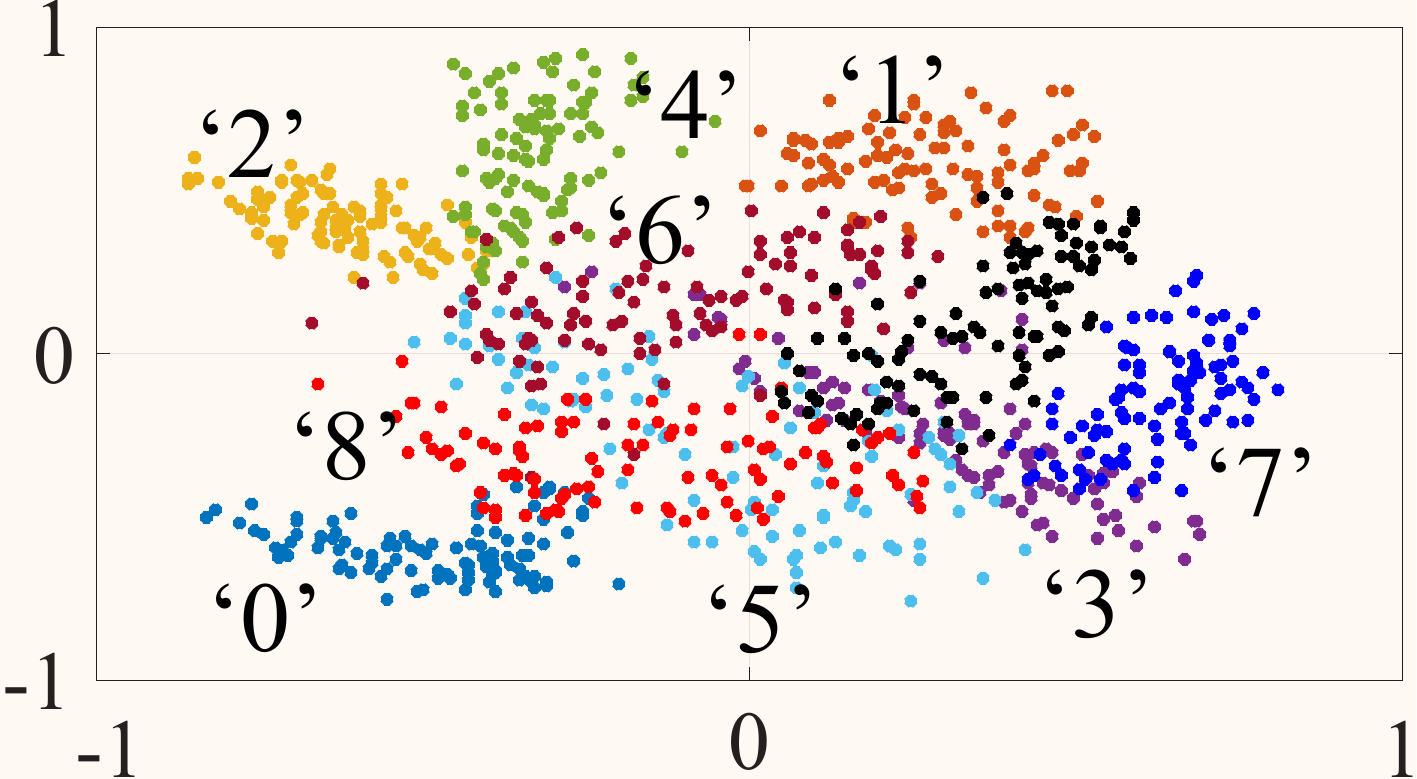}
	    \label{sfig:tSNE_no_adv}
    }
    \hspace{-0.5em}
	\subfigure[w/ adversarial learning.]{
		\includegraphics[width=.48\linewidth]{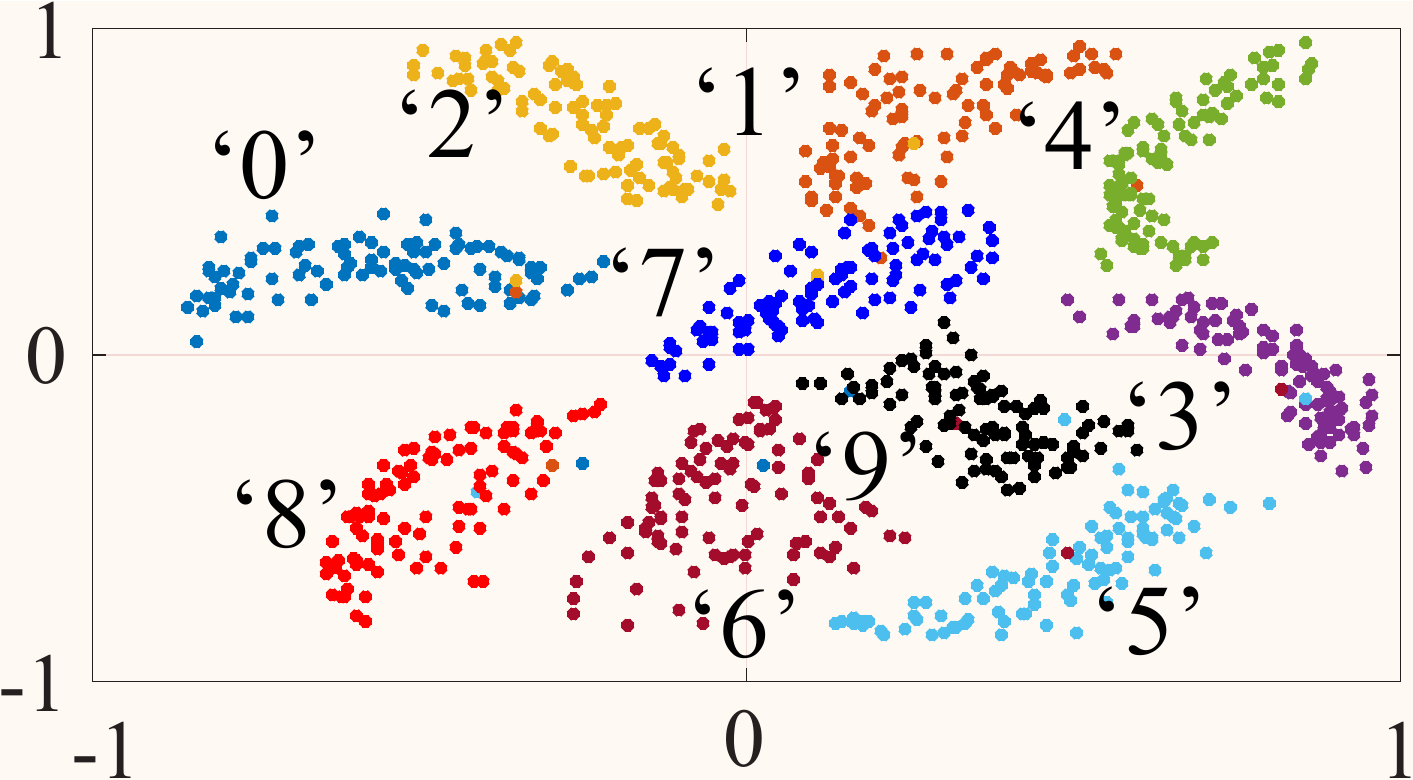}
		\label{sfig:tSNE_adv}
	}
\caption{t-SNEs of the features output by the feature extractor $G_\mathrm{f}$ evidently confirm that adversarial learning results in domain-invariant representations.}
\label{fig:t_SNE}
\vspace{-.5ex}
\end{figure}

To demonstrate the effectiveness of \name's adversarial learning framework in Section~\ref{sssec:keystroke_inference}, we use t-SNE (t-Distributed Stochastic Neighbor Embedding)~\cite{maaten2008visualizing} to visualize the feature maps of 10 numerical keys segmented from 100 random passwords in Figure~\ref{fig:t_SNE}. As shown in Figure~\ref{sfig:tSNE_no_adv}, the \newrev{normal feature extractor $G_\mathrm{f}$ fails to find a domain-invariant feature map: features of different keys apparently
get mixed together due to domain interference.}
In contrast, Figure~\ref{sfig:tSNE_adv} demonstrates that, with adversarial learning, the features of the same keys are consistent across domains and form distinct clusters, indicating that domain-invariant representations have been successfully learned.

\subsubsection{Sparse Recovery}
\begin{figure}[t]
    \setlength\abovecaptionskip{3pt}
    \centering
    \hspace{-0.5em}
    \subfigure[Illustration of sparse recovery.]{
	    \includegraphics[width=.48\linewidth]{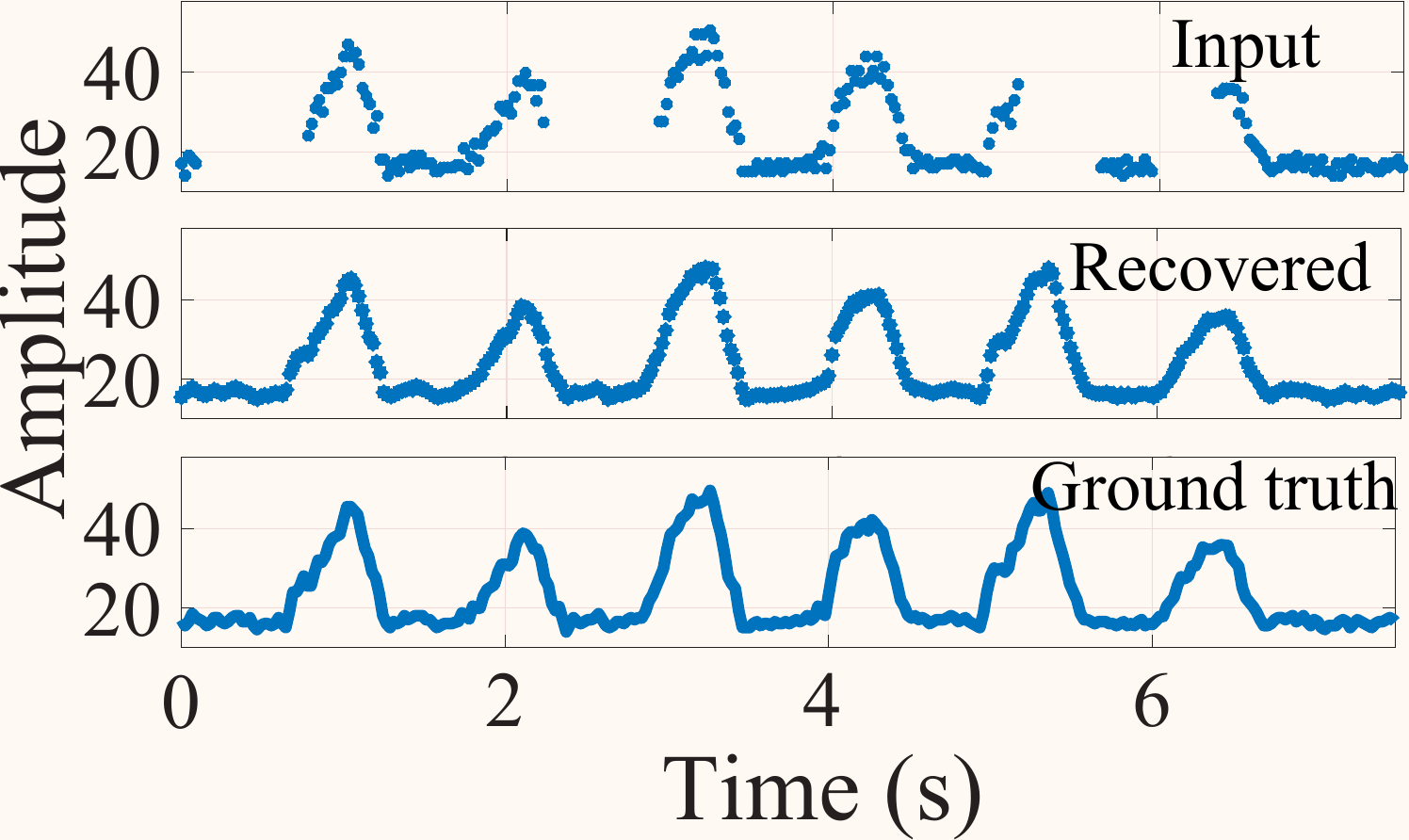}
	    \label{sfig:spa_rec}
    }
    \hfill
    \hspace{-0.5em}
	\subfigure[Relative RMSE of sparse recovery.]{
		\includegraphics[width=.48\linewidth]{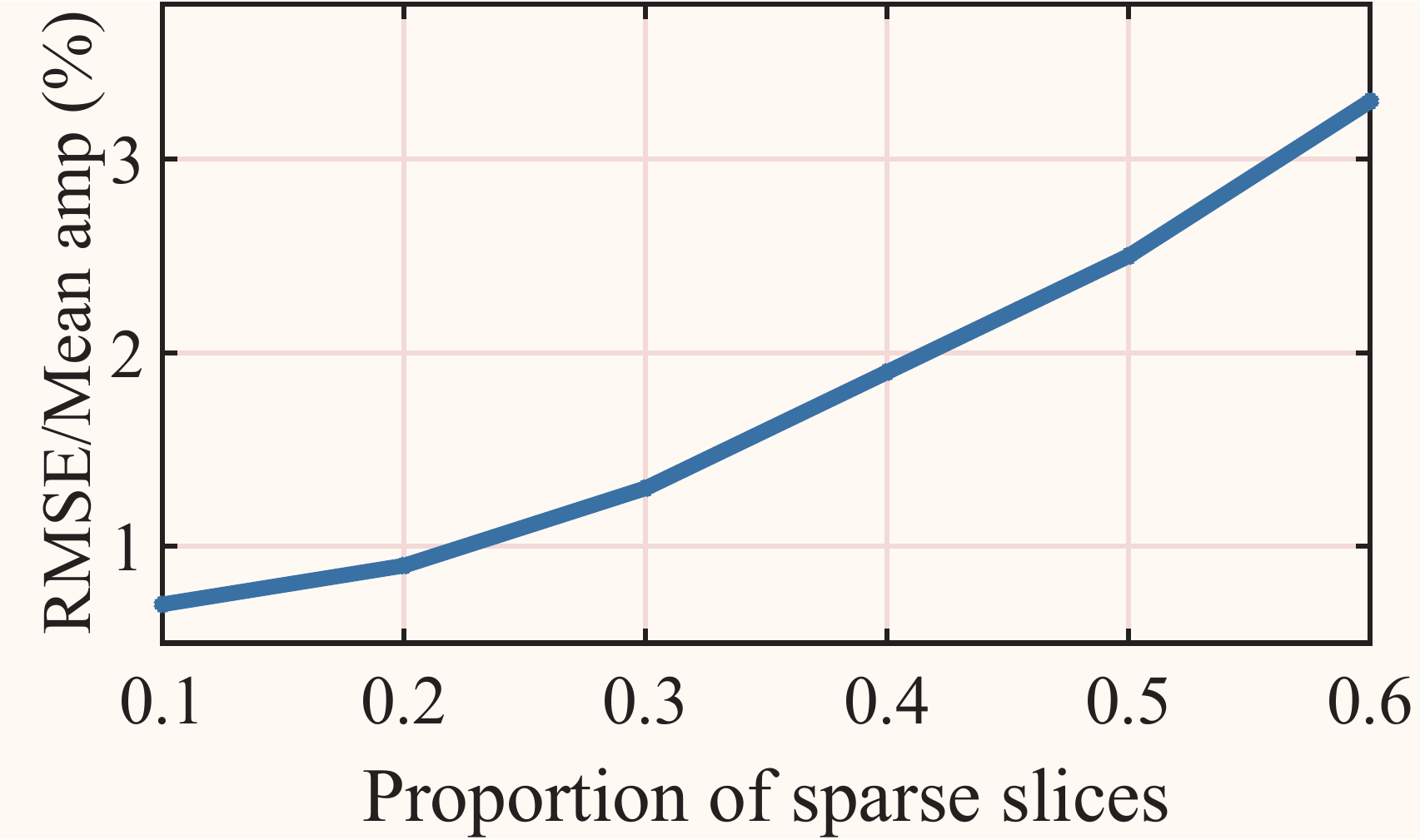}
		\label{sfig:mse_loss}
	}
\caption{Performance of sparse recovery.}
\label{fig:sparse_recovery}
\vspace{-1em}
\end{figure}

We apply SRA to recover the non-sparse BFI time series from \newrev{passwords typed by a subject,} and evaluate its effectiveness using the Root Mean Squared Error (RMSE) between the recovered and the ground truth series. As shown in Figure~\ref{sfig:spa_rec}, our TCN-AE enabled SRA is able to recover a series with high similarity to the ground truth, capturing details when the subject's finger hits on screen as well as during transition periods. Furthermore, Figure~\ref{sfig:mse_loss} illustrates how RMSE changes with the proportion of missing BFI segments: even when 60\% of the BFI segments are missing, 
\newrev{SRA still achieves a sufficiently low RMSE at 3.3\% of the mean amplitude of the ground truth series, indicating a successful recovery and providing a solid basis for \name's ultimate password inference.}

\subsection{Overall Performance}
\subsubsection{Classification Accuracy} 
\label{sssec:classify_acc}
\begin{figure}[b]
    \setlength\abovecaptionskip{3pt}
    \centering
    \subfigure[\hjy{Overall accuracy.}]{
	    \includegraphics[width=.465\linewidth]{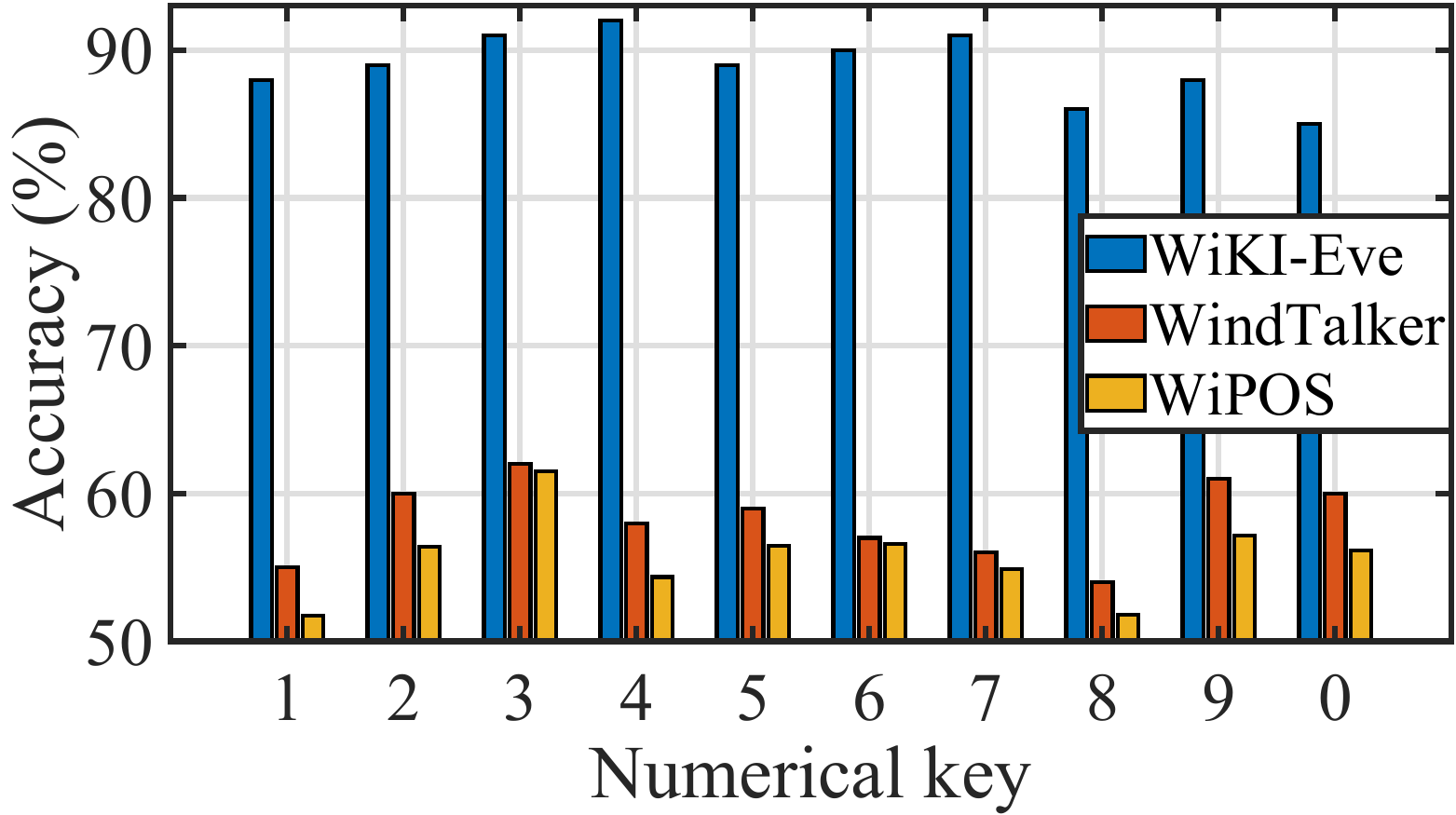}
	    \label{sfig:clas_acc}
    }
    \hspace{0.5ex}
    \subfigure[Confusion matrix.]{
        \includegraphics[width=.465\linewidth]{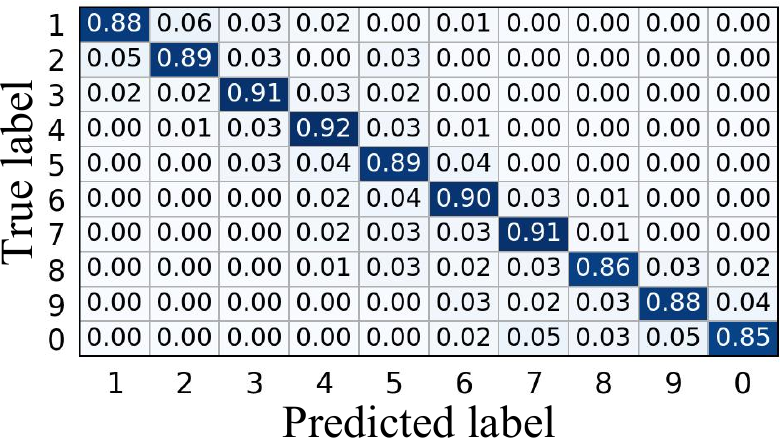}
        \label{sfig:conf_matrix1}
    }
\caption{\hjy{Comparing the classification accuracy of \name\ with WindTalker and WiPOS.}}
\label{fig:classification_accuracy}
\vspace{-.5ex}
\end{figure}

In this section, we present the accuracy of classifying numerical keys of \name, and compare it with \hjy{two baseline methods WindTalker~\cite{li2016csi} and WiPOS~\cite{zhang2020wipos}}. We do not compare \name\ with WINK~\cite{yang2022wink} in terms of keystroke classification accuracy because WINK is based on series learning that predicts the password as a whole. As shown in Figure~\ref{sfig:clas_acc}, the classification accuracy of \name\ for keys `0' to `9' remains steady at around 88.9\%, while WindTalker \hjy{and WiPOS  achieve} an average accuracy of only 58.2\% \hjy{and 55.1\%, respectively,} which is significantly lower than \newrev{that reported in~\cite{li2016csi}, and should hence be further explained in Section~\ref{sssec:password_infer_acc}.} To further analyze the classification accuracy for each key, we present the confusion matrix of \name\ in Figures~\ref{sfig:conf_matrix1}. It is intuitive that each key is most commonly confused with adjacent keys (e.g., the key `5' is most commonly confused with `2', `4', `6', and `8'). Despite the inevitable confusion, the high success rate of classifying individual keys lays a solid foundation for later password inference.

The superiority of \name\ over WindTalker and WiPOS can be attributed to two reasons. As discussed in Section~\ref{ssec: motiv bfi}, BFI \newrev{dampens}
the close impact of IKI from \newrev{its} on-screen keystrokes, making \name\ more stable than CSI-based \newrev{approaches}. This allows \name\ to extract consistent features 
\newrev{effectively learnable by its neural models.} WindTalker and WiPOS, on the \newrev{contrary}, suffers from CSI \newrev{noises 
possibly} confused with useful features. \newrev{Moreover}, the overlapping segmentation technique proposed in Section~\ref{ssec:segmentation} \newrev{endows \name\ with richer domain ``context’’ for its} adversarial learning framework.
In contrast, WindTalker's and WiPOS's rule-based segmentation \newrev{potentially discards certain} essential parts of the CSI \newrev{features already overwhelmed by noises, thus failing to obtain full representations for individual keystrokes.}






\begin{figure}[t]
    \setlength\abovecaptionskip{3pt}
    \centering
    \hspace{-0.5em}
    \subfigure[Top-1 to top-10.]{
	    \includegraphics[width=.479\linewidth]{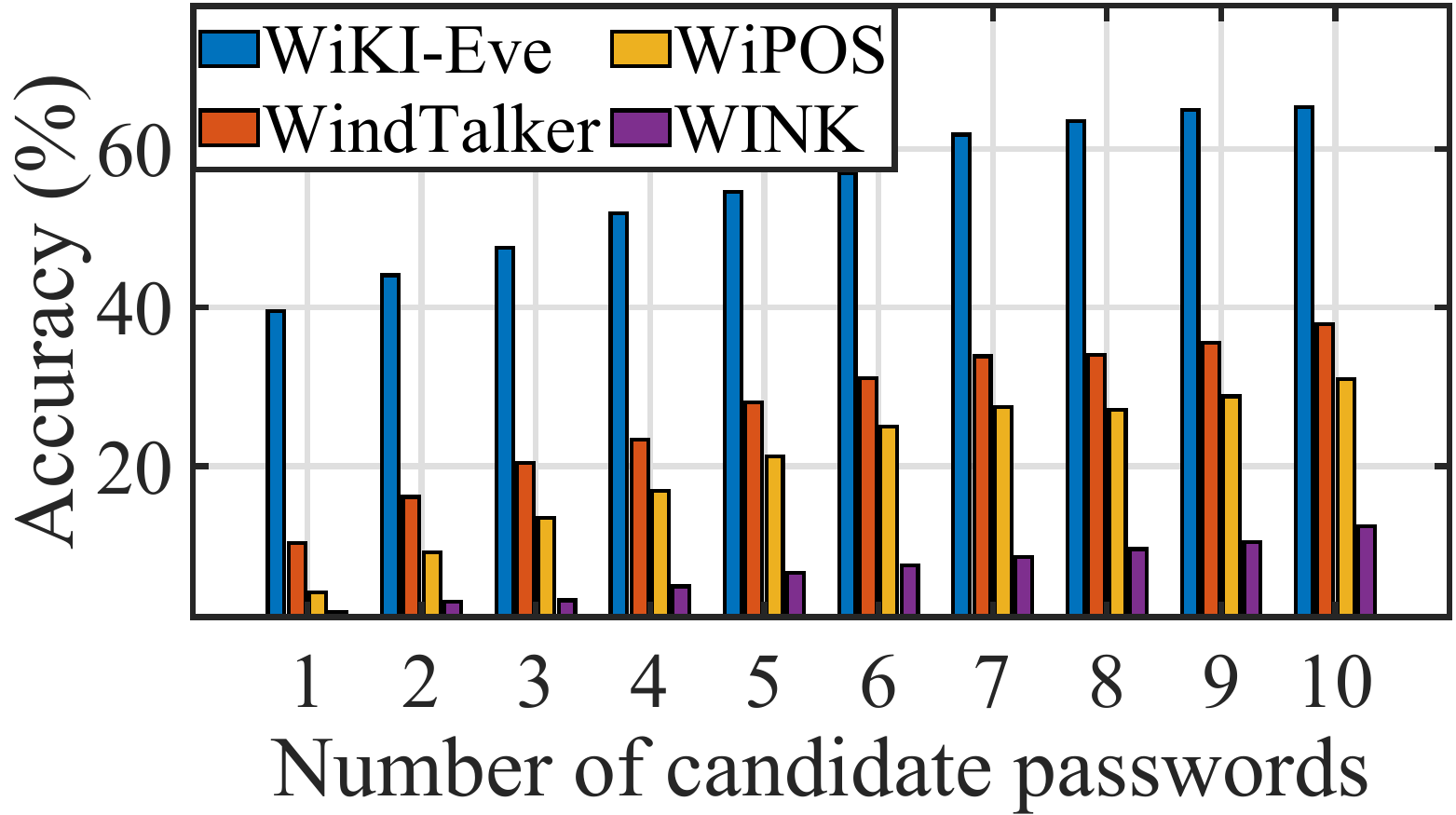}
	    \label{sfig:top10_can}
    }
    \hfill
    \hspace{-0.5em}
	\subfigure[Top-10 to top-100.]{
		\includegraphics[width=.477\linewidth]{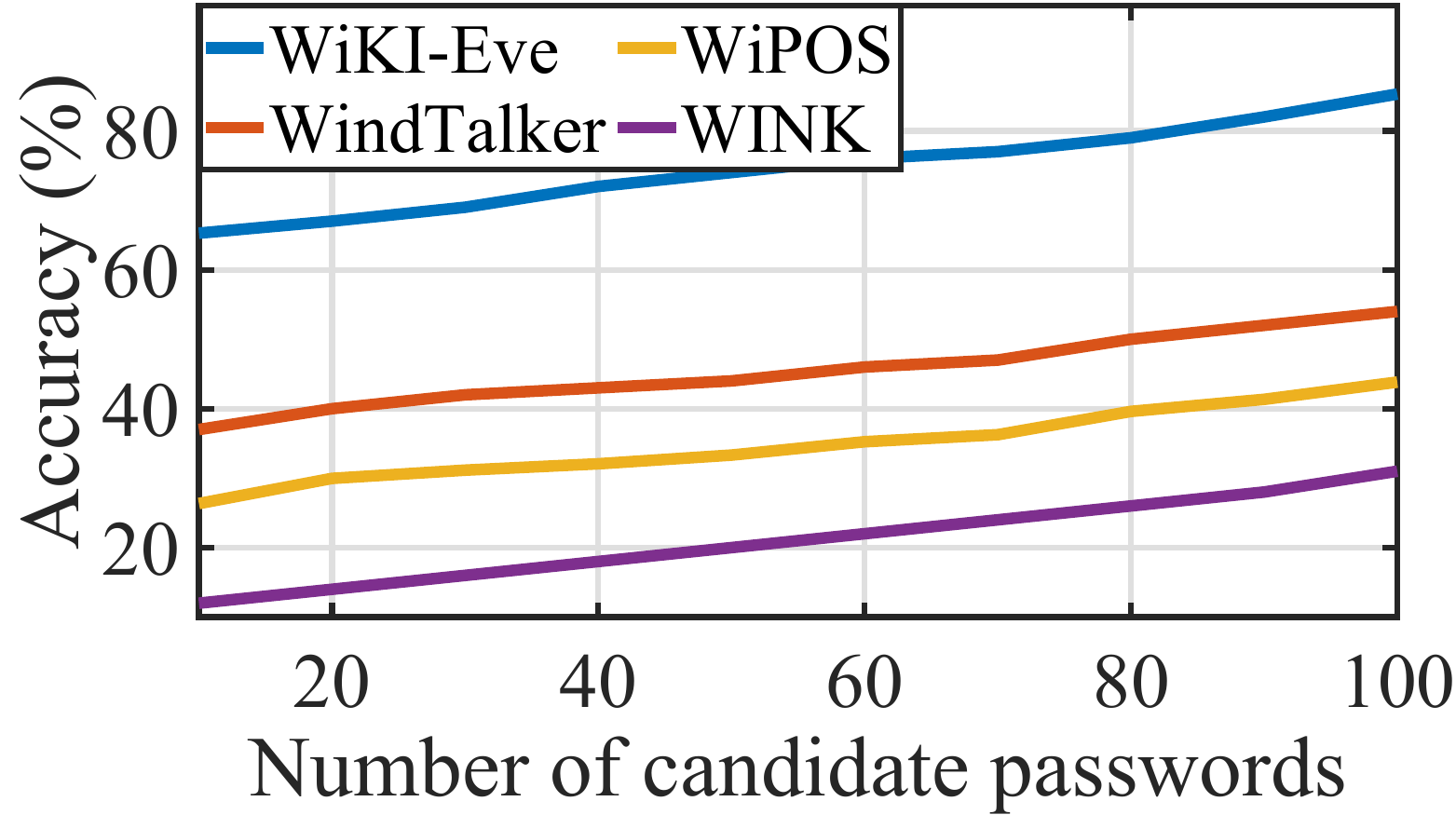}
		\label{sfig:top100_can}
	}
\caption{\hjy{Comparison for password inference accuracy under different numbers of password candidates.}}
\label{fig:pass_inf_acc}
\vspace{-1em}
\end{figure}

\subsubsection{Password Inference Accuracy} \label{sssec:password_infer_acc}
%
\newrev{Let us} further evaluate \name's password inference capability, focusing on 6-digit numerical passwords due to their widespread usage in daily scenarios, but \newrev{leaving the performance assessment 
for different password lengths to} Section~\ref{sssec:length}. Figure~\ref{sfig:top10_can} compares the top-1 to -10 accuracy of \name, WindTalker, \hjy{WiPOS,} and WINK: while \name's accuracy varies from 40\% to 65\% for top-1 to -10 candidates, WindTalker's, \hjy{WiPOS's,} and WINK's only reach 37\%, \hjy{32\%,} and 12\% for top-10 accuracy, respectively. Figure~\ref{sfig:top100_can} indicates that \name\ can infer passwords with an 85\% success rate in 100 attempts, yet WindTalker, \hjy{WiPOS,} and WINK can only achieve a rate of 54\%, \hjy{42\%,} and 31\% at the same number of attempts.

The reasons for \name's superiority in Section~\ref{sssec:classify_acc} also apply to explain \name's \newrev{much better} performance in password inference than \hjy{all baselines.} Additionally, \name\ has an edge over \hjy{WiPOS} and WINK because \model\ has a higher SNR than OKI, and the digital nature of BFI prevents fidelity loss of sensing signal. 
\newrev{One may notice the performance discrepancy of \hjy{all baselines} 
from that reported in~\cite{li2016csi, zhang2020wipos, yang2022wink}, as also highlighted in Section~\ref{sssec:classify_acc} for WindTalker. This may stem from their designs failing to properly take into account the influence of domain,} thereby limiting their ability to effectively handle diverse data collected from various domains in our experiment setup.





\vspace{-.5ex}
\subsection{Impact of Practical Factors}

\subsubsection{Environments and Subjects}
\begin{figure}[t]
    \setlength\abovecaptionskip{3pt}
    \centering
    \hspace{-0.5em}
    \subfigure[Different environments.]{
	    \includegraphics[width=.482\linewidth]{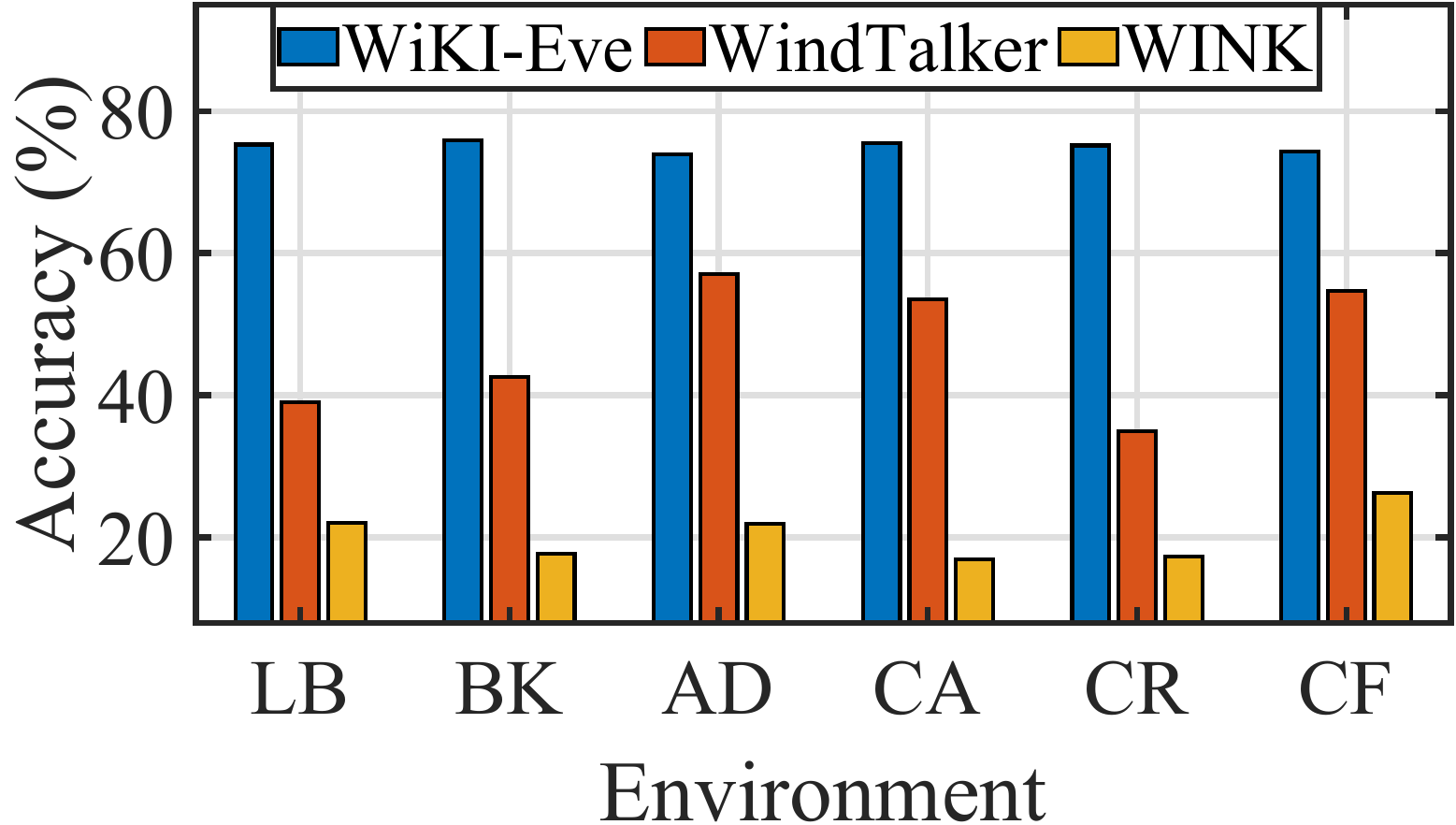}
	    \label{sfig:scenarios}
    }
    \hfill
    \hspace{-0.5em}
	\subfigure[Different subjects.]{
		\includegraphics[width=.482\linewidth]{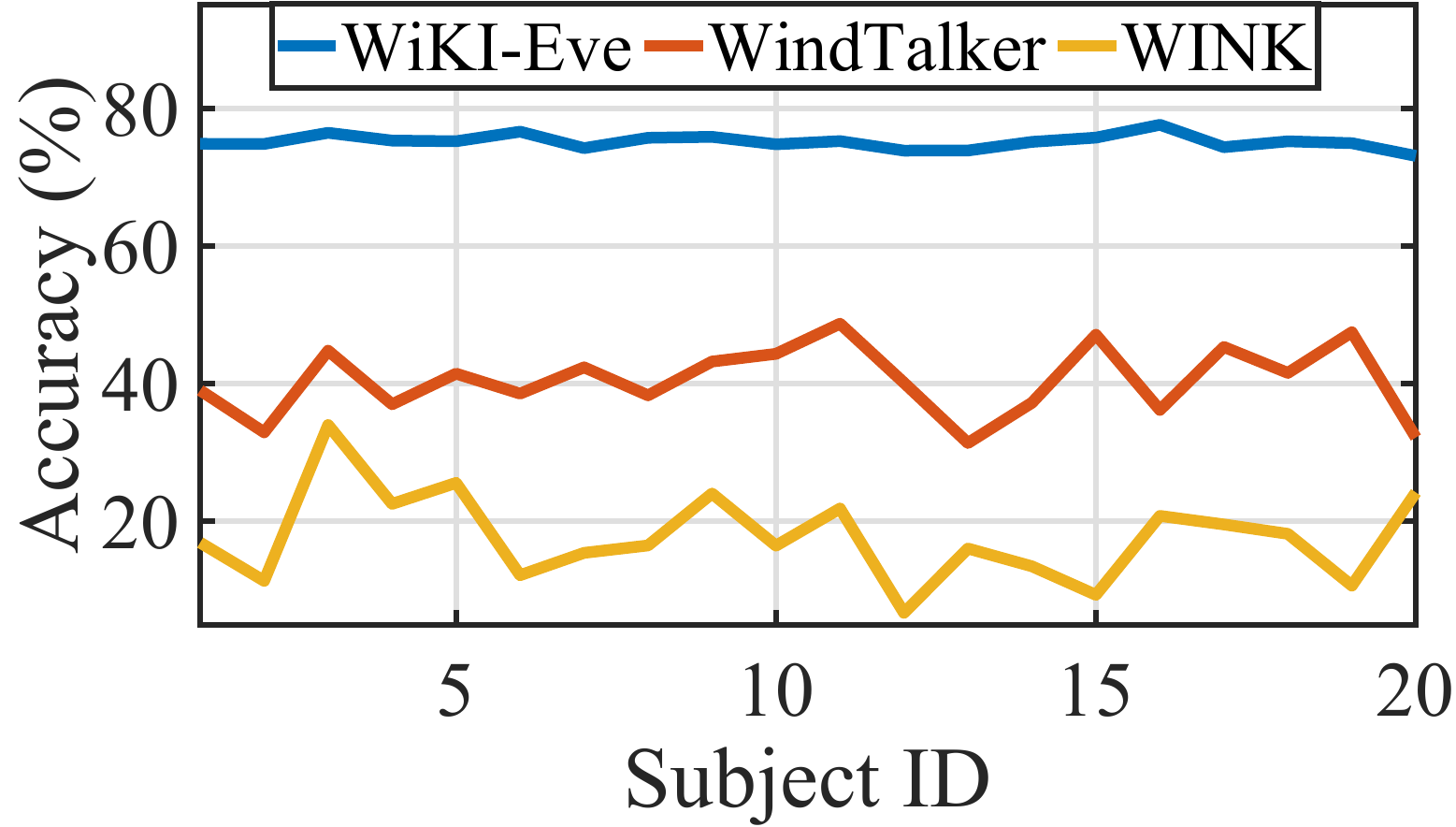}
		\label{sfig:subjects}
	}
\caption{Impact of environments and subjects.}
\label{fig:scenarios_subjects}
\end{figure}

We use the ``leave-one-out'' strategy~\cite{wong2015performance} to study the impacts of different environments and subjects. This means that the test set consists of all data from one of the 6 environments or one of the 20 subjects, \newrev{leaving the rest to the training set.} Figure~\ref{sfig:scenarios} and ~\ref{sfig:subjects} respectively show the top-100 password inference accuracy for each environment and each subject. Although the testing environments and subjects are unseen during training, \name's top-100 accuracy across all cases is consistently above 75\%, thanks to the generalizability of the adversarial learning. Moreover, \name\ is robust across environments since \model\ relies on the diffraction pattern around the phone body \newrev{that are} rarely influenced by environment-specific interference. In contrast, the average top-100 accuracy of WindTalker and WINK drops from \newrev{that} in Figure~\ref{fig:pass_inf_acc} to less than 39\% and 18\%, due to their limited generalizability to unseen environments and subjects.


\subsubsection{Device Diversity}
\begin{figure}[h]
    \setlength\abovecaptionskip{3pt}
    \vspace{-1ex}
    \centering
    \hspace{-0.5em}
    \subfigure[Keystroke classification.]{
	    \includegraphics[width=.482\linewidth]{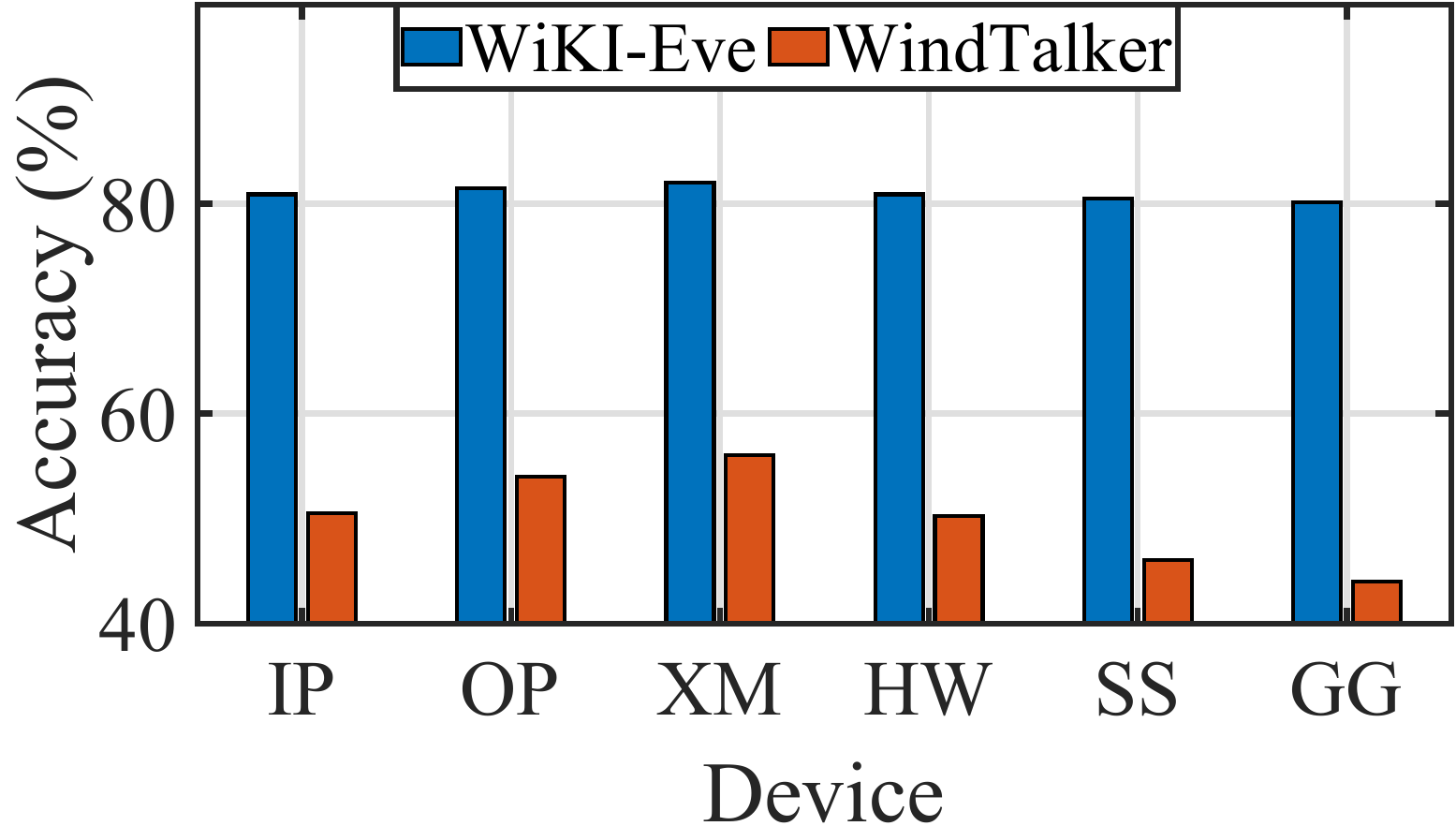}
	    \label{sfig:classify_acc_device_diversity}
    }
    \hfill
    \hspace{-0.5em}
	\subfigure[Password inference.]{
		\includegraphics[width=.482\linewidth]{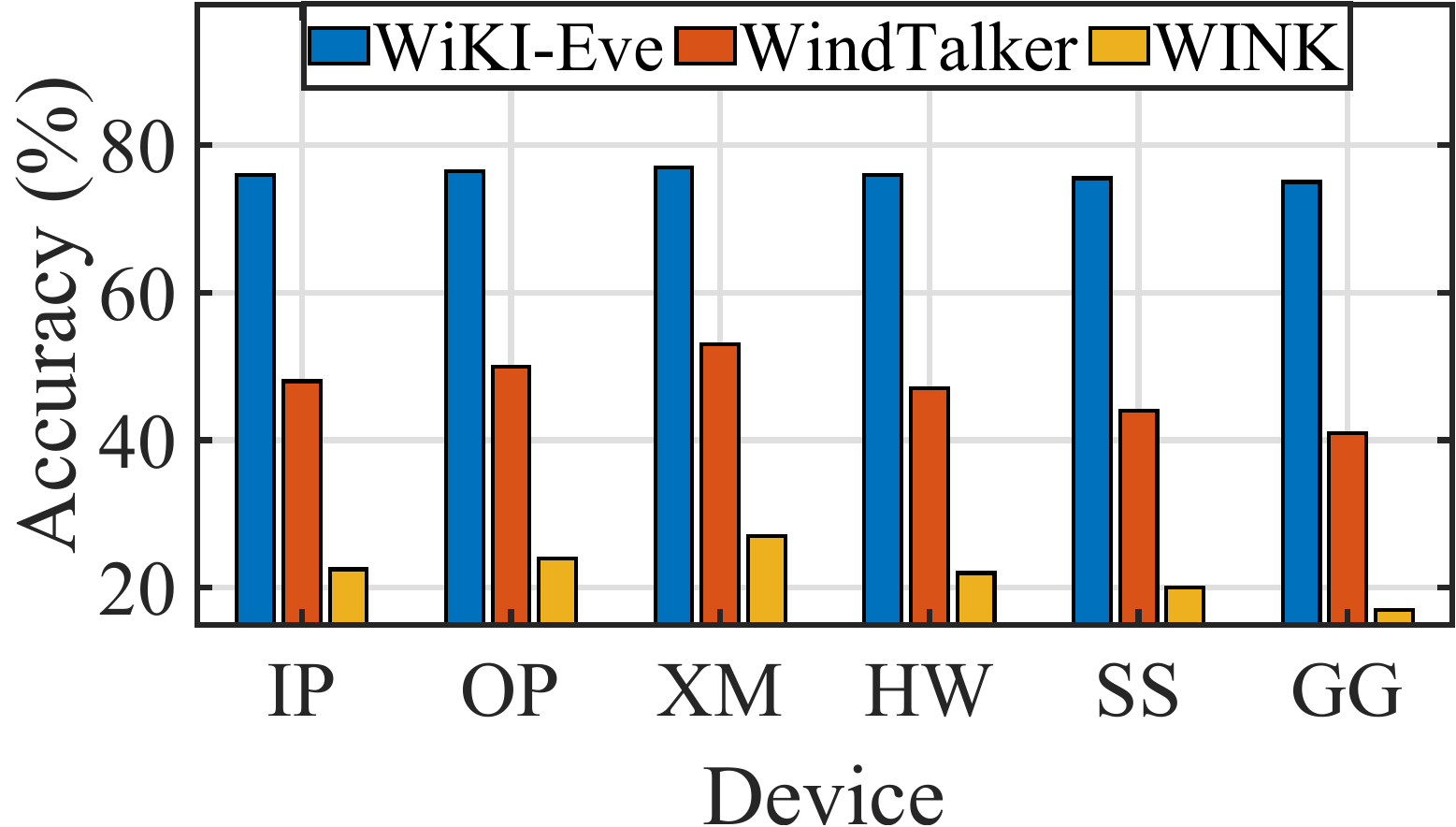}
		\label{sfig:inference_acc_device_diversity}
	}
\caption{Impact of device diversity.}
\label{fig:device_diversity}
\vspace{-1.5ex}
\end{figure}

%
%
\newrev{We again} use the ``leave-one-out'' strategy to evaluate the performance of \name on 6 \newrev{smartphones specified} in Section~\ref{ssec:setup}. Figure~\ref{sfig:classify_acc_device_diversity} shows that \name\ can reliably identify keystrokes on different devices, with an average keystroke classification accuracy of over 80\%, but WindTalker's accuracy is under 58\%. Furthermore, Figure~\ref{sfig:inference_acc_device_diversity} indicates that the top-100 password inference accuracy of \name, WindTalker, and WINK is respectively above 76\%, below 53\%, and below 27\%. The consistently high accuracy of \name\ across different smartphone devices confirms that our adversarial learning framework can generalize to unseen devices. In contrast, the low accuracy of the baselines (\newrev{evidently} worse than the results in Figure~\ref{fig:pass_inf_acc}) highlights their failure on unseen devices. One may also observe some accuracy variations among smartphones, which we attribute to different screen sizes. Specifically, \name\ achieves the highest accuracy on Xiaomi 13 Pro having the largest screen size (6.73~inch), while on Google Pixel 6a, with the smallest screen size (6.1~inch), it achieves the lowest accuracy. A possible explanation is smartphones with larger screens tend to have larger key distances that result in longer transition periods, thus making \newrev{the incurred BFI features} more distinguishable. Due to the consistently worse performance of the baselines, we do not compare \name\ with them in subsequent experiments.
%




\subsubsection{Distance}
\begin{figure}[t]
    \setlength\abovecaptionskip{0pt}
    \centering
    \hspace{-0.5em}
    \subfigure[Bob to the AP.]{
        \includegraphics[width=.478\linewidth]{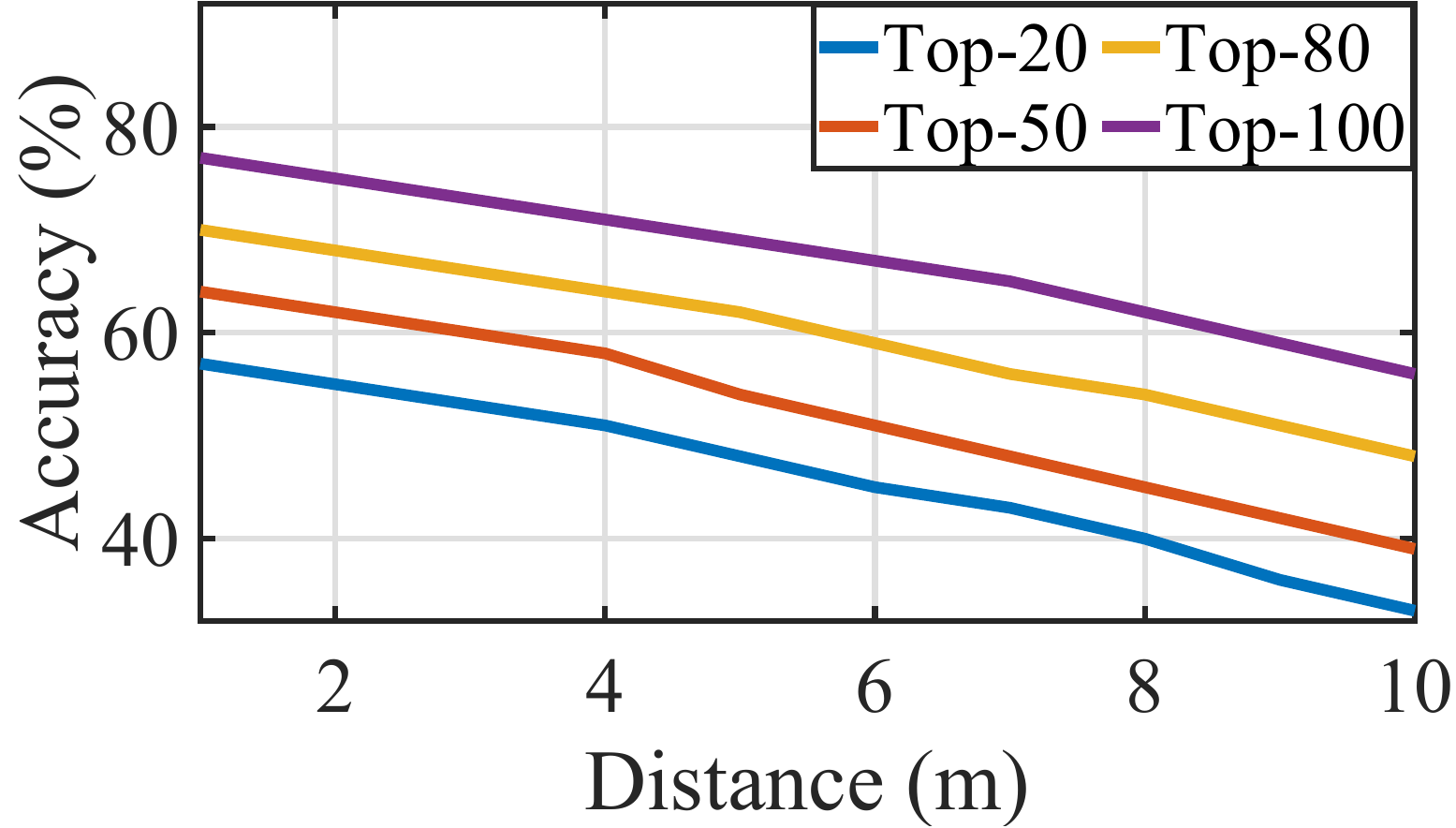}
        \label{sfig:dist_bob_to_ap}
    }
    \hfill
    \hspace{-0.5em}
    \subfigure[Eve to Bob.]{
	    \includegraphics[width=.478\linewidth]{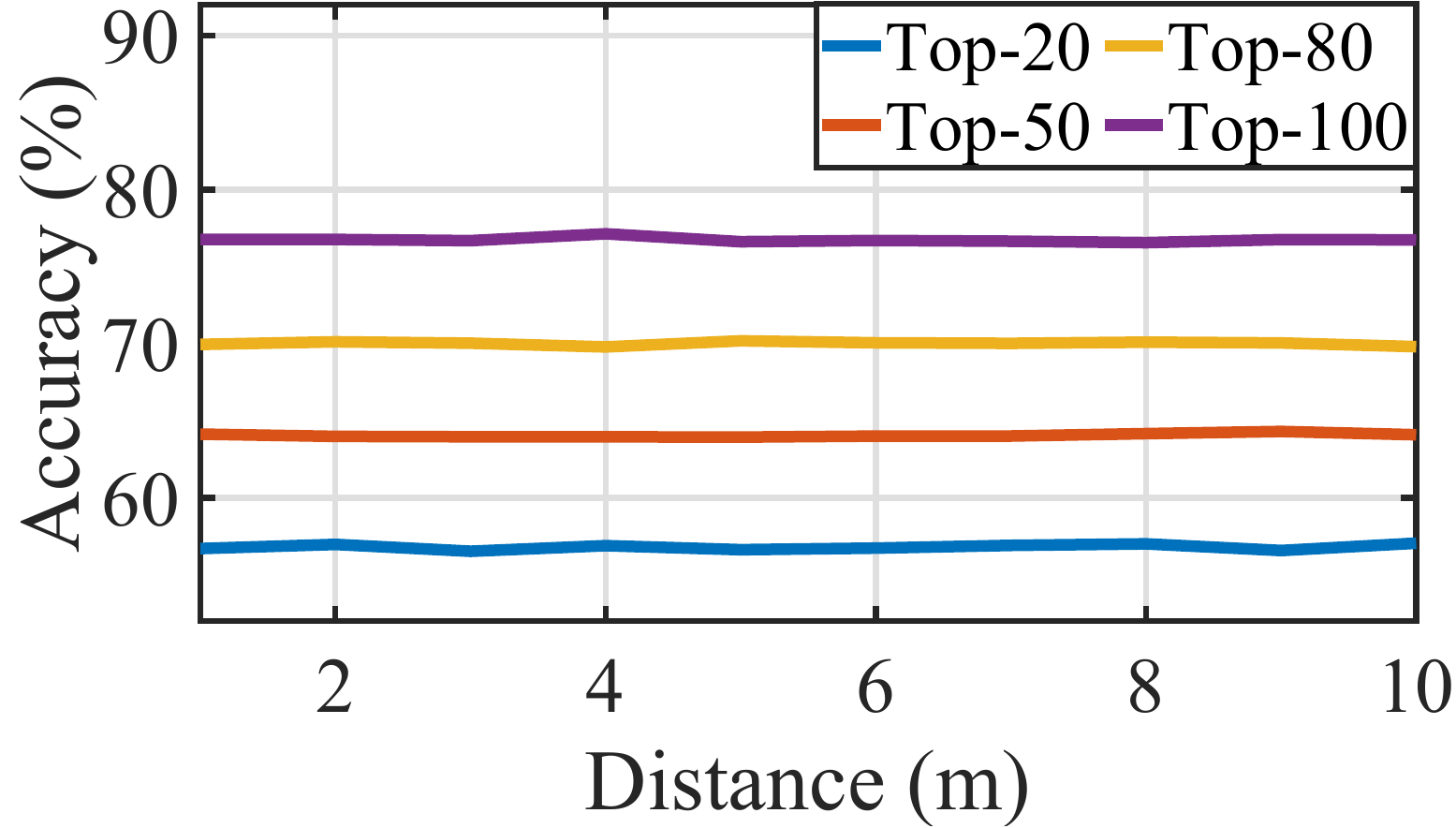}
	    \label{sfig:dist_eve_to_bob}
    }
\caption{Impact of different distances.}
\label{fig:distance}
\vspace{-1em}
\end{figure}

We evaluate the effect of distances on \name, i.e., the distances from Bob to the AP and from Eve to Bob. Figure~\ref{fig:distance} presents the top-20, 50, 80, and 100 password inference accuracy at various distances. Figure~\ref{sfig:dist_bob_to_ap} shows that the average accuracy decreases by about 23\% as the distance between Bob and the AP increases from 1~\!m to 10~\!m, because a longer distance from Bob to the AP weakens the Wi-Fi signal and takes in more interference. On the contrary, Figure~\ref{sfig:dist_eve_to_bob} confirms that the distance between Eve and Bob barely affects the performance of \name, as the digital nature of BFI makes it robust to long-range transmission. Consequently, Eve can eavesdrop stealthily from a long distance without compromising inference accuracy, clearly demonstrating the advantage of \name's \model\ method.


\subsubsection{Typing Speed}
\begin{figure}[h]
    \setlength\abovecaptionskip{0pt}
    \vspace{-.5ex}
    \centering
    \hspace{-0.5em}
    \subfigure[Keystroke classification.]{
	    \includegraphics[width=.485\linewidth]{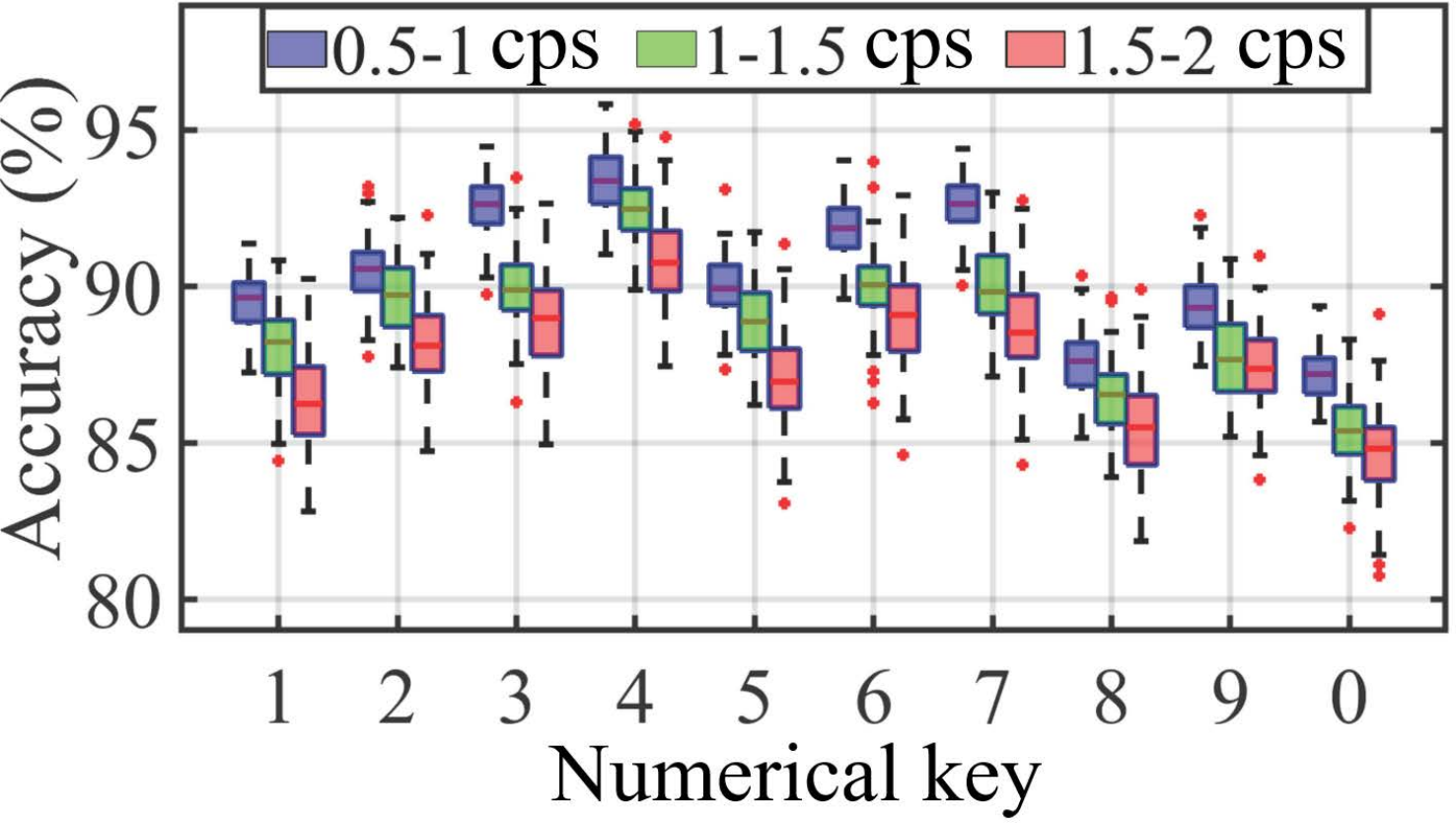}
	    \label{sfig:classify_acc_speed}
    }
    \hfill
    \hspace{-0.5em}
	\subfigure[Password inference.]{
		\includegraphics[width=.48\linewidth]{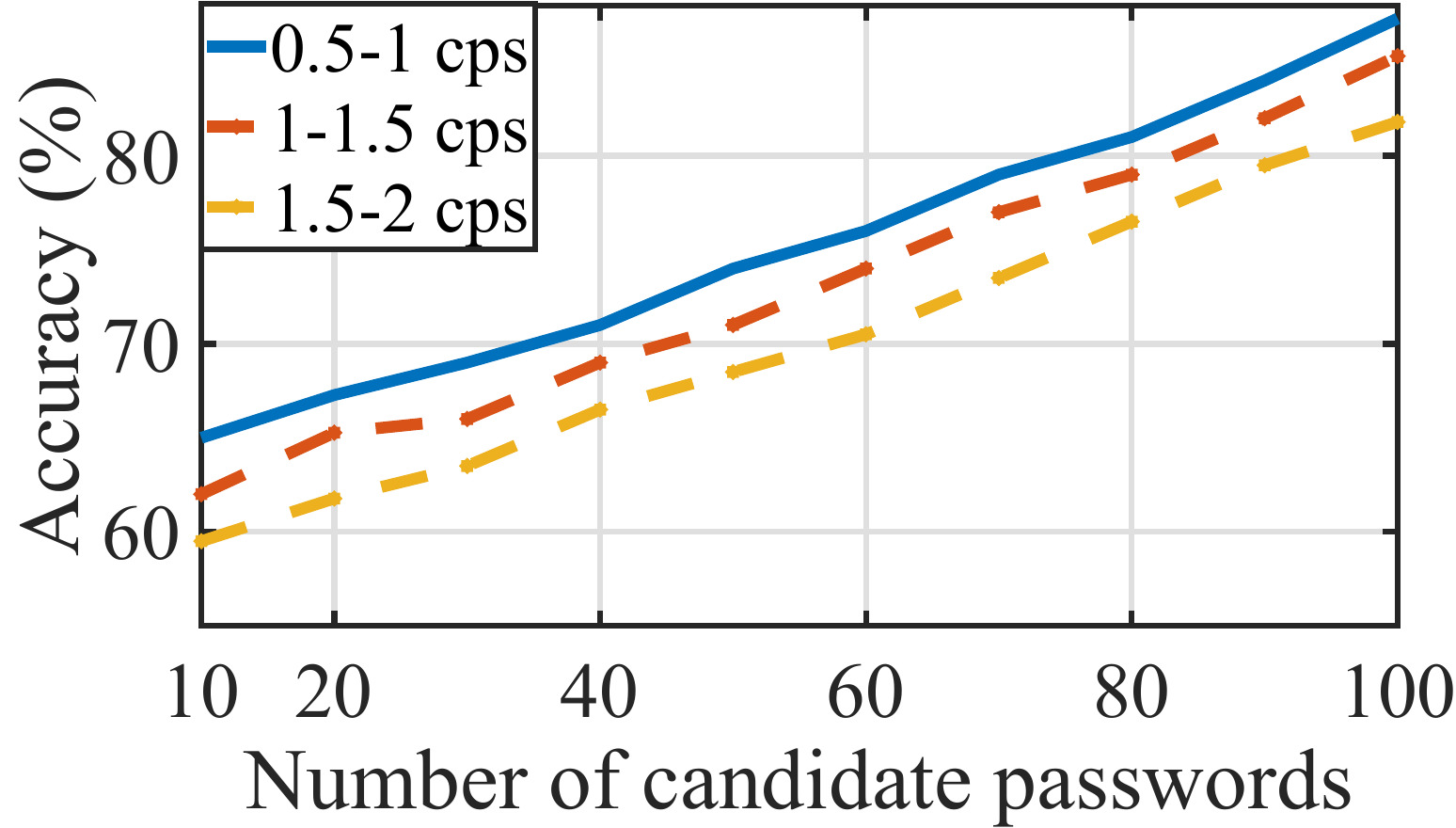}
		\label{sfig:inference_acc_speed}
	}
\caption{
Impact of typing speed.}
\label{fig:type_speed}
\vspace{-1.5ex}
\end{figure}

In this section, we examine how \name's performance varies with typing speeds. Figures~\ref{sfig:classify_acc_speed} and \ref{sfig:inference_acc_speed} respectively show the keystroke classification and top-$[1,100]$ accuracy for tying speed ranges of $[0.5,1.0]$, $[1.0,1.5]$, and $[1.5,2.0]$~\!cps. As expected, both metric values decrease with higher typing speeds, probably due to stronger inter-typing irregularities. Nevertheless, 
\name still \newrev{achieves sufficiently} good performance in fast typing case with speed from $[1.5,2.0]$~\!cps, with only a minor decrease of around 3\% in keystroke classification and \newrev{less than} 7\% in password inference accuracy when compared with those in slow typing case with speed from $[0.5,1.0]$~\!cps. \newrev{The relatively consistent performance of \name\ across different typing speeds is also the consequence of adopting adversarial learning.}
%
%




\subsubsection{Typing Scenarios}
\begin{figure}[b]
    \setlength\abovecaptionskip{3pt}
    \centering
    \hspace{-0.5em}
    \subfigure[Keystroke classification.]{
	    \includegraphics[width=.485\linewidth]{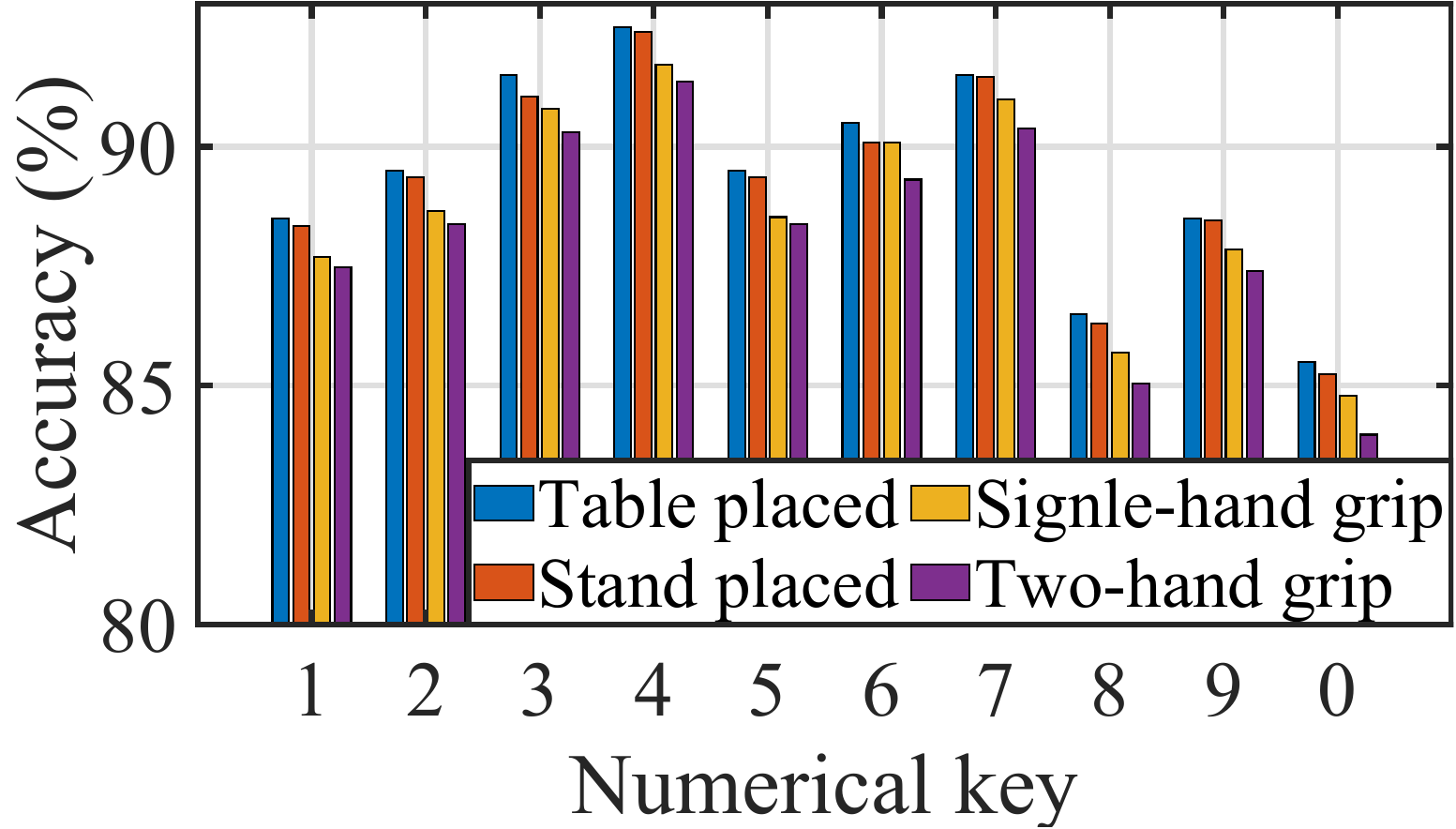}
	    \label{sfig:classify_acc_type_scenario}
    }
    \hfill
    \hspace{-0.5em}
	\subfigure[Password inference.]{
		\includegraphics[width=.482\linewidth]{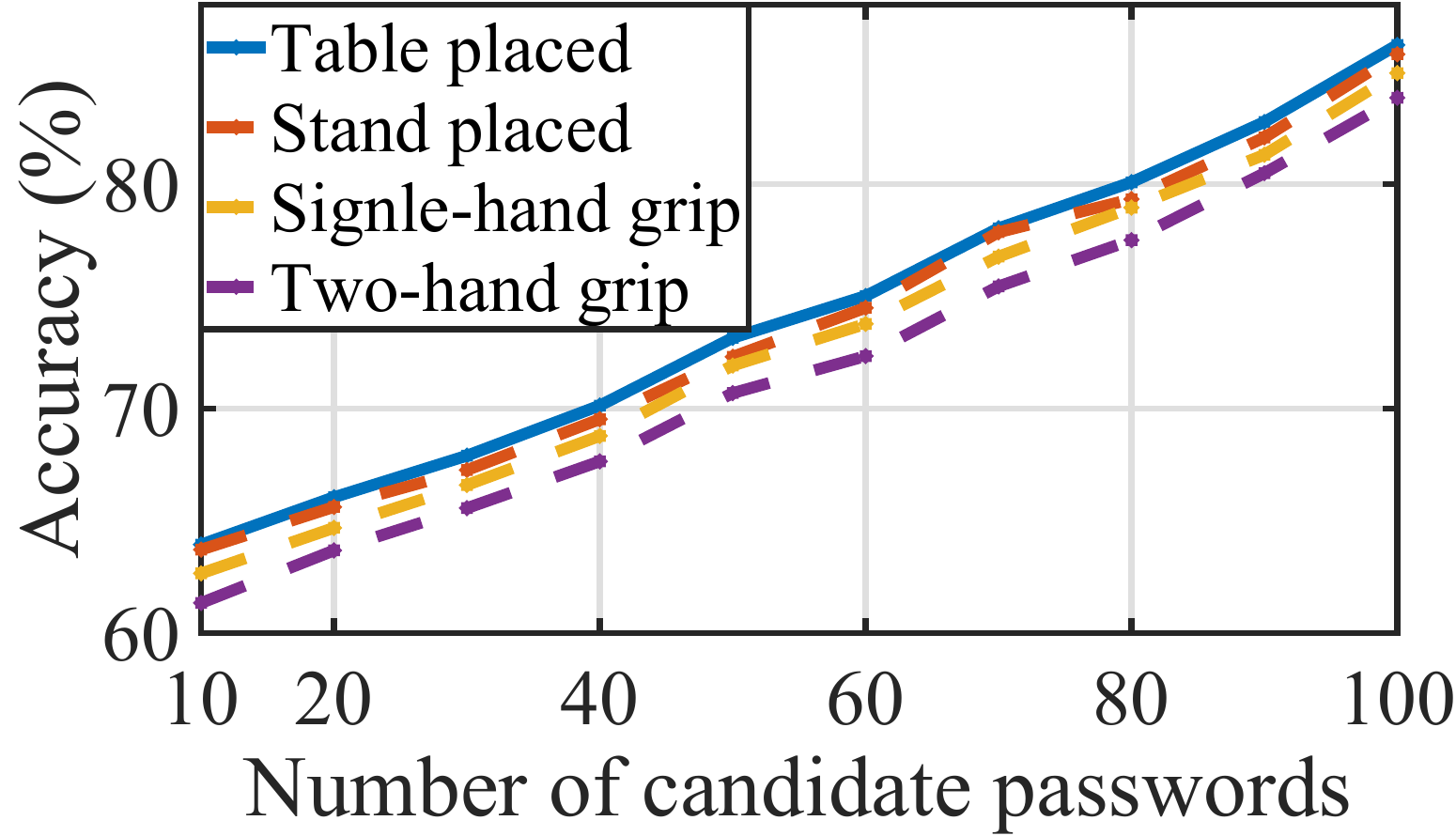}
		\label{sfig:inference_acc_type_scenario}
	}
\caption{Comparison of 4 different typing scenarios.}
\label{fig:type_scenario}
\vspace{-.5ex}
\end{figure}

We further investigate the performance of \name across different typing scenarios, \newrev{including} holding the phone with one or both hands and placing the phone on a stand or a table. Figures~\ref{sfig:classify_acc_type_scenario} and \ref{sfig:inference_acc_type_scenario} show that when the smartphone is placed on a stand or a table, \name achieves higher keystroke classification and password inference accuracy, likely due to the \newrev{stability inherent to} these scenarios. Despite the accuracy differences across scenarios, keystroke classification and password inference accuracy variations are less than 2.5\%. These consistent results demonstrate that \name is robust to various occlusions and different typing scenarios, further validating the effectiveness of our adversarial learning framework.

%






\subsubsection{Password Length} \label{sssec:length}
\begin{figure}[t]
    \setlength\abovecaptionskip{3pt}
    \centering
    \hspace{-0.5em}
    \subfigure[Keystroke classification.]{
	    \includegraphics[width=.483\linewidth]{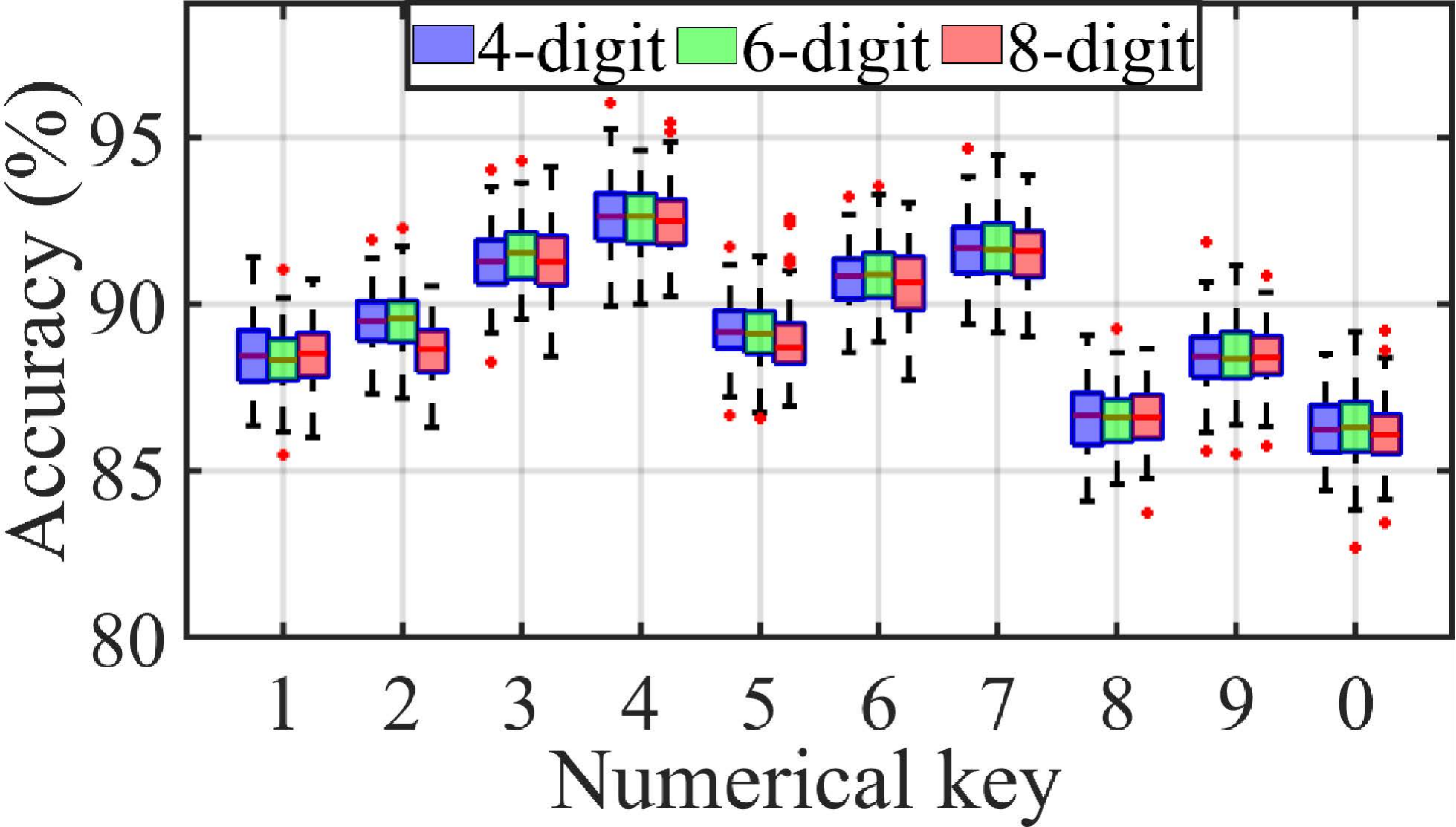}
	    \label{sfig:classiy_acc_length}
    }
    \hfill
	\subfigure[Password inference.]{
		\includegraphics[width=.48\linewidth]{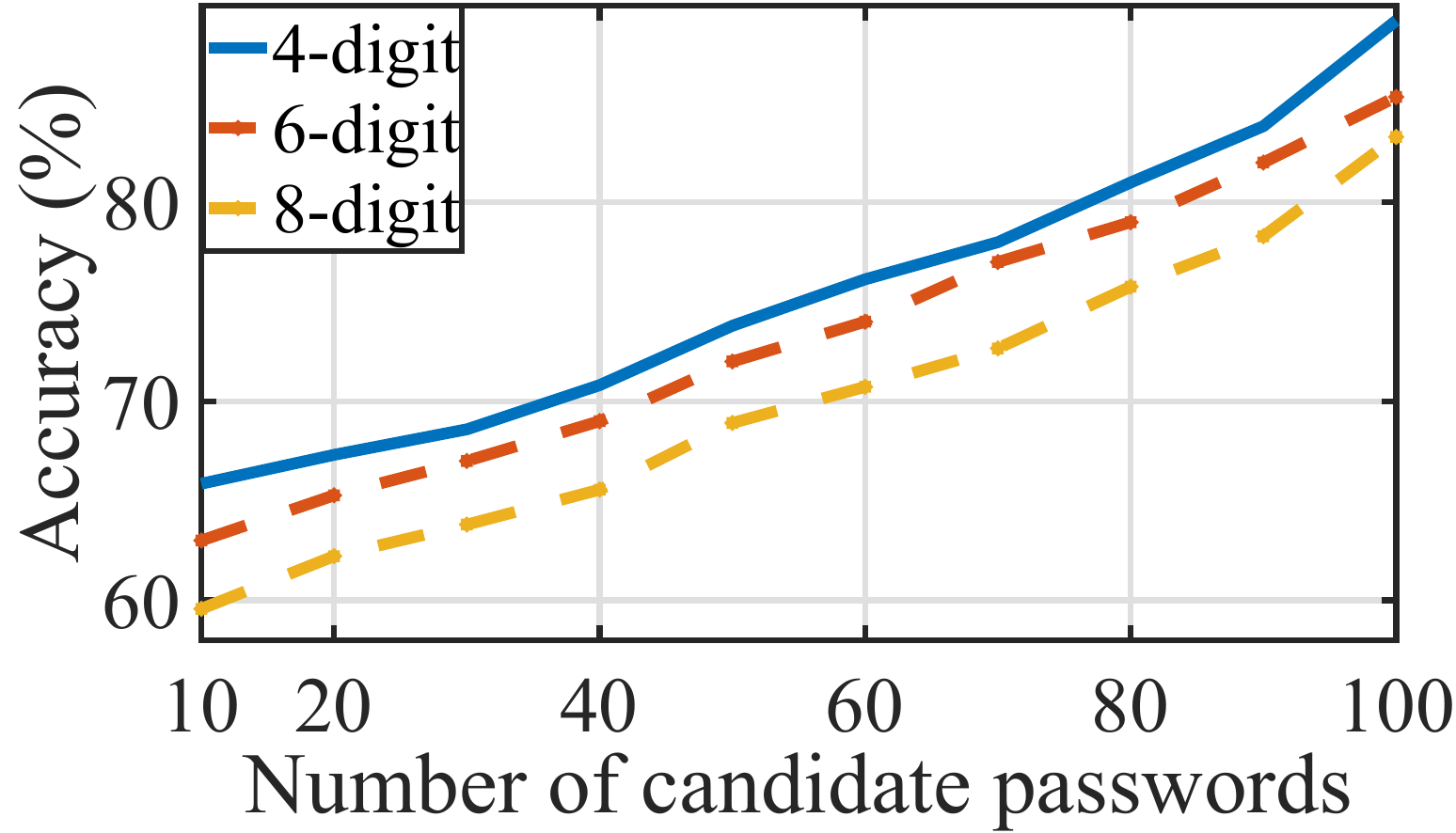}
		\label{sfig:inference_acc_length}
	}
\caption{Impact of 3 different password length.}
\label{fig:pass_len}
\vspace{-1em}
\end{figure}

We finally examine how password length affects \name's performance. Figure~\ref{sfig:classiy_acc_length} demonstrates that the password length does not affect the accuracy of keystroke classification because \name\ \newrev{treats} each keystroke \newrev{independently}
regardless of how many keys are \newrev{typed}. However, it significantly impacts password inference accuracy, as shown in Figure~\ref{sfig:inference_acc_length}. For instance, the top-20 and top-100 accuracy for 4-digit numerical passwords is 69\% and 89\%, respectively, \newrev{yet it becomes 64\% and 83\%, respectively, for 8-digit numerical passwords. The accuracy loss is attributed to the increased uncertainty caused by involving more keys.} Nevertheless, even for an 8-digit numerical password, the remarkable success rate of 64\% after 20 attempts still poses a severe threat to smartphone users.

\subsection{Real-World Experiment}
\begin{figure}[b]
    \setlength\abovecaptionskip{3pt}
    \vspace{-1ex}
    \centering
    \subfigure[Attack timing identification.]{
	    \includegraphics[width=.472\linewidth]{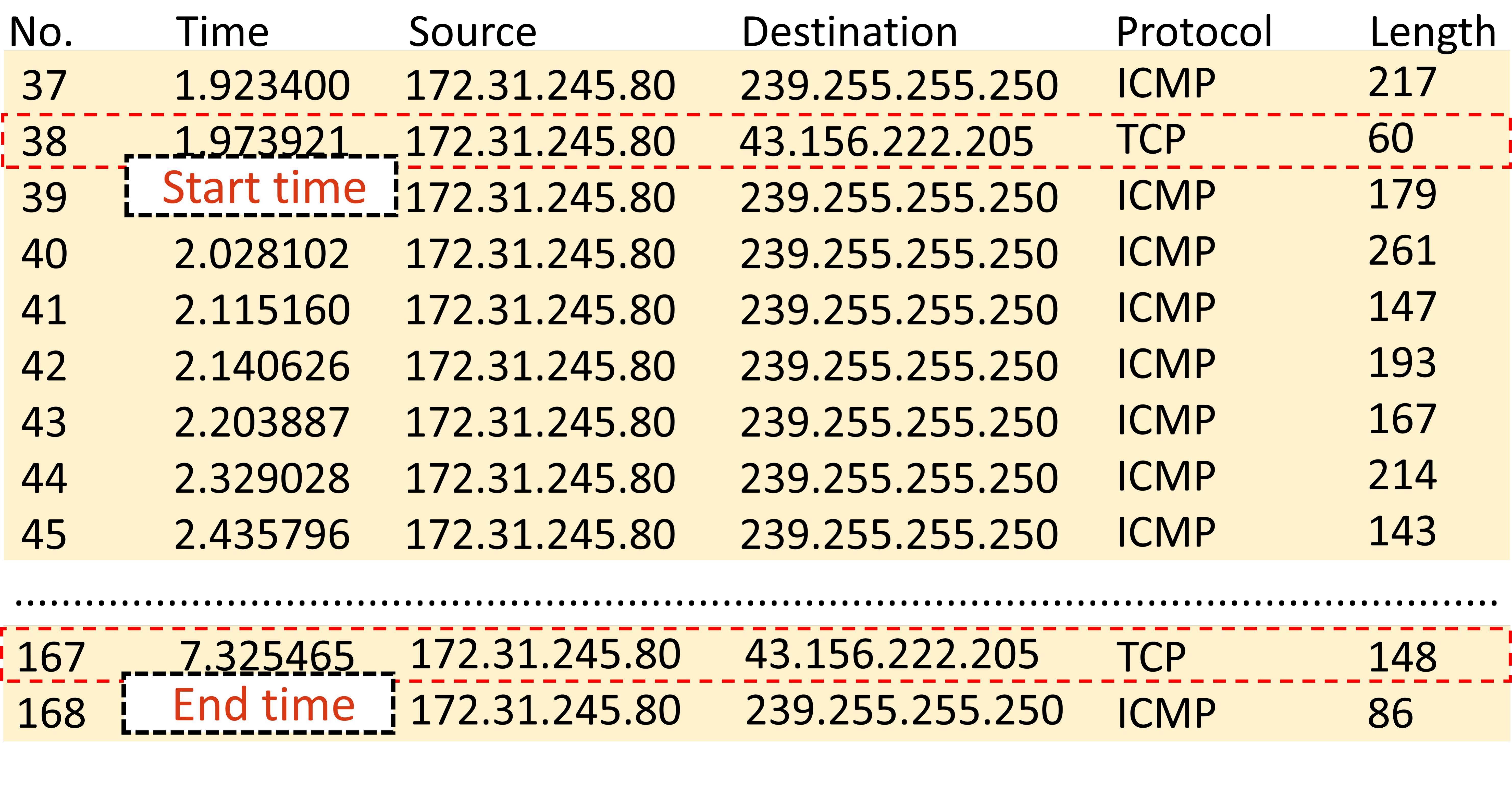}
	    \label{sfig:attack_ip}
    }
    \hspace{-1ex}
	\subfigure[Targeted keystroke extraction.]{
		\includegraphics[width=.468\linewidth]{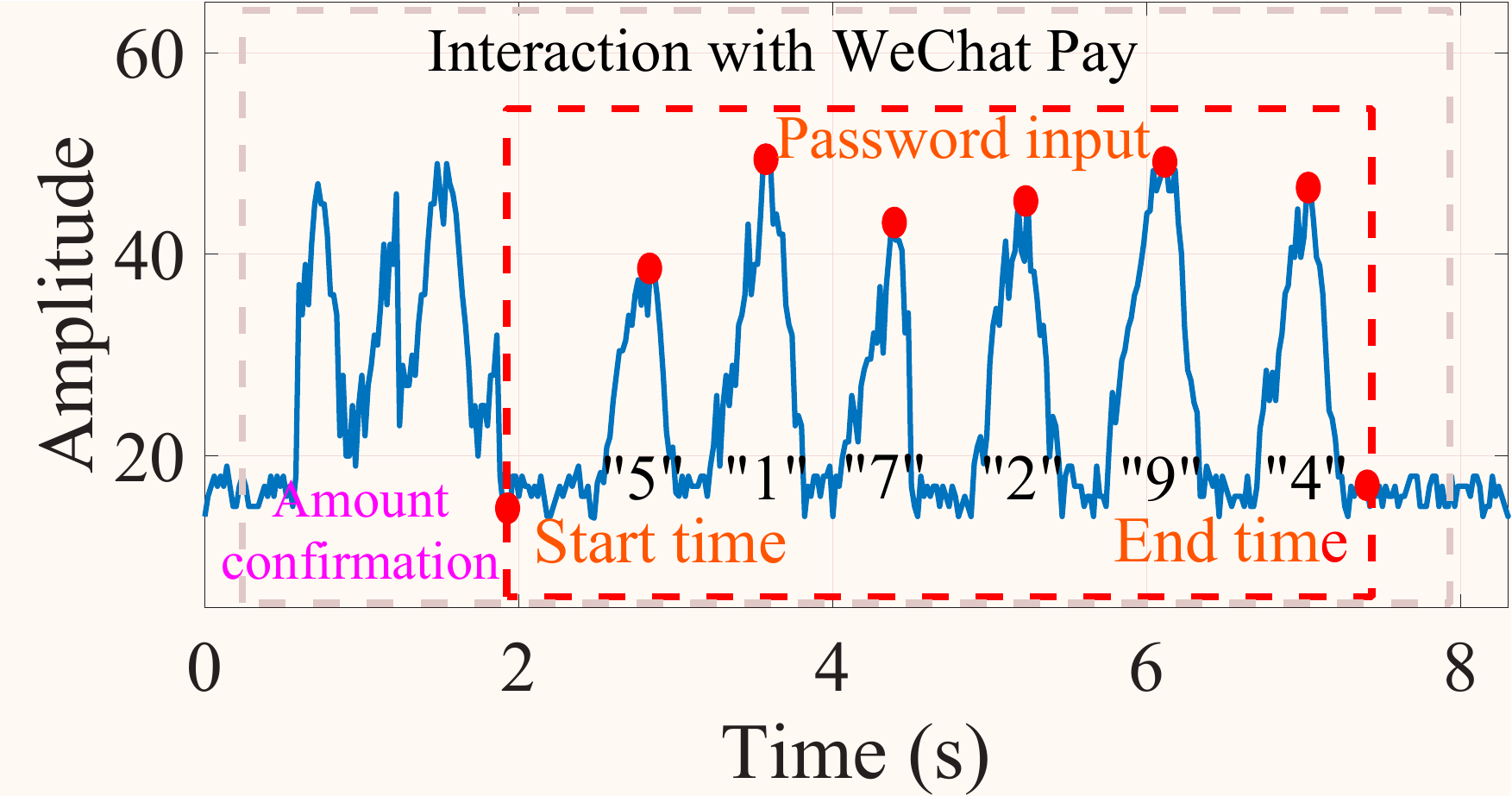}
		\label{sfig:key_extra}
	}
\caption{Real-world experiment with WeChat Pay.}
\label{fig:eve_loc}
\vspace{-.5ex}
\end{figure}

\subsubsection{WeChat Pay Password Inference}
To showcase the practicality of \name, we conduct a real-world experiment by acting as Eve to steal password from WeChat Pay, a digital payment service integrated into WeChat~\cite{wechat}. The victim Bob uses an iPhone 13 for his daily activities, typically including WeChat usage, and he is supposed to make a mobile payment transaction with WeChat Pay, for which a numerical password is required, in a conference room of size 5~\!m $\times$ 8~\!m. The AP is placed on a table and the distance between Bob and the AP ranges from 1.5 to 5~\!m, as confined by the room layout. Meanwhile, Eve leverages \name\ to achieve a stealthy 
eavesdropping at a distance of 3~\!m from Bob.




Following the method in Section~\ref{ssec:identify}, \name\ first identifies Bob's Wi-Fi traffic; this is followed by detecting an IP address ``43.156.222.205'' coinciding with an entry in a pre-recorded IP database, as shown in Figure~\ref{sfig:attack_ip}, which in turns starts BFI recording. The recording is stopped once no more requests to that address are made. Subsequently, \name\ performs SRA on the BFI time series, and the resulting non-sparse BFI series is shown in Figure~\ref{sfig:key_extra}. It appears that the BFI series includes not only the 6-digit numerical password but also other keys entered beforehand  
(e.g., the transfer amount and confirmation), so we extract the last six peaks corresponding to the password specifically for WeChat Pay, as highlighted by the red box. 
%

After segmenting the signal, \name\ initiates the password inference. Since WeChat Pay freezes after five incorrect password inputs, we focus on identifying correct passwords among the top 5 candidates. In the experiment shown in Figure~\ref{sfig:key_extra}, the actual password entered by Bob is ``517294'', and the top 5 candidates are ``547294'', ``517204'', ``\textbf{517294}'', ``517594'', and ``517394'', indicating a successful password stealing. 
We conduct 50 such experiments in total, each with a different password. The results indicate that, out of these 50 input passwords, \name achieves a top-5 accuracy of 50\%, which is quite close to that shown in Figure~\ref{sfig:top10_can}, albeit with a potentially biased statistics given only a small amount of trials. These experiments evidently demonstrate the practicality of \name in real-world scenarios. 

\begin{figure}[b]
    \setlength\abovecaptionskip{3pt}
    \vspace{-1em}
    \centering
    \subfigure[Keystroke classification.]{
	    \includegraphics[width=.466\linewidth]{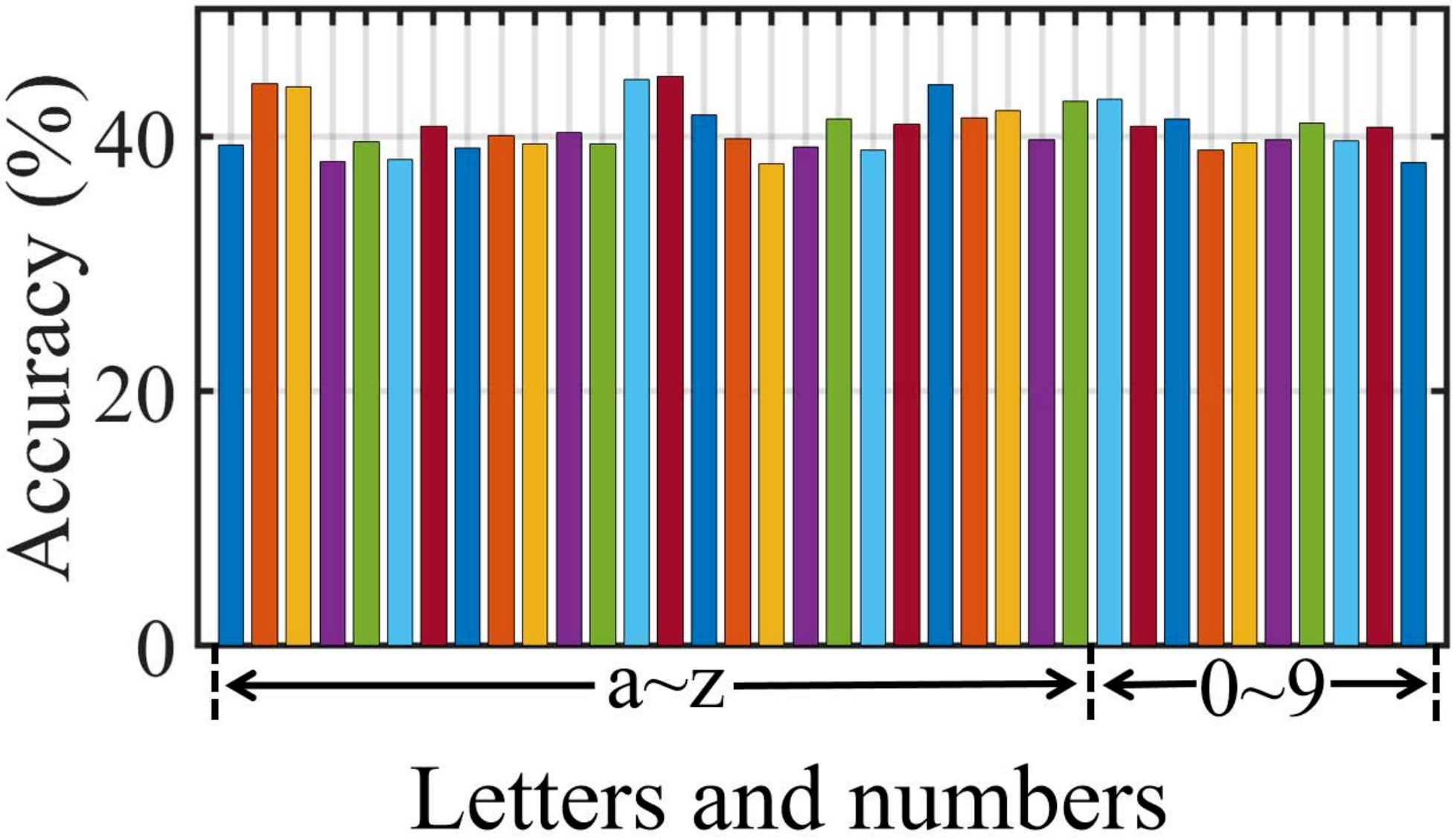}
	    \label{sfig:full_classify}
    }
    \hfill
	\subfigure[Password inference.]{
		\includegraphics[width=.475\linewidth]{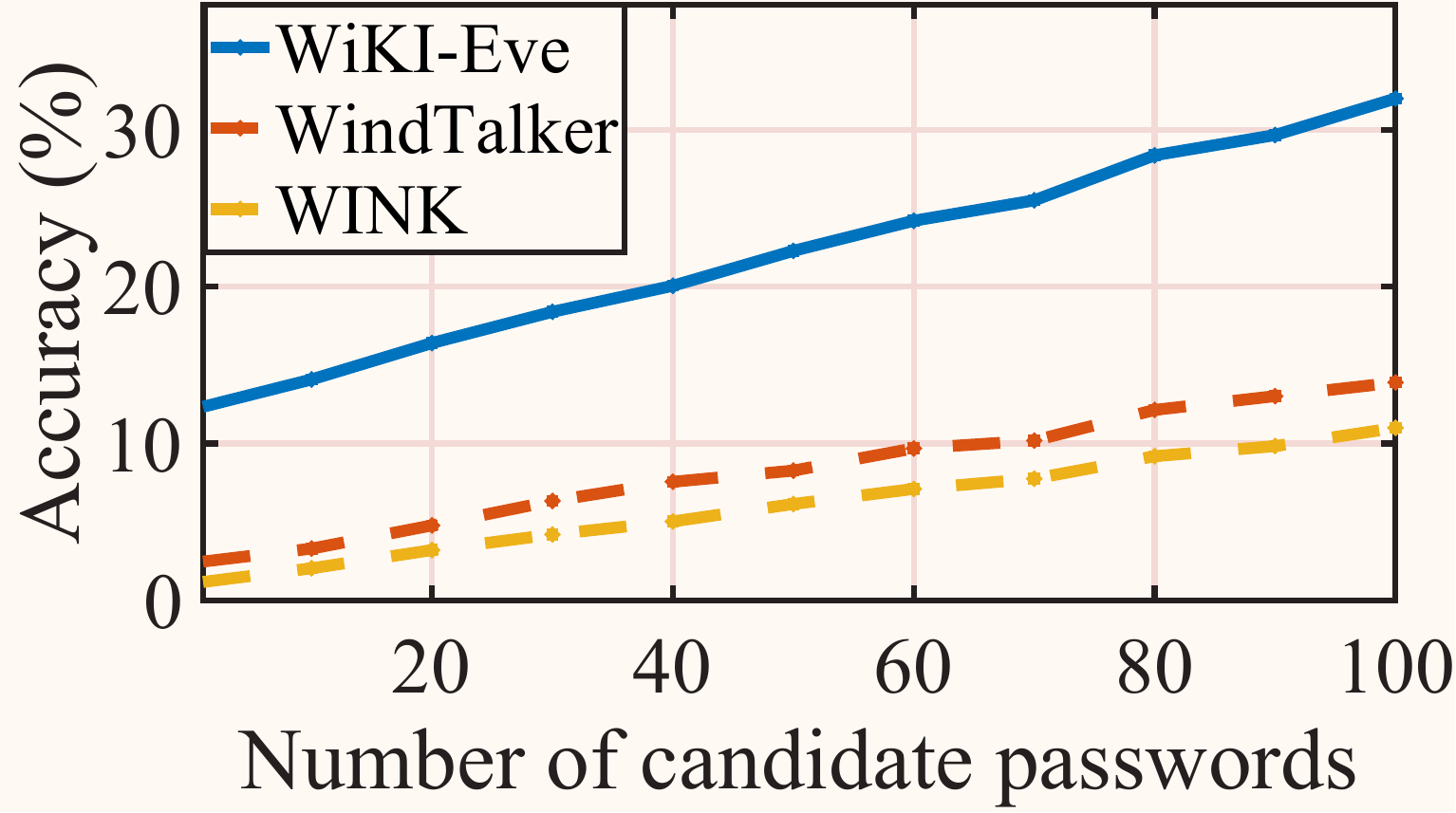}
		\label{sfig:full_inference}
	}
\caption{Performance on QWERTY keyboards.}
\label{fig:full_keyboard}
\vspace{-.5ex}
\end{figure}

\begin{figure}[t]
    \setlength\abovecaptionskip{3pt}
    \centering
    \hspace{-0.5em}
    \subfigure[Keystroke classification.]{
	    \includegraphics[width=.487\linewidth]{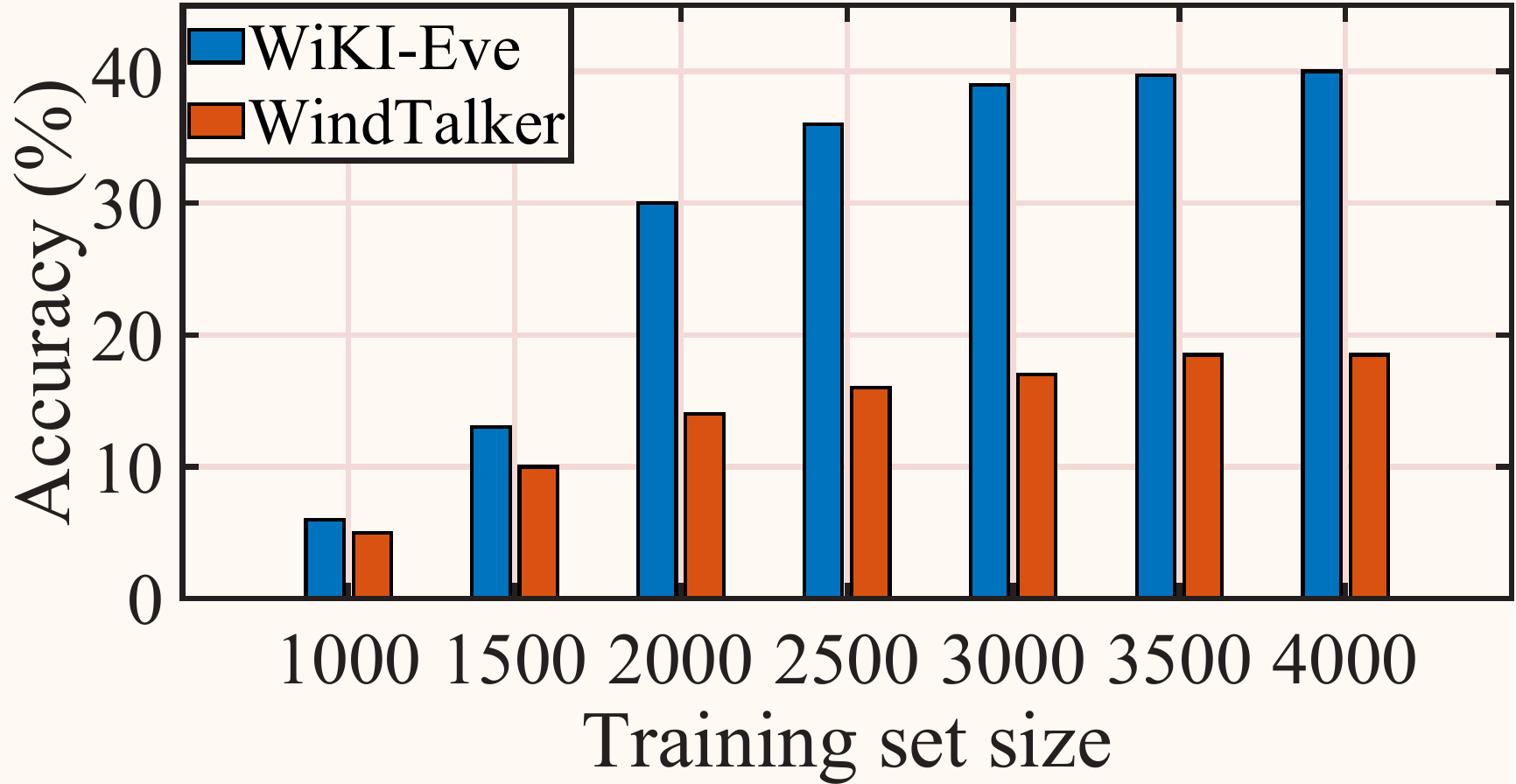}
	    \label{sfig:full_classify_training}
    }
    \hfill
    \hspace{-0.5em}
	\subfigure[Password inference.]{
		\includegraphics[width=.484\linewidth]{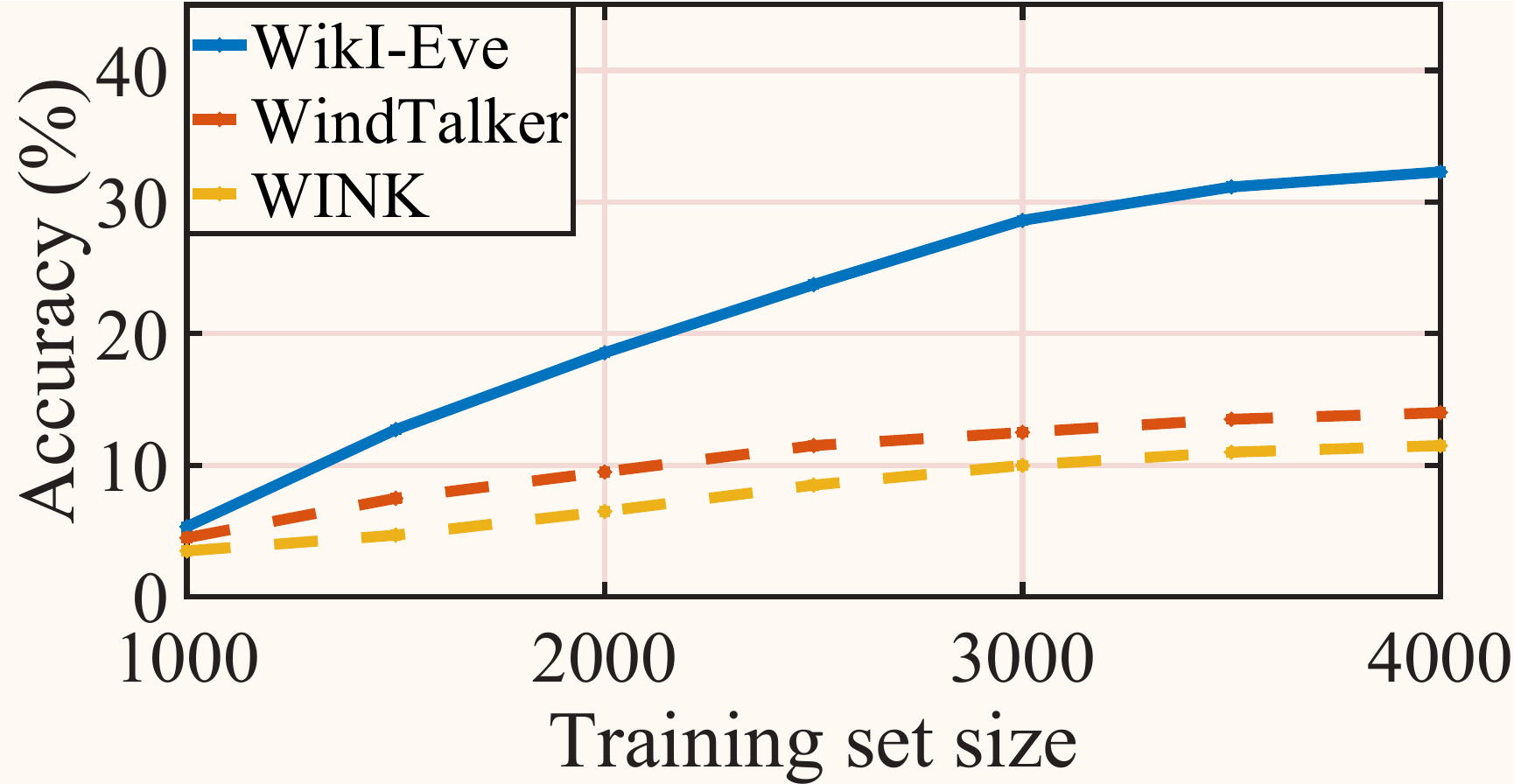}
		\label{sfig:full_inference_training}
	}
\caption{Extending \name\ to QWERTY keyboards requires more training data.}
\label{fig:full_keyboard_training_data}
\vspace{-.5ex}
\end{figure} 

\subsubsection{Extending to Virtual QWERTY Keyboard} \label{sssec:qwerty}
Many applications need more diversified characters than what a numerical keyboard can offer. Typically, banking applications (e.g., the popular Chase Mobile~\cite{chase}) handling sensitive financial transactions and identity information demand using a virtual (on-screen) QWERTY keyboard for users to create more secure \textit{alphanumeric} passwords. To test the applicability of \name\ in such scenario, we conduct keystroke classification experiments on the QWERTY keyboard of Chase Mobile. We collect a dataset of 4,000 pre-defined passwords with varying lengths: 1,500 with 6 characters, 1,500 with 8 characters, and 1,000 with 10 characters. The passwords consist of lowercase letters from `a' to `z’ and numbers from `0' to `9’. Except for the larger dataset size, we adopt the same experiment settings in Section~\ref{sec:implementation}. 

Figure~\ref{sfig:full_classify} shows that \name\ achieves an average keystroke classification accuracy of 40\%. Additionally, Figure~\ref{sfig:full_inference} indicates that \name's top-$[1,100]$ accuracy of 6-character alphanumeric password ranges from 12\% to 32\%, surpassing WindTalker and WINK whose top-100 accuracy is only 11\% and 14\%, respectively. Although the accuracy is lower than those in Section~\ref{sssec:classify_acc} and~\ref{sssec:password_infer_acc}, it still poses a severe threat to smartphone users. The performance drop on QWERTY keyboards can be attributed to these keyboards having approximately four times more keys than numerical keyboards within the same area. Consequently, the BFI features of clicking different keys are less distinguishable due to their proximity. Additionally, shorter distances (hence shorter transition periods) among keys increase inter-keystroke interference, thereby decreasing KI accuracy. 

We also find that KI on a QWERTY keyboard demands a much larger training dataset than on a numerical keyboard. According to Figure~\ref{fig:full_keyboard_training_data}, \name\ performs similarly to the baselines when the training set is small. Fortunately, as the training set size increases from 1,000 to 4,000, \name\ begins to show its strengths: it improves the keystroke classification accuracy from 6\% to 40\%, and the top-100 accuracy from 6\% to 32\%, outperforming the baselines by large margins. The need for a large training set can be explained (again) by the drastic increase in the number of keys on QWERTY keyboards, along with the corresponding increase in the number of domains. By employing the adversarial learning framework, \name\ can fully utilize the training data and perform adequate KI 
under domain interference. 
In contrast, WindTalker and WINK struggle with interference and artifacts caused by a large number of domains, barely improving KI performance. 
%

This experiment also reveals a few challenges to be tackled in future for general KI on QWERTY keyboards. First, more diversified password length should be considered, as 
over 20\% of user may have passwords longer than 10 characters~\cite{length}. 
Second, handling more general passwords containing special characters and uppercase letters is also a crucial aspect:
typing these characters may require combinations of multiple keys (e.g., ``shift’’ and its paired keys) and thus complicating the BFI series.
Third, certain applications have separate keyboard layouts for distinct groups of
keys, requiring users to switch between layouts while entering passwords. Performing KI for such applications requires accurate detection of the layout switching, as well as training two separate neural models for each layout, potentially increasing system complexity. Instead of increasing training data in a brute-force manner, 
%
other side-channel attacks and social engineering techniques~\cite{hadnagy2010social} may be combined with \name\ to enhance its KI capability in tackling these challenges. 


\begin{figure*} [t]
	\centering
	\subfigure[Saturated traffic.]{
        \label{sfig:CSISaturated}
	\begin{minipage}[b]{0.18\textwidth}
	\includegraphics[scale=0.12]{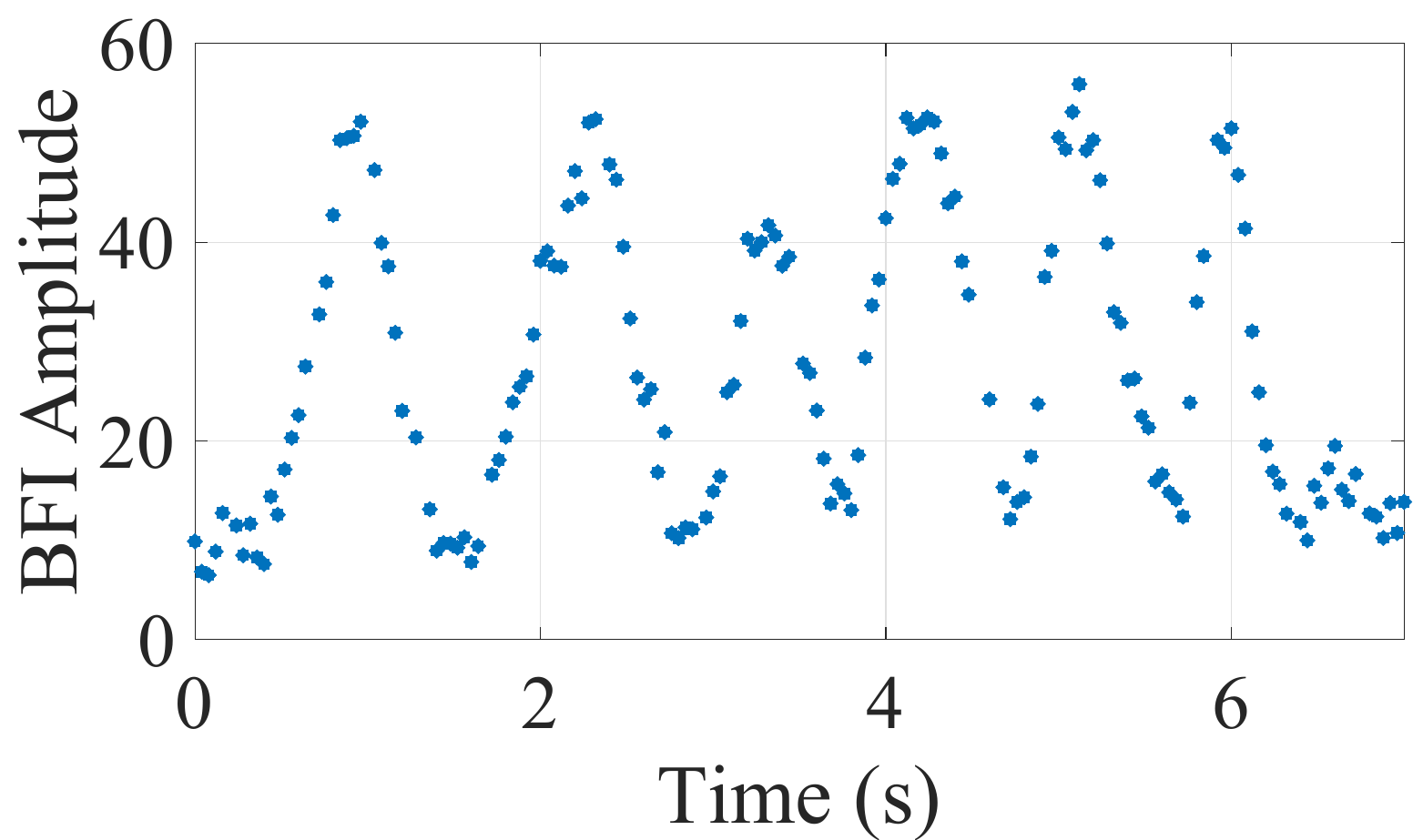}\\
		\includegraphics[scale=0.12]{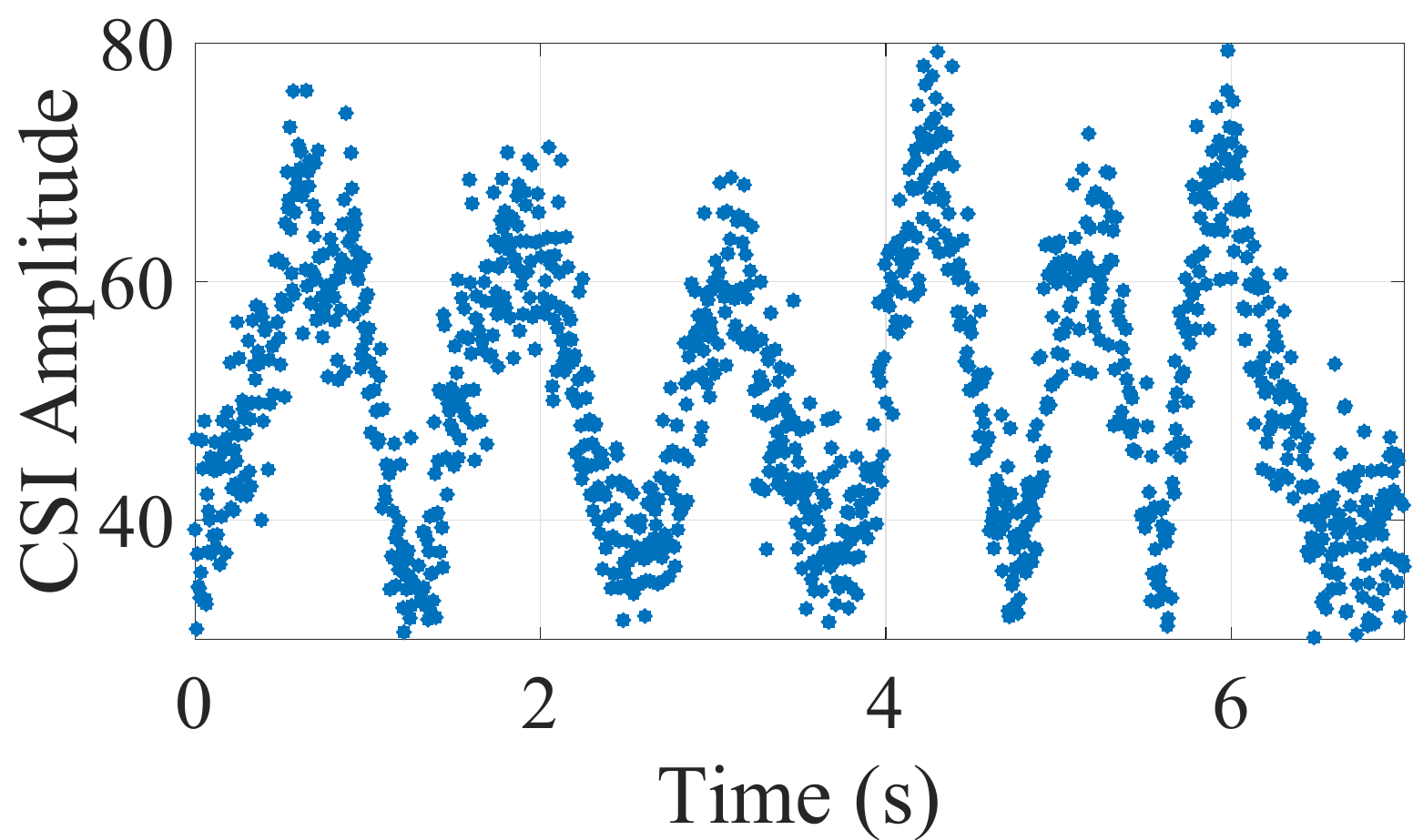}
	\end{minipage}
}
	\subfigure[Video conferencing.]{
        \label{sfig:CSIVideo}
	\begin{minipage}[b]{0.18\textwidth}
		\includegraphics[scale=0.12]{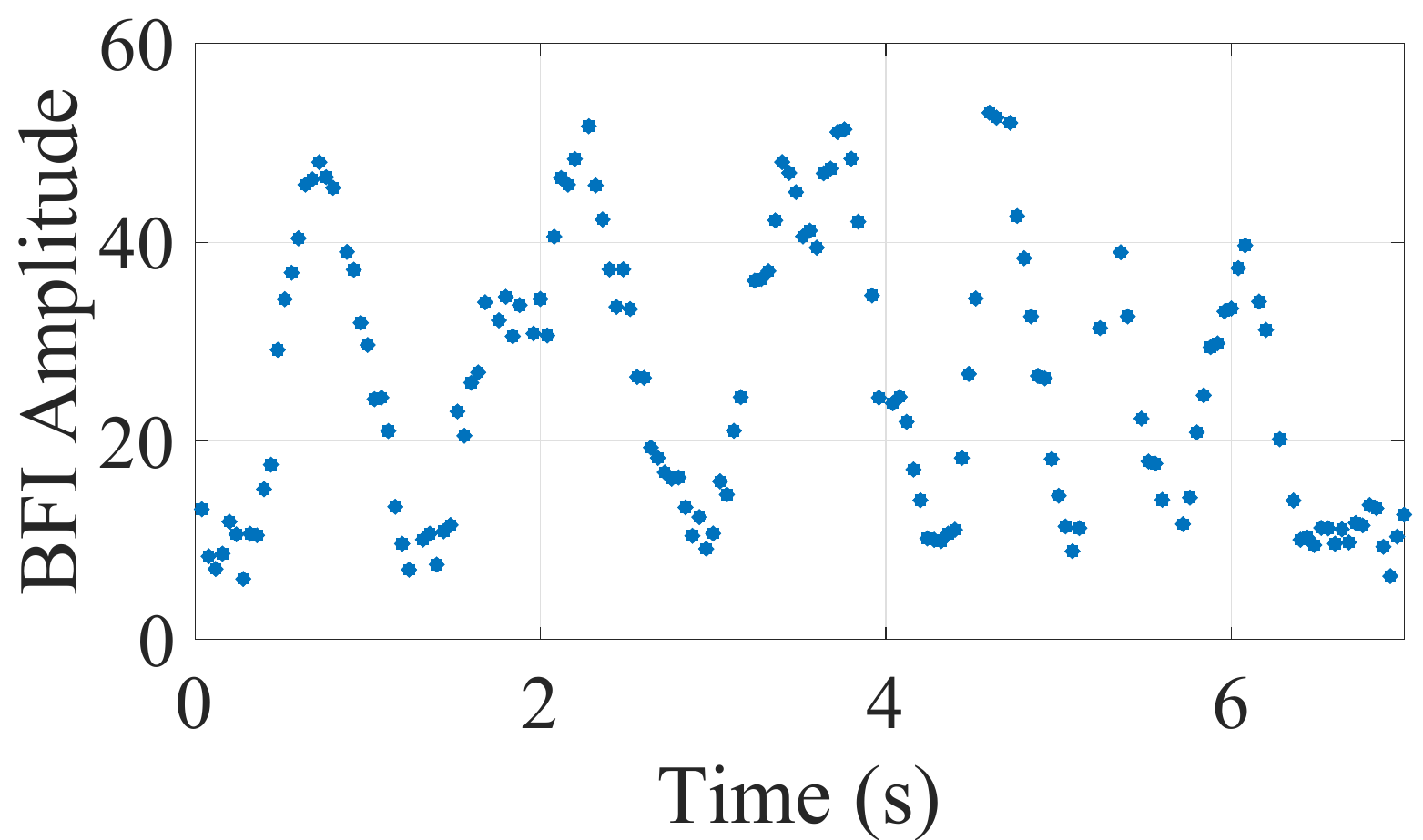} \\
	\includegraphics[scale=0.12]{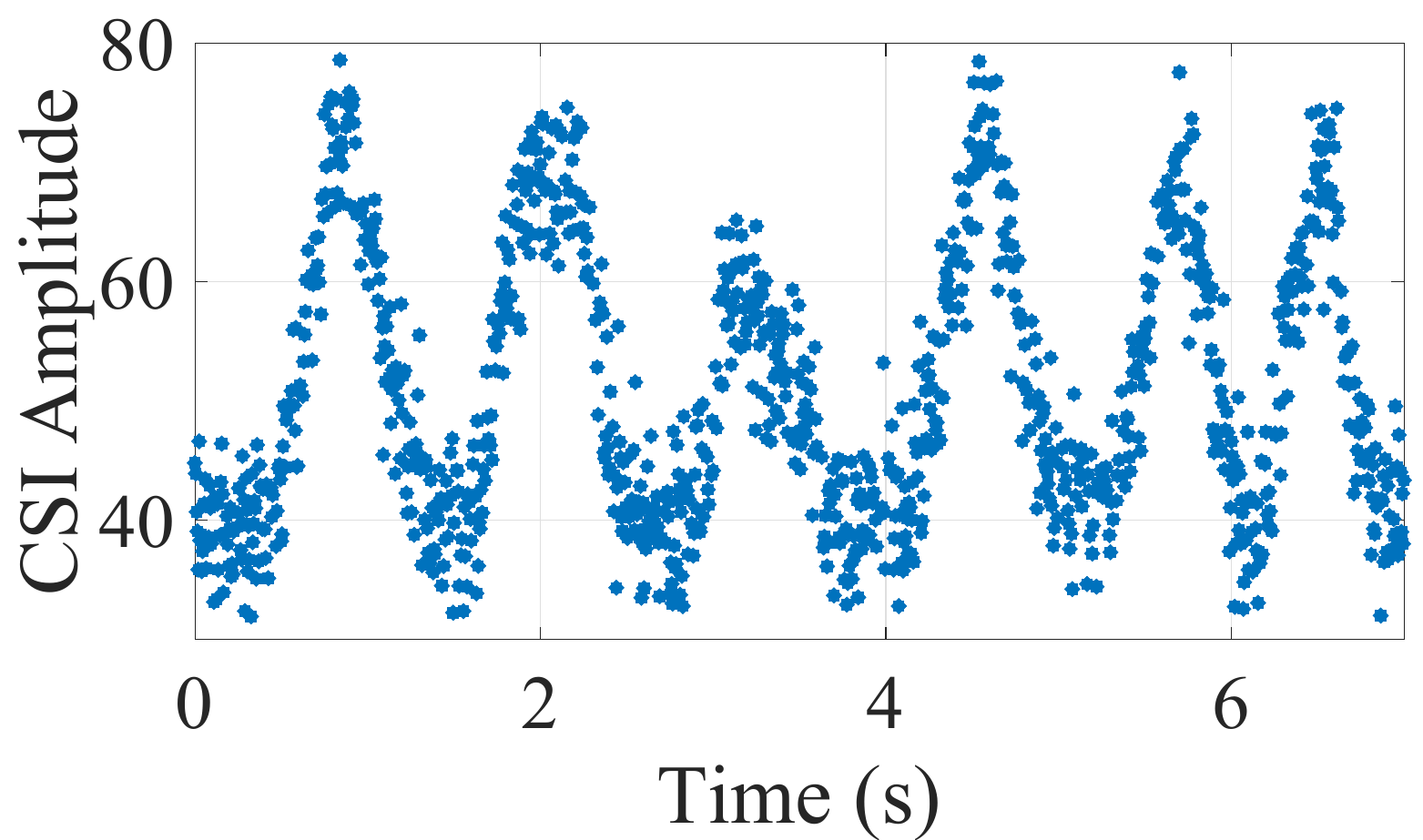}
	\end{minipage}
}
	\subfigure[Software update.]{
        \label{sfig:CSISoftware}
	\begin{minipage}[b]{0.18\textwidth}
		\includegraphics[scale=0.12]{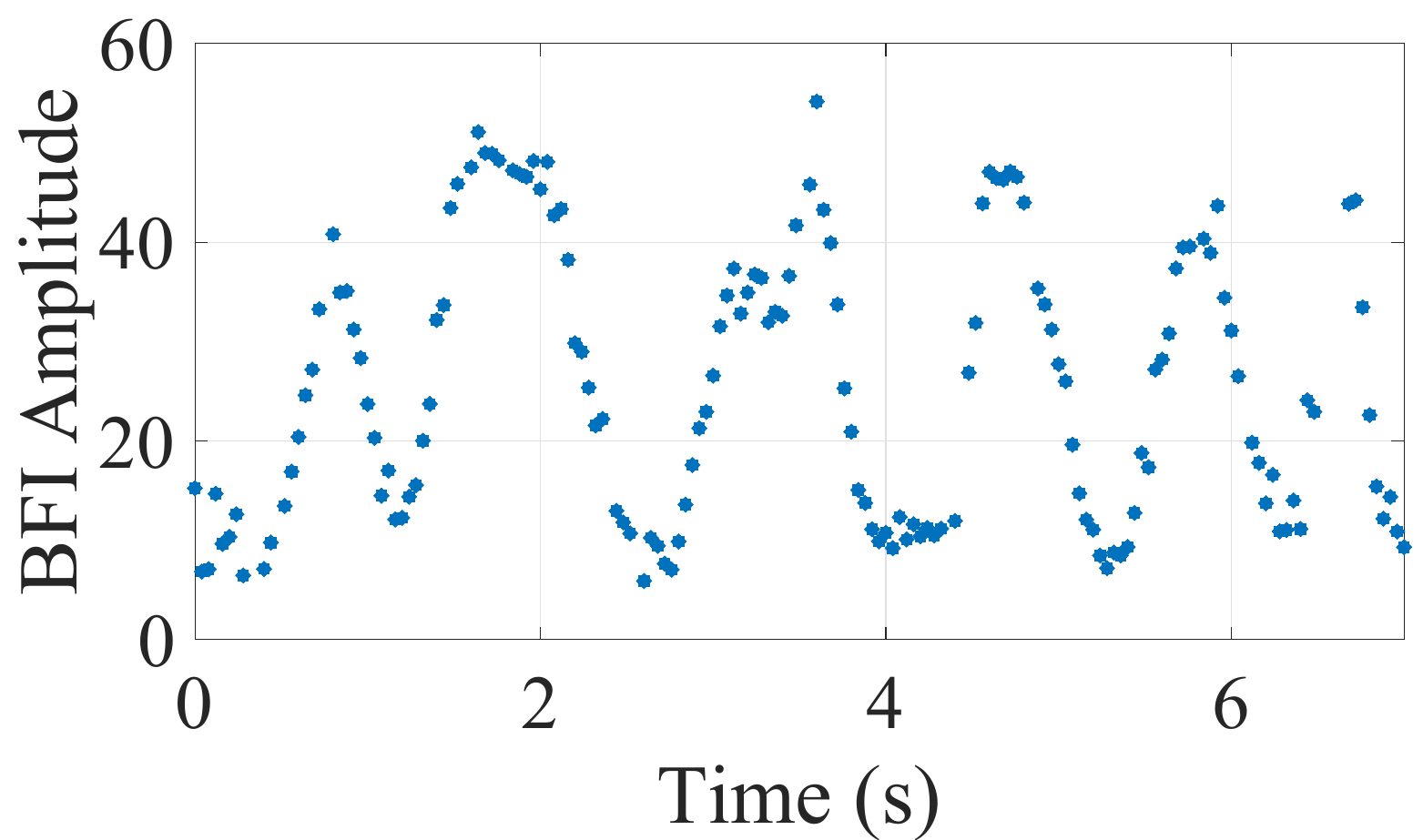} \\
		\includegraphics[scale=0.12]{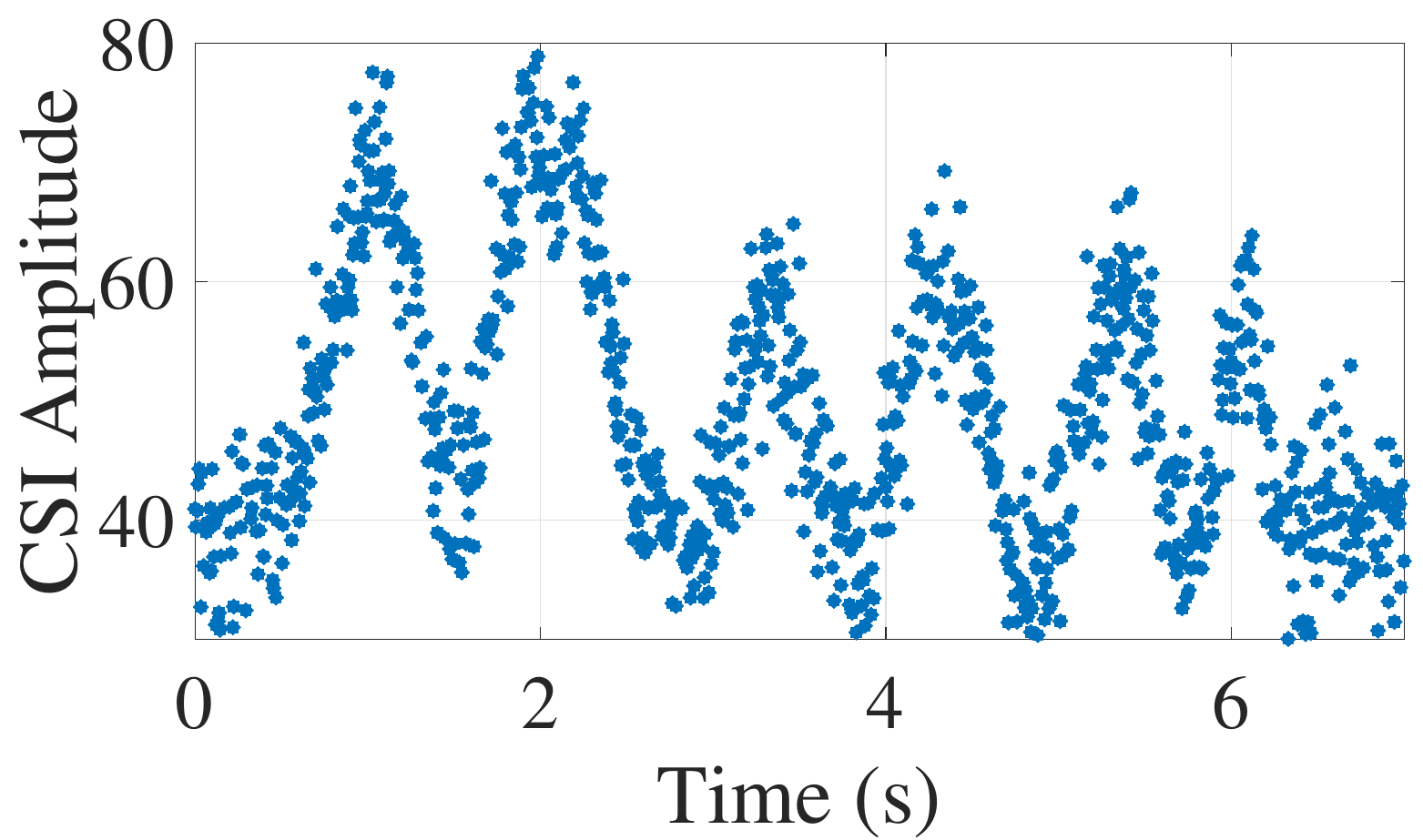}
	\end{minipage}
}
	\subfigure[Music streaming.]{
        \label{sfig:CSIMusic}
	\begin{minipage}[b]{0.18\textwidth}
		\includegraphics[scale=0.12]{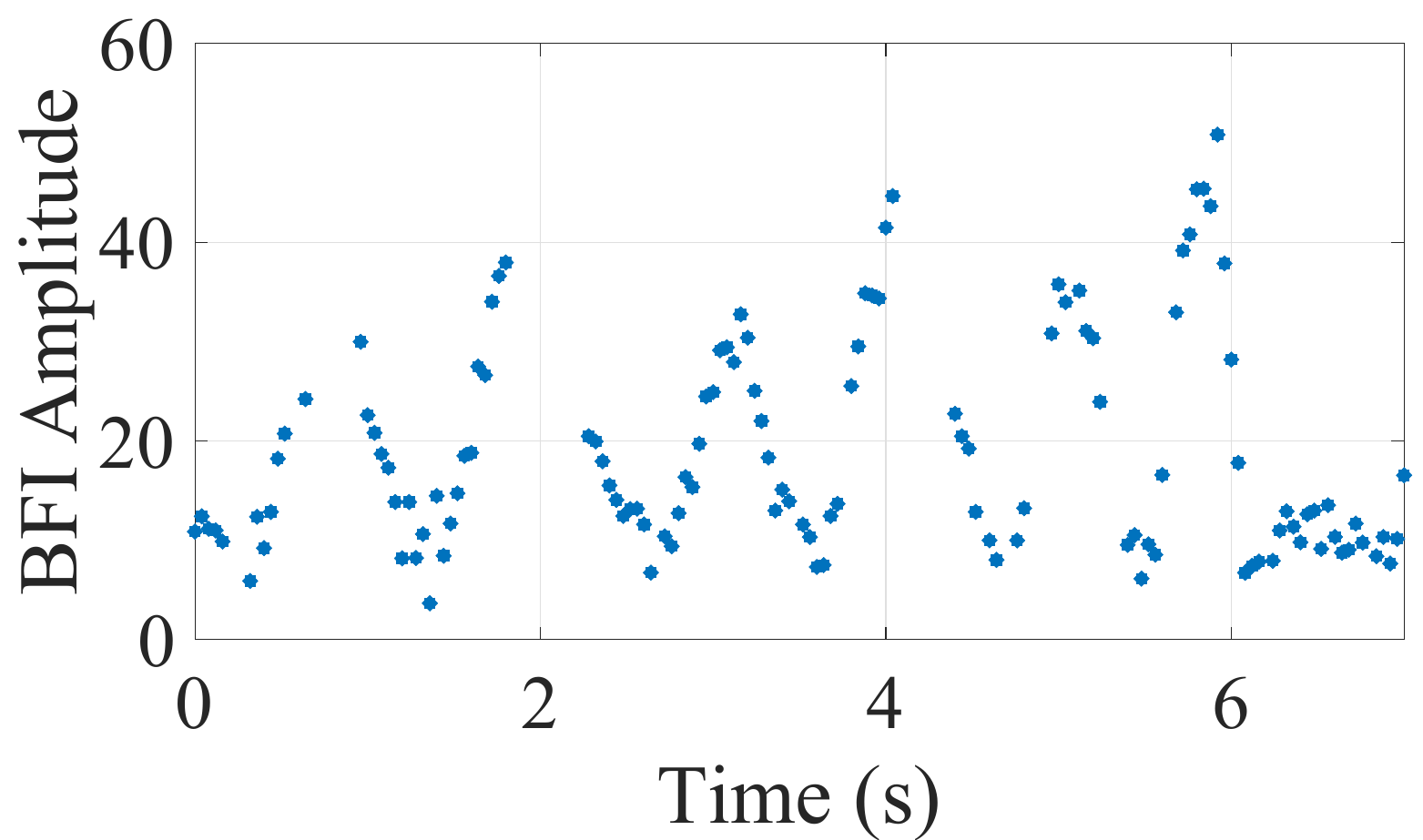} \\
		\includegraphics[scale=0.12]{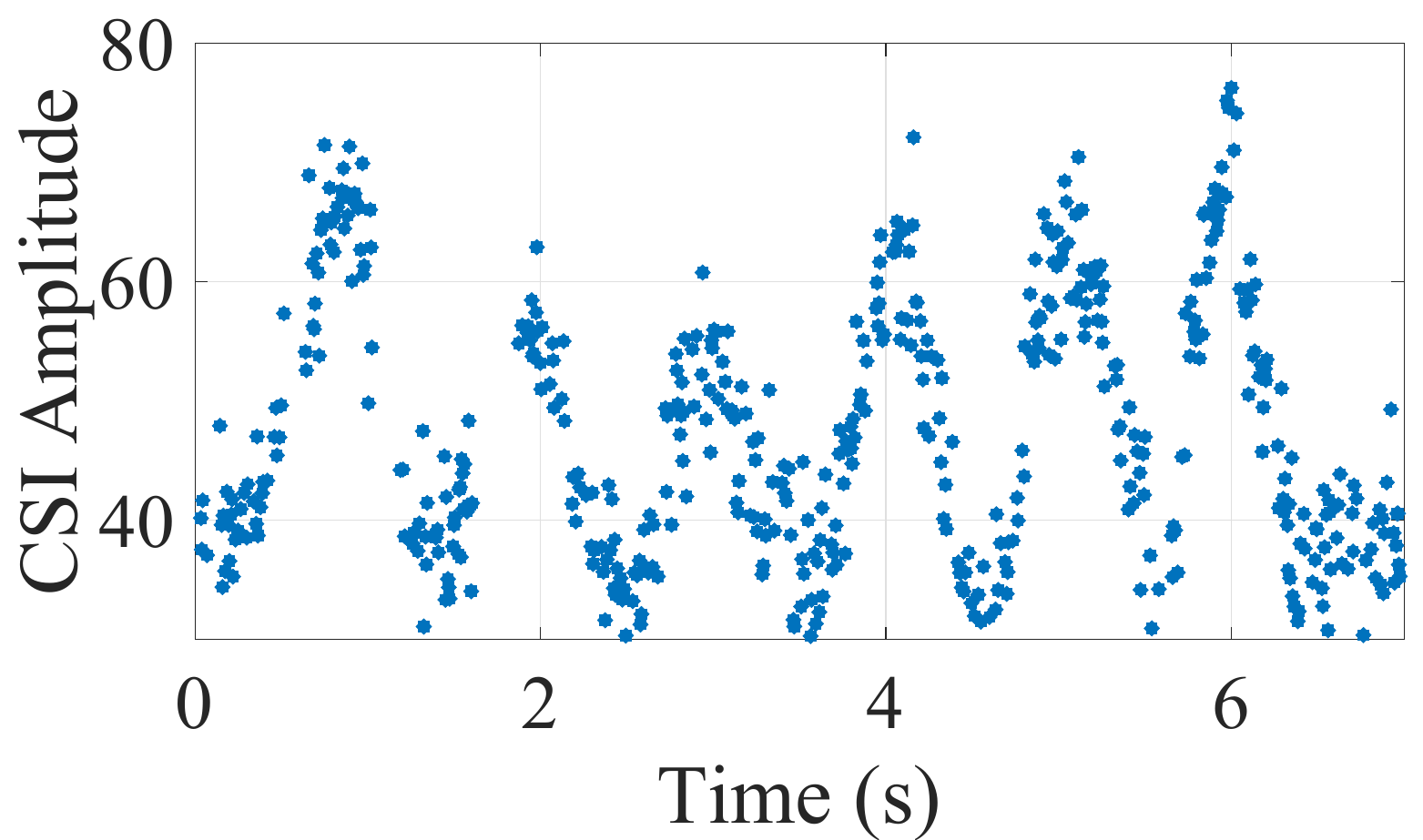}
	\end{minipage}
}
	\subfigure[Background chat.]{
        \label{sfig:CSIchat}
	\begin{minipage}[b]{0.18\textwidth}
		\includegraphics[scale=0.12]{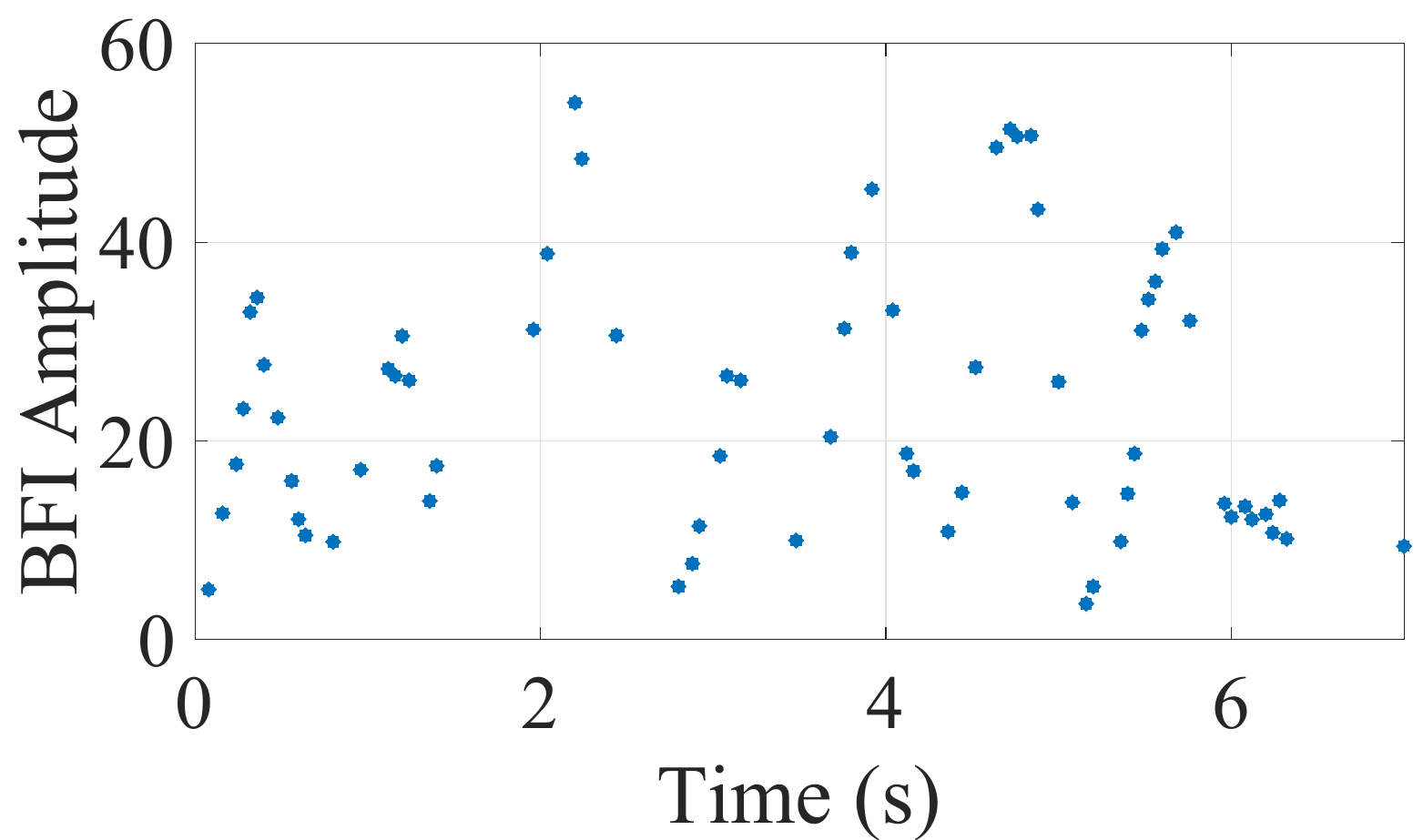} \\
		\includegraphics[scale=0.12]{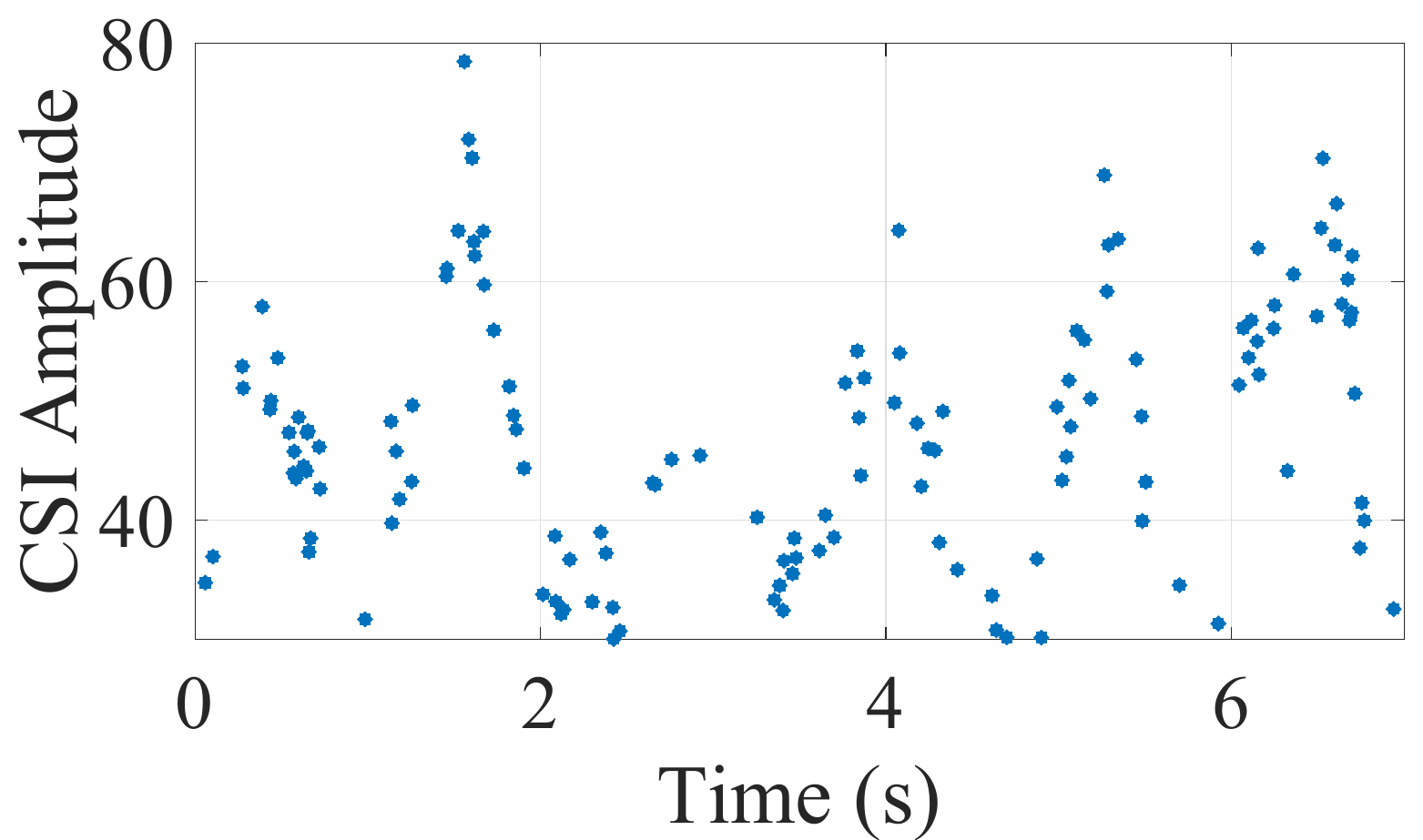}
	\end{minipage}
}
\vspace{-2ex}
\caption{\hjy{BFI (first row) and CSI (second row) time series under real-life background traffic. }}
\label{DiffBFI:traffic}
\vspace{-.5ex}
\end{figure*}

\vspace{-.2em}
\section{Traffic Impact and Defense}\label{sec:Discussion}
In this section, we first study the impact of five different background traffic on the sparsity of BFI (and CSI) time series, then we propose four different defense strategies against \name.

\vspace{-.5em}
\subsection{Background Traffic Analysis}
\label{sec:Traffic_Analysis}
%
To showcase how real-life background traffic intensities affect the sparsity of BFI (and CSI), we set five types of background traffic in (rate) descending order: (artificially) saturated traffic, video conferencing, software update, music streaming, and background chat.
Take the 6-digit password as an example, the sparsity of the BFI (and CSI) time series varies with the intensity of the background traffic as shown in Figure~\ref{DiffBFI:traffic}.
Figure~\ref{sfig:CSISaturated} depicts the dense and continuous time series under saturated traffic. 
Since video conferencing and software update have their traffic patterns go very close to saturated ones, the resulting time series, as shown in Figures~\ref{sfig:CSIVideo} and~\ref{sfig:CSISoftware}, again exhibit dense and continuous nature, which 
can be directly used for KI without enhanced by SRA.
On the contrary, the time series under two types of low-rate background traffic, as shown in Figures~\ref{sfig:CSIMusic} and~\ref{sfig:CSIchat}, apparently behave much sparser, potentially demanding the assistance of SRA.

It is worth noting that, although the BFI (and CSI) time series can be sparse and bursty under real-life background traffic, we barely observe cases where a whole keystroke goes missing due to traffic sparsity.
As a result, the BFI time series collected under real-life background traffic, even when sparse and bursty, can almost always be accommodated (hence enhanced) by our SRA presented in Section~\ref{ssec:recovery}.
In fact, all existing Wi-Fi-based password eavesdropping scheme~\cite{li2016csi, zhang2020wipos, yang2022wink} have to face the challenge of sparse background traffic, yet we are the first in rising to this challenge, enabling Wi-Fi based password eavesdropping under most real-life traffic conditions. 
The BFI and CSI data collected under real-life background traffic 
are online as specified in Section~\ref{sec:implementation}.
Finally, WiKI-Eve can even steal passwords not sent over the WiFi (e.g., phone unlock) with sufficient background traffic. However, the challenge lies in acquiring the precise timing of the beginning and end of a password input process, for which limited visual cues might be the only feasible approach for now, as discussed in Section~\ref{ssec:identify}.

\vspace{-.5em}
\subsection{Defense Strategies}
Since \name achieves keystroke eavesdropping by overhearing Wi-Fi BFI, the most direct defense strategy is to encrypt data traffic, hence preventing attackers from obtaining BFI in clear text. In fact, this strategy is commonly used in institutional Wi-Fi deployments, which indeed invalidates the basic assumption required by \name as stated in Section~\ref{ssec:atk_scen_mth}.
However, this strategy may cause trouble for scenarios with high user dynamics, 
as frequently performing key exchanges 
substantially increases system complexity.
%
One may consider keyboard randomization~\cite{li2016csi} as an indirect defense strategy, where a randomly keyboard layout is generated whenever a user attempts to enter password. 
By shifting the trouble to user side, this strategy, as indicated by~\cite{jin2021periscope}, forces users to pay more effort when searching for keys on random keyboards, especially affects those used to relying on muscle memory to enter passwords without much visual aid.

A novel strategy against sensing attacks is signal obfuscation. In particular, IRShield~\cite{staat2022irshield} 
leverages IRSs (intelligent reflecting surfaces) installed beside an AP to physically scramble CSI, so as to thwart all sensing attempts. 
%
Unfortunately, this proposal goes against the current trend of evovling Wi-Fi towards ISAC (Integrated Sensing And Communications)~\cite{MUSE-Fi-MobiCom23} where legitimate sensing users should be catered.
To this end, we suggest to exploit MIMO (multiple-in multiple-out) technology adopted by Wi-Fi hardware to scramble Wi-Fi channels~\cite{mimoCrypt-SP24}. 
Acting at physical layer, this strategy can be void of limitations inherent to earlier upper-level digital strategies (e.g., no need for per-user key generation), hence potentially applicable to 
a much wider range of application scenarios.
Of course, this strategy requires hardware or firmware reconstruction, resulting in extra cost compared with digital strategyies. 
Fortunately, realizing ISAC framework by revising Wi-Fi architecture~\cite{isacot} is more and more recognized as a desirable development path.

      

\section{Related works}
\label{sec:rela_work}
We classify existing KI proposals related to \name\ into the following five different categories:

\vspace{-1.5ex}
\paragraph{Radio-Frequency} 
WiKey~\cite{ali2015keystroke} pioneers in leveraging Wi-Fi CSI distortions induced by keystrokes to conduct KI, but the OKI mode used by WiKey is soon exceeded (in SNR) by the IKI mode introduced by WindTalker~\cite{li2016csi} for password inference,
which is followed by Fang et al.~\cite{fang2018no} who exploit English linguistic structure to infer (non-password) keystrokes from CSI obtained via IKI. \hjy{Recently, WiPOS~\cite{zhang2020wipos} uses the OKI model for POS (point of sale) terminal keystroke eavesdropping. SpiderMon~\cite{ling2020spidermon} attempts to perform passive serial keystroke eavesdropping using signals transmitted by commercial cell towers.} WINK~\cite{yang2022wink} also leverages OKI but claims that spatiotemporal analysis could enhance the performance of password inference.

\vspace{-1.5ex}
\paragraph{Acoustic} 
Liu et al.~\cite{liu2015snooping} propose to classify keys on a keyboard based on the time difference of arrival of the acoustic signals (generated by pressing and releasing a key) at the two microphones on a smartphone. Similarly, KeyListener~\cite{lU2019listener} performs KI on touchscreen based on different attenuation of the signals (generated by phone speaker) at the two microphones. PatternListener~\cite{zhou2018patternlistener} compromises pattern locks by using acoustic signals reflected from fingertips to measure their relative movement and infer the pattern lines.
\newrev{These methods can be deemed as acoustic version of OKI.}


\vspace{-1.5ex}
\paragraph{Vision}
Early vision-based KI attacks depend on directly observing the contents displayed on a screen~\cite{maggi2011fast, yue2014my}. 
To make it more practical, later works explore 
side-channel visual cues. KI can be achieved by analyzing changes in the device's physical appearance, such as shadows and deformations on the screen~\cite{yue2014blind}, as well as backside motions of tablet computers~\cite{sun2016visible}. Moreover, capturing videos of the victim's biometric features during typing, such as finger~\cite{shukla2014beware} and eye~\cite{chen2018eyetell} movements, may also enable KI. 
\hjy{Recent work~\cite{cardaioli2022hand} claims to achieve KI
even when victims cover the typing hand with the other hand. 
Although vision-based side-channel attacks have shown a high success rate, the corresponding defense strategies~\cite{liu2023camradar,sami2021lapd} have also grown mature and effective.
Compared with the action features required by vision-based KI attacks, \name only requires visual hints (e.g., actions before starting input) rather than the complete input process, as explained in Section~\ref{ssec:identify}.
}

\vspace{-1.5ex}
\paragraph{Motion Sensors}
TouchLogger~\cite{cai2011touchlogger} uses the accelerometer and gyroscope on smartphones to capture phone body movement and infer numerical keys typed on its touchscreen. (sp)iPhone~\cite{marquardt2011sp} leverages the accelerometer on a nearby phone to detect vibrations from a physical keyboard for enabling KI. Liu et al.~\cite{liu2015good} further exploit the accelerometer on a smartwatch to capture hand movement and infer keystrokes 
on POS terminals or QWERTY keyboards. 

\vspace{-1.5ex}
\paragraph{Electromagnetic Emission} 
Vuagnoux et al.~\cite{vuagnoux2009compromising} propose to eavesdrop on keystrokes from wired and wireless keyboards by capturing electromagnetic emissions during their 
communications. A later work Periscope~\cite{jin2021periscope} extends this idea to a broader range of mobile devices by exploiting human-coupled emission from touchscreens to estimate finger movement trajectories and infer numerical passwords. 
\hjy{Vulnerabilities in USB data transfers have also been exploited for password-stealing~\cite{monaco2018sok} and malicious command execution~\cite{tian2018attention}. Charger-Surfing~\cite{cronin2021charger} further demonstrates that, even without any data transfer over USB, variations of consumed power can be exploited to extract private information such as user passwords.} 




\section{Conclusion}
\label{sec:conclusion}
In this paper, we have proposed \name\ as the first Wi-Fi based KI attack 
with no need for hacking or specialized hardware, making it widely applicable to diversified Wi-Fi devices and attack scenarios. Moreover, \name's adversarial learning framework enables KI to be generalized towards unseen domains, further lifting its practical significance. Finally, we propose SRA to restore the sparse BFI series.
Our extensive evaluations confirm that \name achieves sufficiently high inference accuracy for both individual keystrokes and numerical passwords, and we also tentatively explore extensions to general keyboards.
Our results expose critical vulnerabilities in widely-used applications (e.g., WeChat) and hence
underscore an urgent need for 
enhanced security measures against such risks.

\section*{Acknowledgement}
We are grateful to anonymous reviewers for their constructive suggestions. 
This research is support by National Research Foundation, Singapore and Infocomm Media Development Authority under its Future Communications Research \& Development Programme grant FCP-NTU-RG-2022-015, as well as MoE Tier 1 grant RG16/22. We also thank ERI@N and NTU-IGP for supporting the PhD scholarship of Hongbo Wang.

\balance
\bibliographystyle{ACM-Reference-Format}
\bibliography{WiKI-Eve-arXiv}


\begin{thebibliography}{73}


\ifx \showCODEN    \undefined \def \showCODEN     #1{\unskip}     \fi
\ifx \showDOI      \undefined \def \showDOI       #1{#1}\fi
\ifx \showISBNx    \undefined \def \showISBNx     #1{\unskip}     \fi
\ifx \showISBNxiii \undefined \def \showISBNxiii  #1{\unskip}     \fi
\ifx \showISSN     \undefined \def \showISSN      #1{\unskip}     \fi
\ifx \showLCCN     \undefined \def \showLCCN      #1{\unskip}     \fi
\ifx \shownote     \undefined \def \shownote      #1{#1}          \fi
\ifx \showarticletitle \undefined \def \showarticletitle #1{#1}   \fi
\ifx \showURL      \undefined \def \showURL       {\relax}        \fi
\providecommand\bibfield[2]{#2}
\providecommand\bibinfo[2]{#2}
\providecommand\natexlab[1]{#1}
\providecommand\showeprint[2][]{arXiv:#2}

\bibitem[567(2010)]%
        {5677290}
 \bibinfo{year}{2010}\natexlab{}.
\newblock \showarticletitle{{IEEE Standard for Information Technology--Telecommunications and Information Exchange between Systems--Local and Metropolitan Area Networks--Specific Requirements Part 11: Wireless LAN Medium Access Control (MAC) and Physical Layer (PHY) Specifications Amendment 10: Mesh Networking}}.
\newblock \bibinfo{journal}{\emph{{IEEE P802.11s/D8.0, December 2010}}} (\bibinfo{year}{2010}), \bibinfo{pages}{1--350}.
\newblock


\bibitem[Ali et~al\mbox{.}(2015)]%
        {ali2015keystroke}
\bibfield{author}{\bibinfo{person}{Kamran Ali}, \bibinfo{person}{Alex~X. Liu}, \bibinfo{person}{Wei Wang}, {and} \bibinfo{person}{Muhammad Shahzad}.} \bibinfo{year}{2015}\natexlab{}.
\newblock \showarticletitle{{Keystroke Recognition using WiFi Signals}}. In \bibinfo{booktitle}{\emph{Proc. of the 21st ACM MobiCom}}. \bibinfo{pages}{90--102}.
\newblock


\bibitem[{Apple Inc.}(2023)]%
        {iphone}
\bibfield{author}{\bibinfo{person}{{Apple Inc.}}} \bibinfo{year}{2023}\natexlab{}.
\newblock \bibinfo{title}{{Buy iPhone 13}}.
\newblock \bibinfo{howpublished}{\url{https://www.apple.com/sg/shop/buy-iphone/iphone-13}}.
\newblock
\newblock
\shownote{Online; accessed 12 February 2023}.


\bibitem[Bai et~al\mbox{.}(2018)]%
        {bai2018empirical}
\bibfield{author}{\bibinfo{person}{Shaojie Bai}, \bibinfo{person}{J.~Zico Kolter}, {and} \bibinfo{person}{Vladlen Koltun}.} \bibinfo{year}{2018}\natexlab{}.
\newblock \showarticletitle{{An Empirical Evaluation of Generic Convolutional and Recurrent Networks for Sequence Modeling}}.
\newblock \bibinfo{journal}{\emph{arXiv preprint arXiv:1803.01271}} (\bibinfo{year}{2018}).
\newblock


\bibitem[Bank(2023)]%
        {id4d}
\bibfield{author}{\bibinfo{person}{The~World Bank}.} \bibinfo{year}{2023}\natexlab{}.
\newblock \bibinfo{title}{{Mobile ID}}.
\newblock \bibinfo{howpublished}{\url{https://id4d.worldbank.org/guide/mobile-id}}.
\newblock
\newblock
\shownote{Online; accessed 25 March 2023}.


\bibitem[Ben-David et~al\mbox{.}(2010)]%
        {ben2010theory}
\bibfield{author}{\bibinfo{person}{Shai Ben-David}, \bibinfo{person}{John Blitzer}, \bibinfo{person}{Koby Crammer}, \bibinfo{person}{Alex Kulesza}, \bibinfo{person}{Fernando Pereira}, {and} \bibinfo{person}{Jennifer~Wortman Vaughan}.} \bibinfo{year}{2010}\natexlab{}.
\newblock \showarticletitle{{A Theory of Learning from Different Domains}}.
\newblock \bibinfo{journal}{\emph{Machine Learning}}  \bibinfo{volume}{79} (\bibinfo{year}{2010}), \bibinfo{pages}{151--175}.
\newblock


\bibitem[Beyah and Venkataraman(2011)]%
        {beyah2011rogue}
\bibfield{author}{\bibinfo{person}{Raheem Beyah} {and} \bibinfo{person}{Aravind Venkataraman}.} \bibinfo{year}{2011}\natexlab{}.
\newblock \showarticletitle{{Rogue-access-point Detection: Challenges, Solutions, and Future Directions}}.
\newblock \bibinfo{journal}{\emph{IEEE Security \& Privacy}} \bibinfo{volume}{9}, \bibinfo{number}{5} (\bibinfo{year}{2011}), \bibinfo{pages}{56--61}.
\newblock


\bibitem[Bullock and Parker(2017)]%
        {bullock2017wireshark}
\bibfield{author}{\bibinfo{person}{Jessey Bullock} {and} \bibinfo{person}{Jeff~T. Parker}.} \bibinfo{year}{2017}\natexlab{}.
\newblock \bibinfo{booktitle}{\emph{{Wireshark for Security Professionals: Using Wireshark and the Metasploit Framework}}}.
\newblock \bibinfo{publisher}{John Wiley \& Sons}.
\newblock


\bibitem[Cai and Chen(2011)]%
        {cai2011touchlogger}
\bibfield{author}{\bibinfo{person}{Liang Cai} {and} \bibinfo{person}{Hao Chen}.} \bibinfo{year}{2011}\natexlab{}.
\newblock \showarticletitle{{TouchLogger: Inferring Keystrokes on Touch Screen from Smartphone Motion}}. In \bibinfo{booktitle}{\emph{Proc. of the 6th USENIX Security HotSec}}. \bibinfo{pages}{1--9}.
\newblock


\bibitem[Cardaioli et~al\mbox{.}(2022)]%
        {cardaioli2022hand}
\bibfield{author}{\bibinfo{person}{Matteo Cardaioli}, \bibinfo{person}{Stefano Cecconello}, \bibinfo{person}{Mauro Conti}, \bibinfo{person}{Simone Milani}, \bibinfo{person}{Stjepan Picek}, {and} \bibinfo{person}{Eugen Saraci}.} \bibinfo{year}{2022}\natexlab{}.
\newblock \showarticletitle{{Hand Me Your PIN! Inferring ATM PINs of Users Typing with a Covered Hand}}. In \bibinfo{booktitle}{\emph{Proc. of the 31st USENIX Security}}. \bibinfo{pages}{1687--1704}.
\newblock


\bibitem[Chen et~al\mbox{.}(2018)]%
        {chen2018eyetell}
\bibfield{author}{\bibinfo{person}{Yimin Chen}, \bibinfo{person}{Tao Li}, \bibinfo{person}{Rui Zhang}, \bibinfo{person}{Yanchao Zhang}, {and} \bibinfo{person}{Terri Hedgpeth}.} \bibinfo{year}{2018}\natexlab{}.
\newblock \showarticletitle{{EyeTell: Video-assisted Touchscreen Keystroke Inference from Eye Movements}}. In \bibinfo{booktitle}{\emph{Proc. of the 39th IEEE S \& P}}. \bibinfo{pages}{144--160}.
\newblock


\bibitem[Chen et~al\mbox{.}(2023)]%
        {isacot}
\bibfield{author}{\bibinfo{person}{Zhe Chen}, \bibinfo{person}{Tianyue Zheng}, \bibinfo{person}{Chao Hu}, \bibinfo{person}{Hangcheng Cao}, \bibinfo{person}{Yanbing Yang}, \bibinfo{person}{Hongbo Jiang}, {and} \bibinfo{person}{Jun Luo}.} \bibinfo{year}{2023}\natexlab{}.
\newblock \showarticletitle{{ISACoT: Integrating Sensing with Data Traffic for Ubiquitous IoT Devices}}.
\newblock \bibinfo{journal}{\emph{IEEE Communications Magazine}} \bibinfo{volume}{61}, \bibinfo{number}{5} (\bibinfo{year}{2023}), \bibinfo{pages}{98--104}.
\newblock


\bibitem[Cisco~Systems(2023)]%
        {cisco}
\bibfield{author}{\bibinfo{person}{Inc. Cisco~Systems}.} \bibinfo{year}{2023}\natexlab{}.
\newblock \bibinfo{title}{{Cisco Wireless Controller Configuration Guide, Release 8.4}}.
\newblock \bibinfo{howpublished}{\url{https://www.cisco.com/c/en/us/td/docs/wireless/controller/8-4/config-guide/b_cg84/wireless_intrusion_detection_system.html\#rogue-ap-classification}}.
\newblock
\newblock
\shownote{Online; accessed 25 March 2023}.


\bibitem[Co.(2023)]%
        {chase}
\bibfield{author}{\bibinfo{person}{JPMorgan Chase~\& Co.}} \bibinfo{year}{2023}\natexlab{}.
\newblock \bibinfo{title}{{Mobile Banking Features with Chase Mobile App}}.
\newblock \bibinfo{howpublished}{\url{https://www.chase.com/digital/mobile-banking}}.
\newblock
\newblock
\shownote{Online; accessed 25 March 2023}.


\bibitem[Corporation(2008)]%
        {5300}
\bibfield{author}{\bibinfo{person}{Intel Corporation}.} \bibinfo{year}{2008}\natexlab{}.
\newblock \bibinfo{title}{{Intel Ultimate N WiFi Link 5300}}.
\newblock \bibinfo{howpublished}{\url{https://www.intel.com/content/dam/www/public/us/en/documents/product-briefs/ultimate-n-wifi-link-5300-brief.pdf}}.
\newblock
\newblock
\shownote{Online; accessed 28 March 2023}.


\bibitem[Corporation(2023)]%
        {intel}
\bibfield{author}{\bibinfo{person}{Intel Corporation}.} \bibinfo{year}{2023}\natexlab{}.
\newblock \bibinfo{title}{{Intel® Wi-Fi 6 AX201}}.
\newblock \bibinfo{howpublished}{\url{https://www.intel.sg/content/www/xa/en/products/sku/130293/intel-wifi-6-ax201-gig/specifications.html}}.
\newblock
\newblock
\shownote{Online; accessed 25 March 2023}.


\bibitem[Cover(2015)]%
        {cover2015digital}
\bibfield{author}{\bibinfo{person}{Rob Cover}.} \bibinfo{year}{2015}\natexlab{}.
\newblock \bibinfo{booktitle}{\emph{{Digital Identities: Creating and Communicating the Online Self}}}.
\newblock \bibinfo{publisher}{Academic Press}.
\newblock


\bibitem[Cronin et~al\mbox{.}(2021)]%
        {cronin2021charger}
\bibfield{author}{\bibinfo{person}{Patrick Cronin}, \bibinfo{person}{Xing Gao}, \bibinfo{person}{Chengmo Yang}, {and} \bibinfo{person}{Haining Wang}.} \bibinfo{year}{2021}\natexlab{}.
\newblock \showarticletitle{Charger-Surfing: Exploiting a Power Line Side-Channel for Smartphone Information Leakage}. In \bibinfo{booktitle}{\emph{Proc. of the 30th USENIX Security}}. \bibinfo{pages}{681--698}.
\newblock


\bibitem[Fang et~al\mbox{.}(2018)]%
        {fang2018no}
\bibfield{author}{\bibinfo{person}{Song Fang}, \bibinfo{person}{Ian Markwood}, \bibinfo{person}{Yao Liu}, \bibinfo{person}{Shangqing Zhao}, \bibinfo{person}{Zhuo Lu}, {and} \bibinfo{person}{Haojin Zhu}.} \bibinfo{year}{2018}\natexlab{}.
\newblock \showarticletitle{{No Training Hurdles: Fast Training-agnostic Attacks to Infer Your Typing}}. In \bibinfo{booktitle}{\emph{Proc. of the 25th ACM CCS}}. \bibinfo{pages}{1747--1760}.
\newblock


\bibitem[Ganin and Lempitsky(2015)]%
        {ganin2015unsupervised}
\bibfield{author}{\bibinfo{person}{Yaroslav Ganin} {and} \bibinfo{person}{Victor Lempitsky}.} \bibinfo{year}{2015}\natexlab{}.
\newblock \showarticletitle{{Unsupervised Domain Adaptation by Backpropagation}}. In \bibinfo{booktitle}{\emph{Proc. of the 32nd ACM ICML}}. \bibinfo{pages}{1180--1189}.
\newblock


\bibitem[Gast(2013)]%
        {survivalguide}
\bibfield{author}{\bibinfo{person}{Matthew~S. Gast}.} \bibinfo{year}{2013}\natexlab{}.
\newblock \bibinfo{booktitle}{\emph{{802.11ac A Survival Guide: Wi-Fi at Gigabit and Beyond}}}.
\newblock \bibinfo{publisher}{O'Reilly Media, Inc.}
\newblock


\bibitem[Goodfellow et~al\mbox{.}(2014)]%
        {gan}
\bibfield{author}{\bibinfo{person}{Ian Goodfellow}, \bibinfo{person}{Jean Pouget-Abadie}, \bibinfo{person}{Mehdi Mirza}, \bibinfo{person}{Bing Xu}, \bibinfo{person}{David Warde-Farley}, \bibinfo{person}{Sherjil Ozair}, \bibinfo{person}{Aaron Courville}, {and} \bibinfo{person}{Yoshua Bengio}.} \bibinfo{year}{2014}\natexlab{}.
\newblock \showarticletitle{{Generative Adversarial Nets}}. In \bibinfo{booktitle}{\emph{Proc. of 28th NeurIPS}}. \bibinfo{pages}{2672--2680}.
\newblock


\bibitem[{Google LLC}(2023)]%
        {google}
\bibfield{author}{\bibinfo{person}{{Google LLC}}.} \bibinfo{year}{2023}\natexlab{}.
\newblock \bibinfo{title}{{Pixel 6a}}.
\newblock \bibinfo{howpublished}{\url{https://store.google.com/product/pixel_6a?hl=en-GB}}.
\newblock
\newblock
\shownote{Online; accessed 10 April 2023}.


\bibitem[Hadnagy(2010)]%
        {hadnagy2010social}
\bibfield{author}{\bibinfo{person}{Christopher Hadnagy}.} \bibinfo{year}{2010}\natexlab{}.
\newblock \bibinfo{booktitle}{\emph{{Social Engineering: The Art of Human Hacking}}}.
\newblock \bibinfo{publisher}{John Wiley \& Sons}.
\newblock


\bibitem[Halperin et~al\mbox{.}(2011)]%
        {csi}
\bibfield{author}{\bibinfo{person}{Daniel Halperin}, \bibinfo{person}{Wenjun Hu}, \bibinfo{person}{Anmol Sheth}, {and} \bibinfo{person}{David Wetherall}.} \bibinfo{year}{2011}\natexlab{}.
\newblock \showarticletitle{{Tool Release: Gathering 802.11n Traces with Channel State Information}}.
\newblock \bibinfo{journal}{\emph{ACM SIGCOMM Comput. Commun. Rev.}} \bibinfo{volume}{41}, \bibinfo{number}{1} (\bibinfo{year}{2011}), \bibinfo{pages}{53}.
\newblock


\bibitem[He et~al\mbox{.}(2015)]%
        {he2015spatial}
\bibfield{author}{\bibinfo{person}{Kaiming He}, \bibinfo{person}{Xiangyu Zhang}, \bibinfo{person}{Shaoqing Ren}, {and} \bibinfo{person}{Jian Sun}.} \bibinfo{year}{2015}\natexlab{}.
\newblock \showarticletitle{{Spatial Pyramid Pooling in Deep Convolutional Networks for Visual Recognition}}.
\newblock \bibinfo{journal}{\emph{IEEE Transactions on Pattern Analysis and Machine Intelligence}} \bibinfo{volume}{37}, \bibinfo{number}{9} (\bibinfo{year}{2015}), \bibinfo{pages}{1904--1916}.
\newblock


\bibitem[Hu et~al\mbox{.}(2023)]%
        {MUSE-Fi-MobiCom23}
\bibfield{author}{\bibinfo{person}{Jingzhi Hu}, \bibinfo{person}{Tianyue Zheng}, \bibinfo{person}{Zhe Chen}, \bibinfo{person}{Hongbo Wang}, {and} \bibinfo{person}{Jun Luo}.} \bibinfo{year}{2023}\natexlab{}.
\newblock \showarticletitle{{MUSE-Fi: Contactless MUti-person SEnsing Exploiting Near-field Wi-Fi Channel Variation}}. In \bibinfo{booktitle}{\emph{Proc. of the 29th ACM MobiCom}}. \bibinfo{pages}{1--15}.
\newblock


\bibitem[{Huawei Device Co., Ltd.}(2023)]%
        {huawei}
\bibfield{author}{\bibinfo{person}{{Huawei Device Co., Ltd.}}} \bibinfo{year}{2023}\natexlab{}.
\newblock \bibinfo{title}{{HUAWEI P40 Pro}}.
\newblock \bibinfo{howpublished}{\url{https://consumer.huawei.com/en/phones/p40-pro/}}.
\newblock
\newblock
\shownote{Online; accessed 10 April 2023}.


\bibitem[Inc.(2023)]%
        {acer}
\bibfield{author}{\bibinfo{person}{Acer Inc.}} \bibinfo{year}{2023}\natexlab{}.
\newblock \bibinfo{title}{{Acer TravelMate Laptops for Business}}.
\newblock \bibinfo{howpublished}{\url{https://www.acer.com/sg-en/laptops/travelmate}}.
\newblock
\newblock
\shownote{Online; accessed 25 March 2023}.


\bibitem[Jiang et~al\mbox{.}(2021)]%
        {jiang2021eliminating}
\bibfield{author}{\bibinfo{person}{Zhiping Jiang}, \bibinfo{person}{Tom~H Luan}, \bibinfo{person}{Xincheng Ren}, \bibinfo{person}{Dongtao Lv}, \bibinfo{person}{Han Hao}, \bibinfo{person}{Jing Wang}, \bibinfo{person}{Kun Zhao}, \bibinfo{person}{Wei Xi}, \bibinfo{person}{Yueshen Xu}, {and} \bibinfo{person}{Rui Li}.} \bibinfo{year}{2021}\natexlab{}.
\newblock \showarticletitle{{Eliminating the Barriers: Demystifying wi-fi Baseband Design and Introducing the Picoscenes Wi-Fi sensing Platform}}.
\newblock \bibinfo{journal}{\emph{IEEE Internet of Things Journal}} \bibinfo{volume}{9}, \bibinfo{number}{6} (\bibinfo{year}{2021}), \bibinfo{pages}{4476--4496}.
\newblock


\bibitem[Jin et~al\mbox{.}(2021)]%
        {jin2021periscope}
\bibfield{author}{\bibinfo{person}{Wenqiang Jin}, \bibinfo{person}{Srinivasan Murali}, \bibinfo{person}{Huadi Zhu}, {and} \bibinfo{person}{Ming Li}.} \bibinfo{year}{2021}\natexlab{}.
\newblock \showarticletitle{{Periscope: A Keystroke Inference Attack Using Human Coupled Electromagnetic Emanations}}. In \bibinfo{booktitle}{\emph{Proc. of the 28th ACM CCS}}. \bibinfo{pages}{700--714}.
\newblock


\bibitem[Kingman(1992)]%
        {kingman1992poisson}
\bibfield{author}{\bibinfo{person}{John Frank~Charles Kingman}.} \bibinfo{year}{1992}\natexlab{}.
\newblock \bibinfo{booktitle}{\emph{Poisson Processes}}. Vol.~\bibinfo{volume}{3}.
\newblock \bibinfo{publisher}{Clarendon Press}.
\newblock


\bibitem[Kiranyaz et~al\mbox{.}(2019)]%
        {kiranyaz20191}
\bibfield{author}{\bibinfo{person}{Serkan Kiranyaz}, \bibinfo{person}{Turker Ince}, \bibinfo{person}{Osama Abdeljaber}, \bibinfo{person}{Onur Avci}, {and} \bibinfo{person}{Moncef Gabbouj}.} \bibinfo{year}{2019}\natexlab{}.
\newblock \showarticletitle{{1-D Convolutional Neural Networks for Signal Processing Applications}}. In \bibinfo{booktitle}{\emph{Proc. of the 44th IEEE ICASSP}}. \bibinfo{pages}{8360--8364}.
\newblock


\bibitem[Li et~al\mbox{.}(2016)]%
        {li2016csi}
\bibfield{author}{\bibinfo{person}{Mengyuan Li}, \bibinfo{person}{Yan Meng}, \bibinfo{person}{Junyi Liu}, \bibinfo{person}{Haojin Zhu}, \bibinfo{person}{Xiaohui Liang}, \bibinfo{person}{Yao Liu}, {and} \bibinfo{person}{Na Ruan}.} \bibinfo{year}{2016}\natexlab{}.
\newblock \showarticletitle{{When CSI Meets Public WiFi: Inferring Your Mobile Phone Password via WiFi Signals}}. In \bibinfo{booktitle}{\emph{Proc. of the 23rd ACM CCS}}. \bibinfo{pages}{1068--1079}.
\newblock


\bibitem[Ling et~al\mbox{.}(2020)]%
        {ling2020spidermon}
\bibfield{author}{\bibinfo{person}{Kang Ling}, \bibinfo{person}{Yuntang Liu}, \bibinfo{person}{Ke Sun}, \bibinfo{person}{Wei Wang}, \bibinfo{person}{Lei Xie}, {and} \bibinfo{person}{Qing Gu}.} \bibinfo{year}{2020}\natexlab{}.
\newblock \showarticletitle{{SpiderMon: Towards Using Cell Towers as Illuminating Sources for Keystroke Monitoring}}. In \bibinfo{booktitle}{\emph{Proc. of the 39th IEEE INFOCOM}}. \bibinfo{pages}{666--675}.
\newblock


\bibitem[Liu et~al\mbox{.}(2015a)]%
        {liu2015snooping}
\bibfield{author}{\bibinfo{person}{Jian Liu}, \bibinfo{person}{Yan Wang}, \bibinfo{person}{Gorkem Kar}, \bibinfo{person}{Yingying Chen}, \bibinfo{person}{Jie Yang}, {and} \bibinfo{person}{Marco Gruteser}.} \bibinfo{year}{2015}\natexlab{a}.
\newblock \showarticletitle{{Snooping Keystrokes with mm-level Audio Ranging on a Single Phone}}. In \bibinfo{booktitle}{\emph{Proc. of the 21st ACM MobiCom}}. \bibinfo{pages}{142--154}.
\newblock


\bibitem[Liu et~al\mbox{.}(2015b)]%
        {liu2015good}
\bibfield{author}{\bibinfo{person}{Xiangyu Liu}, \bibinfo{person}{Zhe Zhou}, \bibinfo{person}{Wenrui Diao}, \bibinfo{person}{Zhou Li}, {and} \bibinfo{person}{Kehuan Zhang}.} \bibinfo{year}{2015}\natexlab{b}.
\newblock \showarticletitle{{When Good Becomes Evil: Keystroke Inference with Smartwatch}}. In \bibinfo{booktitle}{\emph{Proc. of the 22nd ACM CCS}}. \bibinfo{pages}{1273--1285}.
\newblock


\bibitem[Liu et~al\mbox{.}(2023)]%
        {liu2023camradar}
\bibfield{author}{\bibinfo{person}{Ziwei Liu}, \bibinfo{person}{Feng Lin}, \bibinfo{person}{Chao Wang}, \bibinfo{person}{Yijie Shen}, \bibinfo{person}{Zhongjie Ba}, \bibinfo{person}{Li Lu}, \bibinfo{person}{Wenyao Xu}, {and} \bibinfo{person}{Kui Ren}.} \bibinfo{year}{2023}\natexlab{}.
\newblock \showarticletitle{{CamRadar: Hidden Camera Detection Leveraging Amplitude-modulated Sensor Images Embedded in Electromagnetic Emanations}}.
\newblock \bibinfo{journal}{\emph{Proc. of the 23rd ACM UbiComp}} \bibinfo{volume}{6}, \bibinfo{number}{4} (\bibinfo{year}{2023}), \bibinfo{pages}{1--25}.
\newblock


\bibitem[Lu et~al\mbox{.}(2019)]%
        {lU2019listener}
\bibfield{author}{\bibinfo{person}{Li Lu}, \bibinfo{person}{Jiadi Yu}, \bibinfo{person}{Yingying Chen}, \bibinfo{person}{Yanmin Zhu}, \bibinfo{person}{Xiangyu Xu}, \bibinfo{person}{Guangtao Xue}, {and} \bibinfo{person}{Minglu Li}.} \bibinfo{year}{2019}\natexlab{}.
\newblock \showarticletitle{KeyListener: Inferring Keystrokes on QWERTY Keyboard of Touch Screen through Acoustic Signals}. In \bibinfo{booktitle}{\emph{Proc. of the 38th IEEE INFOCOM}}. \bibinfo{pages}{775--783}.
\newblock


\bibitem[Luo et~al\mbox{.}(2024)]%
        {mimoCrypt-SP24}
\bibfield{author}{\bibinfo{person}{Jun Luo}, \bibinfo{person}{Hangcheng Cao}, \bibinfo{person}{Hongbo Jiang}, \bibinfo{person}{Yanbing Yang}, {and} \bibinfo{person}{Zhe Chen}.} \bibinfo{year}{2024}\natexlab{}.
\newblock \showarticletitle{{{\tiny MIMO}Crypt: Multi-User Privacy-Preserving Wi-Fi Sensing via MIMO Encryption}}. In \bibinfo{booktitle}{\emph{Proc. of the 45th IEEE S\&P}}. \bibinfo{pages}{1--19}.
\newblock


\bibitem[Maaten and Hinton(2008)]%
        {maaten2008visualizing}
\bibfield{author}{\bibinfo{person}{Laurens van~der Maaten} {and} \bibinfo{person}{Geoffrey Hinton}.} \bibinfo{year}{2008}\natexlab{}.
\newblock \showarticletitle{{Visualizing Data Using t-SNE}}.
\newblock \bibinfo{journal}{\emph{Journal of Machine Learning Research}} \bibinfo{volume}{9}, \bibinfo{number}{Nov} (\bibinfo{year}{2008}), \bibinfo{pages}{2579--2605}.
\newblock


\bibitem[Maggi et~al\mbox{.}(2011)]%
        {maggi2011fast}
\bibfield{author}{\bibinfo{person}{Federico Maggi}, \bibinfo{person}{Alberto Volpatto}, \bibinfo{person}{Simone Gasparini}, \bibinfo{person}{Giacomo Boracchi}, {and} \bibinfo{person}{Stefano Zanero}.} \bibinfo{year}{2011}\natexlab{}.
\newblock \showarticletitle{{A Fast Eavesdropping Attack against Touchscreens}}. In \bibinfo{booktitle}{\emph{Prof. of the 7th IAS}}. IEEE, \bibinfo{pages}{320--325}.
\newblock


\bibitem[Marquardt et~al\mbox{.}(2011)]%
        {marquardt2011sp}
\bibfield{author}{\bibinfo{person}{Philip Marquardt}, \bibinfo{person}{Arunabh Verma}, \bibinfo{person}{Henry Carter}, {and} \bibinfo{person}{Patrick Traynor}.} \bibinfo{year}{2011}\natexlab{}.
\newblock \showarticletitle{{(sp)iPhone: Decoding Vibrations from Nearby Keyboards using Mobile Phone Accelerometers}}. In \bibinfo{booktitle}{\emph{Proc. of the 18th ACM CCS}}. \bibinfo{pages}{551--562}.
\newblock


\bibitem[Monaco(2018)]%
        {monaco2018sok}
\bibfield{author}{\bibinfo{person}{John~V. Monaco}.} \bibinfo{year}{2018}\natexlab{}.
\newblock \showarticletitle{{SoK: Keylogging Side Channels}}. In \bibinfo{booktitle}{\emph{Proc. of the 39th IEEE S\&P}}. \bibinfo{pages}{211--228}.
\newblock


\bibitem[Nitzberg(1972)]%
        {nitzberg1972constant}
\bibfield{author}{\bibinfo{person}{Ramon Nitzberg}.} \bibinfo{year}{1972}\natexlab{}.
\newblock \showarticletitle{{Constant-false-alarm-rate Signal Processors for Several Types of Interference}}.
\newblock \bibinfo{journal}{\emph{IEEE Trans. Aerospace Electron. Systems}} \bibinfo{number}{1} (\bibinfo{year}{1972}), \bibinfo{pages}{27--34}.
\newblock


\bibitem[{OnePlus}(2023)]%
        {oneplus}
\bibfield{author}{\bibinfo{person}{{OnePlus}}.} \bibinfo{year}{2023}\natexlab{}.
\newblock \bibinfo{title}{{OnePlus 10T 5G}}.
\newblock \bibinfo{howpublished}{\url{https://www.oneplus.com/sg/10t}}.
\newblock
\newblock
\shownote{Online; accessed 10 April 2023}.


\bibitem[Orebaugh et~al\mbox{.}(2006)]%
        {orebaugh2006wireshark}
\bibfield{author}{\bibinfo{person}{Angela Orebaugh}, \bibinfo{person}{Gilbert Ramirez}, {and} \bibinfo{person}{Jay Beale}.} \bibinfo{year}{2006}\natexlab{}.
\newblock \bibinfo{booktitle}{\emph{{Wireshark \& Ethereal Network Protocol Analyzer Toolkit}}}.
\newblock \bibinfo{publisher}{Elsevier}.
\newblock


\bibitem[Pan et~al\mbox{.}(2010)]%
        {pan2010domain}
\bibfield{author}{\bibinfo{person}{Sinno~Jialin Pan}, \bibinfo{person}{Ivor~W. Tsang}, \bibinfo{person}{James~T. Kwok}, {and} \bibinfo{person}{Qiang Yang}.} \bibinfo{year}{2010}\natexlab{}.
\newblock \showarticletitle{{Domain Adaptation via Transfer Component Analysis}}.
\newblock \bibinfo{journal}{\emph{IEEE Transactions on Neural Networks}} \bibinfo{volume}{22}, \bibinfo{number}{2} (\bibinfo{year}{2010}), \bibinfo{pages}{199--210}.
\newblock


\bibitem[Paszke et~al\mbox{.}(2019)]%
        {pytorch}
\bibfield{author}{\bibinfo{person}{Adam Paszke}, \bibinfo{person}{Sam Gross}, \bibinfo{person}{Francisco Massa}, \bibinfo{person}{Adam Lerer}, \bibinfo{person}{James Bradbury}, \bibinfo{person}{Gregory Chanan}, \bibinfo{person}{Trevor Killeen}, \bibinfo{person}{Zeming Lin}, \bibinfo{person}{Natalia Gimelshein}, \bibinfo{person}{Luca Antiga}, {et~al\mbox{.}}} \bibinfo{year}{2019}\natexlab{}.
\newblock \showarticletitle{{PyTorch: An Imperative Style, High-Performance Deep Learning Library}}.
\newblock \bibinfo{journal}{\emph{arXiv preprint arXiv:1912.01703}} (\bibinfo{year}{2019}).
\newblock


\bibitem[Sami et~al\mbox{.}(2021)]%
        {sami2021lapd}
\bibfield{author}{\bibinfo{person}{Sriram Sami}, \bibinfo{person}{Sean Rui~Xiang Tan}, \bibinfo{person}{Bangjie Sun}, {and} \bibinfo{person}{Jun Han}.} \bibinfo{year}{2021}\natexlab{}.
\newblock \showarticletitle{{LAPD: Hidden Spy Camera Detection Using Smartphone Time-of-flight Sensors}}. In \bibinfo{booktitle}{\emph{Proc. of the 19th ACM SenSys}}. \bibinfo{pages}{288--301}.
\newblock


\bibitem[{Samsung}(2023)]%
        {samsung}
\bibfield{author}{\bibinfo{person}{{Samsung}}.} \bibinfo{year}{2023}\natexlab{}.
\newblock \bibinfo{title}{{Samsung Galaxy S20 Series}}.
\newblock \bibinfo{howpublished}{\url{https://www.samsung.com/sg/news/local/galaxy-s20-launch/}}.
\newblock
\newblock
\shownote{Online; accessed 10 April 2023}.


\bibitem[Schulz et~al\mbox{.}(2017)]%
        {nexmon:project}
\bibfield{author}{\bibinfo{person}{Matthias Schulz}, \bibinfo{person}{Daniel Wegemer}, {and} \bibinfo{person}{Matthias Hollick}.} \bibinfo{year}{2017}\natexlab{}.
\newblock \bibinfo{title}{{Nexmon: The C-based Firmware Patching Framework}}.
\newblock
\newblock
\urldef\tempurl%
\url{https://nexmon.org}
\showURL{%
\tempurl}


\bibitem[Shukla et~al\mbox{.}(2014)]%
        {shukla2014beware}
\bibfield{author}{\bibinfo{person}{Diksha Shukla}, \bibinfo{person}{Rajesh Kumar}, \bibinfo{person}{Abdul Serwadda}, {and} \bibinfo{person}{Vir~V. Phoha}.} \bibinfo{year}{2014}\natexlab{}.
\newblock \showarticletitle{{Beware, Your Hands Reveal Your Secrets!}}. In \bibinfo{booktitle}{\emph{Proc. of the 21st ACM CCS}}. \bibinfo{pages}{904--917}.
\newblock


\bibitem[Staat et~al\mbox{.}(2022)]%
        {staat2022irshield}
\bibfield{author}{\bibinfo{person}{Paul Staat}, \bibinfo{person}{Simon Mulzer}, \bibinfo{person}{Stefan Roth}, \bibinfo{person}{Veelasha Moonsamy}, \bibinfo{person}{Markus Heinrichs}, \bibinfo{person}{Rainer Kronberger}, \bibinfo{person}{Aydin Sezgin}, {and} \bibinfo{person}{Christof Paar}.} \bibinfo{year}{2022}\natexlab{}.
\newblock \showarticletitle{IRShield: A Countermeasure Against Adversarial Physical-layer Wireless Sensing}. In \bibinfo{booktitle}{\emph{Proc. of the 43rd IEEE S \& P}}. \bibinfo{pages}{1705--1721}.
\newblock


\bibitem[statista(2023)]%
        {length}
\bibfield{author}{\bibinfo{person}{statista}.} \bibinfo{year}{2023}\natexlab{}.
\newblock \bibinfo{title}{{Average Number of Characters for a Password in the United States in 2021}}.
\newblock \bibinfo{howpublished}{\url{https://www.statista.com/statistics/1305713/average-character-length-of-a-password-us/}}.
\newblock
\newblock
\shownote{Online; accessed 25 March 2023}.


\bibitem[Stewart(1993)]%
        {stewart1993early}
\bibfield{author}{\bibinfo{person}{Gilbert~W. Stewart}.} \bibinfo{year}{1993}\natexlab{}.
\newblock \showarticletitle{{On the Early History of the Singular Value Decomposition}}.
\newblock \bibinfo{journal}{\emph{SIAM Rev.}} \bibinfo{volume}{35}, \bibinfo{number}{4} (\bibinfo{year}{1993}), \bibinfo{pages}{551--566}.
\newblock


\bibitem[Sun et~al\mbox{.}(2016)]%
        {sun2016visible}
\bibfield{author}{\bibinfo{person}{Jingchao Sun}, \bibinfo{person}{Xiaocong Jin}, \bibinfo{person}{Yimin Chen}, \bibinfo{person}{Jinxue Zhang}, \bibinfo{person}{Yanchao Zhang}, {and} \bibinfo{person}{Rui Zhang}.} \bibinfo{year}{2016}\natexlab{}.
\newblock \showarticletitle{{Visible: Video-assisted Keystroke Inference From Tablet Backside Motion}}. In \bibinfo{booktitle}{\emph{Proc. of the 23rd ISOC NDSS}}.
\newblock


\bibitem[Tian et~al\mbox{.}(2018)]%
        {tian2018attention}
\bibfield{author}{\bibinfo{person}{Dave~Jing Tian}, \bibinfo{person}{Grant Hernandez}, \bibinfo{person}{Joseph~I Choi}, \bibinfo{person}{Vanessa Frost}, \bibinfo{person}{Christie Raules}, \bibinfo{person}{Patrick Traynor}, \bibinfo{person}{Hayawardh Vijayakumar}, \bibinfo{person}{Lee Harrison}, \bibinfo{person}{Amir Rahmati}, \bibinfo{person}{Michael Grace}, {et~al\mbox{.}}} \bibinfo{year}{2018}\natexlab{}.
\newblock \showarticletitle{ATtention Spanned: Comprehensive Vulnerability Analysis of AT Commands Within the Android Ecosystem}. In \bibinfo{booktitle}{\emph{Proc. of the 27th USENIX Security}}. \bibinfo{pages}{273--290}.
\newblock


\bibitem[Tirumala(1999)]%
        {tirumala1999iperf}
\bibfield{author}{\bibinfo{person}{Ajay Tirumala}.} \bibinfo{year}{1999}\natexlab{}.
\newblock \showarticletitle{{iPerf: The TCP/UDP Bandwidth Measurement Tool}}.
\newblock \bibinfo{journal}{\emph{http://dast.nlanr.net/Projects/Iperf/}} (\bibinfo{year}{1999}).
\newblock


\bibitem[Vuagnoux and Pasini(2009)]%
        {vuagnoux2009compromising}
\bibfield{author}{\bibinfo{person}{Martin Vuagnoux} {and} \bibinfo{person}{Sylvain Pasini}.} \bibinfo{year}{2009}\natexlab{}.
\newblock \showarticletitle{{Compromising Electromagnetic Emanations of Wired and Wireless Keyboards}}. In \bibinfo{booktitle}{\emph{Proc. of the 18th USENIX Security}}, Vol.~\bibinfo{volume}{8}. \bibinfo{pages}{1--16}.
\newblock


\bibitem[Wang et~al\mbox{.}(2016)]%
        {wang2016locating}
\bibfield{author}{\bibinfo{person}{Chen Wang}, \bibinfo{person}{Xiuyuan Zheng}, \bibinfo{person}{Yingying Chen}, {and} \bibinfo{person}{Jie Yang}.} \bibinfo{year}{2016}\natexlab{}.
\newblock \showarticletitle{{Locating Rogue Access Point using Fine-grained Channel Information}}.
\newblock \bibinfo{journal}{\emph{IEEE Transactions on Mobile Computing}} \bibinfo{volume}{16}, \bibinfo{number}{9} (\bibinfo{year}{2016}), \bibinfo{pages}{2560--2573}.
\newblock


\bibitem[WeChat(2023)]%
        {wechat}
\bibfield{author}{\bibinfo{person}{WeChat}.} \bibinfo{year}{2023}\natexlab{}.
\newblock \bibinfo{title}{{WeChat - Free Messaging and Chatting App}}.
\newblock \bibinfo{howpublished}{\url{https://www.wechat.com/}}.
\newblock
\newblock
\shownote{Online; accessed 28 March 2023}.


\bibitem[WiKI-Eve.(2023)]%
        {WiKI}
\bibfield{author}{\bibinfo{person}{WiKI-Eve.}} \bibinfo{year}{2023}\natexlab{}.
\newblock \bibinfo{howpublished}{\url{https://github.com/Nest-Fi/WiKI-Eve}}.
\newblock
\newblock
\shownote{Online; accessed 6 August 2023}.


\bibitem[Wong(2015)]%
        {wong2015performance}
\bibfield{author}{\bibinfo{person}{Tzu-Tsung Wong}.} \bibinfo{year}{2015}\natexlab{}.
\newblock \showarticletitle{{Performance Evaluation of Classification Algorithms by k-fold and Leave-one-out Cross Validation}}.
\newblock \bibinfo{journal}{\emph{Pattern Recognition}} \bibinfo{volume}{48}, \bibinfo{number}{9} (\bibinfo{year}{2015}), \bibinfo{pages}{2839--2846}.
\newblock


\bibitem[Wu et~al\mbox{.}(2017)]%
        {wu2017forensic}
\bibfield{author}{\bibinfo{person}{Songyang Wu}, \bibinfo{person}{Yong Zhang}, \bibinfo{person}{Xupeng Wang}, \bibinfo{person}{Xiong Xiong}, {and} \bibinfo{person}{Lin Du}.} \bibinfo{year}{2017}\natexlab{}.
\newblock \showarticletitle{{Forensic Analysis of WeChat on Android Smartphones}}.
\newblock \bibinfo{journal}{\emph{Digital Investigation}}  \bibinfo{volume}{21} (\bibinfo{year}{2017}), \bibinfo{pages}{3--10}.
\newblock


\bibitem[{Xiaomi}(2023)]%
        {xiaomi}
\bibfield{author}{\bibinfo{person}{{Xiaomi}}.} \bibinfo{year}{2023}\natexlab{}.
\newblock \bibinfo{title}{{Xiaomi 13 Pro}}.
\newblock \bibinfo{howpublished}{\url{https://www.mi.com/sg/product/xiaomi-13-pro/}}.
\newblock
\newblock
\shownote{Online; accessed 10 April 2023}.


\bibitem[Yang et~al\mbox{.}(2022)]%
        {yang2022wink}
\bibfield{author}{\bibinfo{person}{Edwin Yang}, \bibinfo{person}{Qiuye He}, {and} \bibinfo{person}{Song Fang}.} \bibinfo{year}{2022}\natexlab{}.
\newblock \showarticletitle{{WINK: Wireless Inference of Numerical Keystrokes via Zero-Training Spatiotemporal Analysis}}. In \bibinfo{booktitle}{\emph{Proc. of the 29th ACM CCS}}. \bibinfo{pages}{3033--3047}.
\newblock


\bibitem[Yu et~al\mbox{.}(2019)]%
        {yu2019review}
\bibfield{author}{\bibinfo{person}{Yong Yu}, \bibinfo{person}{Xiaosheng Si}, \bibinfo{person}{Changhua Hu}, {and} \bibinfo{person}{Jianxun Zhang}.} \bibinfo{year}{2019}\natexlab{}.
\newblock \showarticletitle{A Review of Recurrent Neural Networks: LSTM Cells and Network Architectures}.
\newblock \bibinfo{journal}{\emph{Neural computation}} \bibinfo{volume}{31}, \bibinfo{number}{7} (\bibinfo{year}{2019}), \bibinfo{pages}{1235--1270}.
\newblock


\bibitem[Yue et~al\mbox{.}(2014a)]%
        {yue2014blind}
\bibfield{author}{\bibinfo{person}{Qinggang Yue}, \bibinfo{person}{Zhen Ling}, \bibinfo{person}{Xinwen Fu}, \bibinfo{person}{Benyuan Liu}, \bibinfo{person}{Kui Ren}, {and} \bibinfo{person}{Wei Zhao}.} \bibinfo{year}{2014}\natexlab{a}.
\newblock \showarticletitle{{Blind Recognition of Touched Keys on Mobile Devices}}. In \bibinfo{booktitle}{\emph{Proc. of the 21st ACM CCS}}. \bibinfo{pages}{1403--1414}.
\newblock


\bibitem[Yue et~al\mbox{.}(2014b)]%
        {yue2014my}
\bibfield{author}{\bibinfo{person}{Qinggang Yue}, \bibinfo{person}{Zhen Ling}, \bibinfo{person}{Xinwen Fu}, \bibinfo{person}{Benyuan Liu}, \bibinfo{person}{Wei Yu}, {and} \bibinfo{person}{Wei Zhao}.} \bibinfo{year}{2014}\natexlab{b}.
\newblock \showarticletitle{{My Google Glass Sees Your Passwords}}.
\newblock \bibinfo{journal}{\emph{Prof. of the Black Hat USA}} (\bibinfo{year}{2014}).
\newblock


\bibitem[Zhang et~al\mbox{.}(2022)]%
        {zhang2022quantifying}
\bibfield{author}{\bibinfo{person}{Shujie Zhang}, \bibinfo{person}{Tianyue Zheng}, \bibinfo{person}{Hongbo Wang}, \bibinfo{person}{Zhe Chen}, {and} \bibinfo{person}{Jun Luo}.} \bibinfo{year}{2022}\natexlab{}.
\newblock \showarticletitle{{Quantifying the Physical Separability of RF-based Multi-Person Respiration Monitoring via SINR}}. In \bibinfo{booktitle}{\emph{Proc. of the 20th ACM SenSys}}. \bibinfo{pages}{47–60}.
\newblock


\bibitem[Zhang et~al\mbox{.}(2020)]%
        {zhang2020wipos}
\bibfield{author}{\bibinfo{person}{Zijian Zhang}, \bibinfo{person}{Nurilla Avazov}, \bibinfo{person}{Jiamou Liu}, \bibinfo{person}{Bakh Khoussainov}, \bibinfo{person}{Xin Li}, \bibinfo{person}{Keke Gai}, {and} \bibinfo{person}{Liehuang Zhu}.} \bibinfo{year}{2020}\natexlab{}.
\newblock \showarticletitle{{WiPOS: A POS Terminal Password Inference System Based on Wireless Signals}}.
\newblock \bibinfo{journal}{\emph{IEEE Internet of Things Journal}} \bibinfo{volume}{7}, \bibinfo{number}{8} (\bibinfo{year}{2020}), \bibinfo{pages}{7506--7516}.
\newblock


\bibitem[Zhou et~al\mbox{.}(2018)]%
        {zhou2018patternlistener}
\bibfield{author}{\bibinfo{person}{Man Zhou}, \bibinfo{person}{Qian Wang}, \bibinfo{person}{Jingxiao Yang}, \bibinfo{person}{Qi Li}, \bibinfo{person}{Feng Xiao}, \bibinfo{person}{Zhibo Wang}, {and} \bibinfo{person}{Xiaofeng Chen}.} \bibinfo{year}{2018}\natexlab{}.
\newblock \showarticletitle{{PatternListener: Cracking Android Pattern Lock using Acoustic Signals}}. In \bibinfo{booktitle}{\emph{Proc. of the 25th ACM CCS}}. \bibinfo{pages}{1775--1787}.
\newblock


\end{thebibliography}

\end{document}